\documentclass[journal, draftclsnofoot, onecolumn]{IEEEtran}
\usepackage{amsmath,amsfonts}
\usepackage{algorithm}
\usepackage{algorithmic}
\usepackage{array}
\usepackage{cuted}
\usepackage[caption=false,font=normalsize,labelfont=sf,textfont=sf]{subfig}
\usepackage{hyperref}
\usepackage{textcomp}
\usepackage{stfloats}
\usepackage{url}
\usepackage{verbatim}
\usepackage{graphicx}
\usepackage{color}
\usepackage{cleveref}
\usepackage{multirow} 
\usepackage{booktabs} 
\usepackage{makecell} 
\renewcommand{\P}{\mathbb{P}}
\newcommand{\E}{\mathbb{E}}

\newcommand{\spt}{\text{supp}}
\newcommand{\R}{\mathbb{R}}

\newcommand{\e}{\mathrm{e}}
\newcommand{\iid}{\textit{i.i.d.}}
\newcommand{\eg}{\textit{e.g.}}
\newcommand{\ie}{\textit{i.e.}}

\newtheorem{theorem}{Theorem}[section]
\newtheorem{definition}{Definition}[section]

\newtheorem{lemma}{Lemma}[section]
\newtheorem{proposition}{Proposition}[section]
\newtheorem{corollary}{Corollary}[section]

\def\BibTeX{{\rm B\kern-.05em{\sc i\kern-.025em b}\kern-.08em
    T\kern-.1667em\lower.7ex\hbox{E}\kern-.125emX}}
\usepackage{balance}
\begin{document}
\title{Achieving the Fundamental Limit of Lossless Analog Compression via Polarization}
\author{Shuai Yuan, Liuquan Yao, Yuan Li, Huazi Zhang, Jun Wang, Wen Tong and Zhiming Ma}


\maketitle
\bgroup\def\thefootnote{}\footnote{Shuai Yuan, Liuquan Yao and Zhiming Ma are with Academy of Mathematics and Systems Science, CAS and University of Chinese academy and Sciences (email: yuanshuai2020@amss.ac.cn, yaoliuquan20@mails.ucas.ac.cn, mazm@amt.ac.cn).}\egroup
\bgroup\def\thefootnote{}\footnote{Yuan Li, Huazi Zhang, Jun Wang and Wen Tong are with Huawei Technologies Co. Ltd. (email: \{liyuan299, zhanghuazi, justin.wangjun, tongwen\}@huawei.com).}\egroup
\bgroup\def\thefootnote{}\footnote{This work was presented in part at the 2023 IEEE Global Communications Conference.}\egroup

\vspace{-4em}
\begin{abstract}
In this paper, we study the lossless analog compression for \iid\ nonsingular signals via the polarization-based framework. We prove that for nonsingular source, the error probability of maximum a posteriori (MAP) estimation polarizes under the Hadamard transform, which extends the polarization phenomenon to analog domain. Building on this insight, we propose partial Hadamard compression and develop the corresponding analog successive cancellation (SC) decoder. The proposed scheme consists of deterministic measurement matrices and non-iterative reconstruction algorithm, providing benefits in both space and computational complexity. Using the polarization of error probability, we prove that our approach achieves the information-theoretical limit for lossless analog compression developed by Wu and Verd\'{u}.
\end{abstract}


\section{Introduction}

\subsection{Related Works}
Lossless analog compression, developed by Wu and Verd\'{u} \cite{WV2010}, is related to several fields in signal processing \cite{WV2012,SRAB2017} and has drawn more attention recently \cite{ABDK+2019,GS2020}. Let the entries of a high-dimensional analog signal $\mathbf{X}\in\R^N$ be modeled as \iid\ random variables generated from the source $X\sim P_X$. In linear
compression, $\mathbf{X}$ is encoded into $ \mathbf{Z} = \mathsf{A}\mathbf{\mathbf{X}}$ where $\mathsf{A}\in\R^{M\times N}$ denotes the measurement matrix. Then the decompressed signal is obtained by $ \widehat{\mathbf{X}}= \varphi(\mathbf{Z})$ where $\varphi: \R^M\rightarrow \R^N$ stands for the reconstruction algorithm. For example, the noiseless compressed sensing falls into this framework by imposing particular prior $P_X$ to highlight the sparse property \cite{WV2010,WV2012}.

In \cite{WV2010}, Wu and Verd\'{u} established the fundamental limit for lossless analog compression. For a nonsingular source $X$, let $d(X)$ denote the R\'{e}nyi information dimension (RID) of $X$ (see Definition \ref{def-RID}). It was proved in \cite{WV2010} that for any $R > d(X)$, there exists a sequence of measurement matrices $\mathsf{A}_N$ and reconstruction algorithms $\varphi_N$ with $M = R N + o(N)$ such that the probability of precise recovery (\ie,\ $\widehat{\mathbf{X}}=\mathbf{X}$) approaches 1 as $N$ goes to infinity. Conversely, it is necessary to have at least $d(X)N+o(N)$ linear measurements to ensure a lossless recovery. However, the existence of $(\mathsf{A}_N,\varphi_N)$ is guaranteed by the random projection argument without efficient encoding-decoding algorithms. To address this problem, several schemes aiming to achieve the compression limit are proposed. Donoho \textit{et al.} \cite{DJM2013} showed that the limit $d(X)$ can be approached by spatial coupling and approximate message passing (AMP) algorithm. Jalali \textit{et al.} \cite{JP2017} proposed universal algorithms that are proved to be limit-achieving for almost lossless recovery. All of the above works consider random measurement matrices, which require larger storage compared with the deterministic ones.

Polar codes, invented by Ar\i kan \cite{Arikan2009}, are the first capacity-achieving binary error-correcting codes with explicit construction. As code length approaches infinity, subchannels in polar codes become either noiseless or pure-noise, and the fraction of the noiseless subchannels approaches channel capacity. This phenomenon is known as ``channel polarization''. Thanks to polarization, efficient successive cancellation (SC) decoding algorithm can be implemented with complexity of $O(N\log N)$. Polar codes are also generalized to finite fields with larger alphabet \cite{KT2010,MT2014,GYB2016}, and applied to lossless and lossy compression \cite{HKU2009,Arikan2010,KU2010}.

Over the analog domain, the polarization of entropy was studied in \cite{Arikan2021}, where the author pointed out that entropy may not polarize due to the non-uniform integrability issue over $\R$. In fact, the absorption phenomenon was shown in \cite{HAT2012} that the entropy vanishes eventually under the Hadamard transform if the source is discrete with finite support. Although the polarization of entropy might fail, it was proved in \cite{HA2013} that RID polarizes under the Hadamard transform for nonsingular source. Based on this fact, RID was utilized as a measure of compressibility in \cite{HA2017} to construct the partial Hadamard matrices for compressed sensing with Basis Pursuit decoding. Nevertheless, low RID does not imply high probability of exact recovery, because there are discrete distributions over $\R$ with extremely high entropy but the RID of which are 0. The relationship between RID and compressibility is still unclear. Li \textit{et al.} \cite{LMK2012} showed that the partial Hadamard matrices with low-RID rows achieve the compression limit under the model of noiseless compressed sensing considered in \cite{WV2012}. However, their reconstruction needs to exhaustively check all possible nonsingular combinations of the linear measurements, which is intractable. The SC decoding was briefly discussed in \cite{LMK2012} but the authors did not provide further analysis. The optimality of SC decoder for analog compression is still unknown.

\subsection{Contributions}
In this paper, we study the lossless analog compression via the polarization-based framework. We prove that for nonsingular source, the error probability of maximum a posteriori (MAP) estimation polarizes under the Hadamard transform. Specifically, let $\mathsf{H}_n$ denote the Hadamard matrix of order $n=\log N$ (see Section \ref{Probabilistic Model for Polarization under Hadamard Transform} for the definition) and $\mathbf{Y} = \mathsf{H}_n\mathbf{X}$. For each $k\in\{1,2,\dots,N\}$, denote $Y^{k-1} = [Y_1,\dots,Y_{k-1}]^\top$. Consider the MAP estimate of $Y_k$ based on $Y^{k-1}$, which is defined to be
\begin{equation}
Y_k^* = \mathop{\arg\max}\limits_{y\in\R} \P(Y_k = y|Y^{k-1}).
\end{equation}
Define $P_e^{\text{MAP}}(Y_k|Y^{k-1}) = \P(Y_k\neq Y^*_k)$ to be the error probability of the MAP estimation for $Y_k$ given $Y^{k-1}$. In this paper, we prove that for nonsingular source $X$ satisfying some regular conditions, $P_e^{\text{MAP}}(Y_k|Y^{k-1})$ approaches either 0 or 1 as $n$ goes to infinity, and the fraction of $Y_k$ with high $P_e^{\text{MAP}}(Y_k|Y^{k-1})$ approaches $d(X)$. The formal statement is presented in Theorem \ref{polarization-of-Qn}. Our result implies that by applying the Hadamard transform on \iid\ nonsingular source, the resulting distributions $P_{Y_k|Y^{k-1}}$ become either entirely deterministic or completely unpredictable as the dimension $N$ tends to infinity. It also signifies the polarization of compressibility over analog domain, since those $Y_k$ with smaller $P_e^{\text{MAP}}(Y_k|Y^{k-1})$ are more likely to be successfully recovered when the information of $Y^{k-1}$ is given.


Based on the polarization of error probability, we propose partial Hadamard matrices for compression and develop the corresponding analog SC decoding algorithm for reconstruction. Inspired by the polarization of $P_e^{\text{MAP}}(Y_k|Y^{k-1})$, the proposed approach entails a sequential recovery of $\mathbf{Y}$ rather than a direct estimation of $\mathbf{X}$. Once $\widehat{\mathbf{Y}}$ is obtained, the estimated signal is given by $\widehat{\mathbf{X}} = \mathsf{H}_n^{-1}\widehat{\mathbf{Y}}$. The linear measurements are selected as the rows of Hadamard matrices corresponding to high error probability. In other words, those $Y_k$ with high $P_e^{\text{MAP}}(Y_k|Y^{k-1})$ are observed, whereas those with vanishing $P_e^{\text{MAP}}(Y_k|Y^{k-1})$ are discarded. Note that the discarded $Y_k$ are nearly deterministic given the previous entries, suggesting that they can be accurately recovered through a sequential decoding scheme. Consequently, we rebuild $\mathbf{Y}$ by sequential MAP estimation of the discarded $Y_k$ based on the conditional distribution $P_{Y_k|\widehat{Y}^{k-1}}$, which is analogous to the SC decoder for binary polar codes. Thanks to the recursive nature of Hadamard transform, this SC decoding scheme can be implemented with complexity of $O(N\log N)$. Since the Hadamard matrices can be explicitly constructed and the SC decoding is non-iterative, the proposed scheme has advantages in both space and computational complexity. Compared to RID, the error probability of MAP estimation exhibits a more explicit correlation with compressibility. Therefore, the proposed method for constructing the measurement matrix is more reasonable than that in \cite{HA2017}. Through an elaborate analysis of the polarization speed, we prove that the proposed scheme achieves the information-theoretical limit for lossless analog compression established in \cite{WV2010}.

The analysis of polarization over finite fields cannot be directly applied to the analog case due to the fundamental difference between real number field and finite fields. The technical challenges of evaluating analog polarization lies in two aspects. Firstly, it is unclear how to quantify the uncertainty of general random variables over $\R$, since Shannon's entropy is only defined for discrete or continuous random variables. Secondly, even for discrete distributions, the entropy process lacks a clear recursive formula and is neither bounded nor uniformly integrable \cite{Arikan2021}, leading to difficulties in determining the rate of polarization. To address these challenges, we introduce the concept of weighted discrete entropy (see Definition \ref{weighted-discrete-entropy}) to characterize the uncertainty contributed by the discrete component of nonsingular distributions. We show that the weighted discrete entropy vanishes under the Hadamard transform for continuous-discrete-mixed source, which generalizes the absorption of entropy for purely discrete source \cite{HAT2012}. To obtain the polarization rate, we develop martingale methods with stopping time to address the issue of unboundedness, and introduce a novel variant of entropy power inequality (EPI) to establish a recursive relationship for the entropy process. These analyses allow us to obtain the convergence rate for the weighted discrete entropy process.

Our contributions are summarized as follows:
\begin{itemize}
\item We prove that the error probability of MAP estimation polarizes under the Hadamard transform, which extends the polarization phenomenon to analog domain.
\item We propose the partial Hadamard matrices and analog SC decoder for analog compression, and prove that the proposed method achieves the fundamental limit for lossless analog compression.
\item We develop new technical approaches to analyze the polarization over $\R$.
\end{itemize}


\subsection{Notations and Paper Outline}
Random variables are denoted by capital letters such as $X$, and the particular realizations are denoted by lowercase letters such as $x$. $[N]$ denotes the set $\{1,2,\dots,N\}$. We use $x^N$ to denote the $N$-dimensional vector $[x_1,\dots,x_N]^\top$. If the dimension is clear based on the context, we use the boldface letter to represent vectors, such as $\mathbf{x}=x^N$. We further abbreviate $[x_i,x_{i+1},\dots,x_j]^\top$ as $x_i^j$, and $[x_i:i\in\mathcal{A}]^\top$ as $x_\mathcal{A}$ for an index set $\mathcal{A}$.

For a random pair $(U,V)$, we write $\langle U|V\rangle$ to represent the conditional distribution $P_{U|V}$. When the particular realization $v$ is given, we denote $\langle U|V=v\rangle = P_{U|V=v}$. In particular, for a random variable $X$, we denote $\langle X\rangle = P_X$. For a functional $F(\cdot)$ that takes a probability distribution $\mu$ as input, such as the discrete entropy $H(\cdot)$ or the differential entropy $h(\cdot)$, we refer to $F(\mu)$ and $F(X)$ interchangeably if $X\sim \mu$. We also follow the convention that $F(U|V=v) = F(P_{U|V=v})$, in which we treat $F(U|V=v)$ as a function of $v$ and write $\E_V[F(U|V=v)]$ to represent the expectation of $F(U|V=v)$ under the distribution $P_V$.

All logarithms are base 2 throughout this paper. The binary entropy function is defined as $h_2(x) = -x\log x - (1-x)\log(1-x)$, and $h_2^{-1}(y)$ stands for the unique solution of $h_2(x)=y$ over $x\in[0,1/2]$. In addition, $\spt(D)$ denotes the support of the discrete random variable $D$, which is defined as $\spt(D) := \{x\in\R: \P(D=x)>0\}$. The cardinality of a set $\mathcal{A}$ is denoted by $|\mathcal{A}|$. Furthermore, we denote $x\vee y = \max\{x,y\}$ and $x\wedge y = \min\{x,y\}$. The indicator function of an event $A$, denoted as $\mathbf{1}_A$, equals $1$ if $A$ is true and 0 otherwise. Lastly, the dirac measure at point $x$ is denoted by $\delta_x$, and $\mathcal{N}(0,1)$ stands for the standard Gaussian distribution.

We use the standard Bachmann-Landau notations. Specifically, $a_n = o(b_n)$ if $\lim_{n\rightarrow\infty}a_n/b_n = 0$; $a_n = \omega(b_n)$ if $b_n = o(a_n)$; $a_n= O(b_n)$ if $\limsup_{n\rightarrow\infty}a_n/b_n<\infty$;  $a_n = \Theta(b_n)$ if $a_n = O(b_n)$ and $b_n = O(a_n)$.

The remaining sections of this paper are organized as follows. Section \ref{Preliminaries} provides the necessary preliminaries. In Section \ref{Probabilistic Model for Polarization under Hadamard Transform}, we show the polarization of RID and the absorption of weighted discrete entropy, based on which we prove the polarization of error probability for MAP estimation. In Section \ref{The Proposed Encoding-Decoding Scheme}, we propose the partial Hadamard compression and analog SC decoder, and discuss its connections to binary polar codes. Section \ref{Basic Hadamard Transform of Nonsingular Distributions} examines the evolution of nonsingular distributions under the basic Hadamard transform. The proof of the absorption of weighted discrete entropy is presented in section \ref{Proof of absorption of hat Hn}, which contains the most technical portion of this paper. We demonstrate the numerical experiments in Section \ref{Numerical Experiments} and conclude this paper in Section \ref{Conclusion}. The proofs for some technical propositions and lemmas are given in appendices.

\section{Preliminaries} \label{Preliminaries}
\subsection{Binary Source Coding via Polarization}\label{SCvP}
\begin{figure}[htbp]
\centerline{\includegraphics[width=0.4\textwidth,trim = 297 30 523 37, clip]{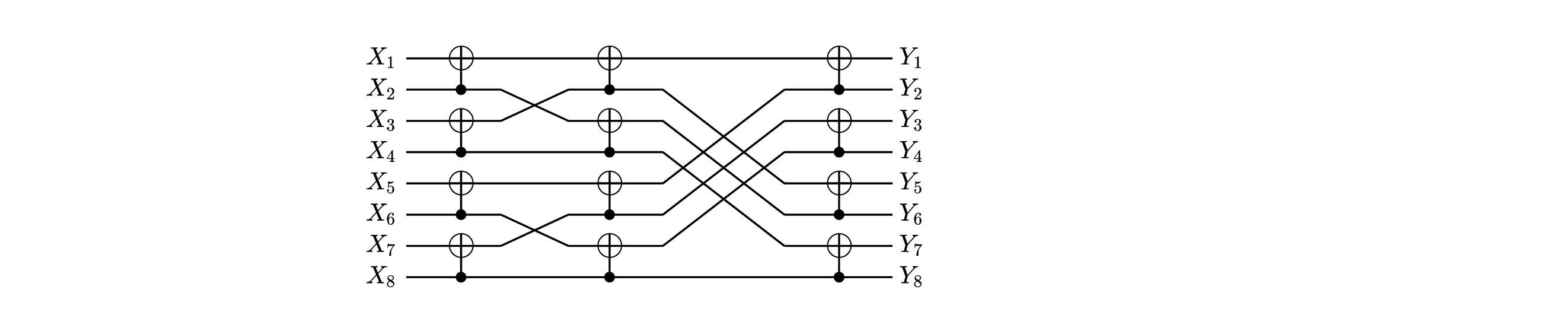}}
\caption{Polar transform for $N=8$.}
\label{PT}
\end{figure}

In this subsection we briefly review the polarization framework for binary source coding \cite{Arikan2010}. Let $\mathbf{X}=[X_1,\dots,X_N]^\top\in\mathbb{F}_2^N$, where $\{X_i\}_{i=1}^N\overset{\iid}{\sim}X$. Denote the polar transform by
\begin{equation}
\mathsf{G}_n =  \mathsf{B}_n\begin{bmatrix}1 & 1\\0 & 1\end{bmatrix}^{\otimes n}\in\mathbb{F}_2^{N \times N},
\end{equation}
where $n = \log N$, $\otimes$ denotes the Kronecker product and $\mathsf{B}_n$ is the bit-reversal permutation matrix of order $n$ \cite{Arikan2009}. Let $\mathbf{Y} = \mathsf{G}_n\mathbf{X}$, where all operations are performed over $\mathbb{F}_2$. The polar transform for $N=8$ is illustrated in Fig. \ref{PT}, where $\oplus$ denotes the sum over $\mathbb{F}_2$.

It was shown in \cite{Arikan2010} that for any $\beta\in(0,1/2)$, the conditional entropy $H(Y_k|Y^{k-1})$ polarizes in the sense that
\begin{equation}
\begin{aligned}
&\lim\limits_{n\rightarrow\infty}\frac{|\{k\in[N]:H(Y_k|Y^{k-1})>1-2^{-2^{\beta n}}\}|}{N} = H(X),\\
&\lim\limits_{n\rightarrow\infty}\frac{|\{k\in[N]:H(Y_k|Y^{k-1})<2^{-2^{\beta n}}\}|}{N} = 1-H(X).
\end{aligned}
\end{equation}
This implies that $H(Y_k|Y^{k-1})$ approaches either 0 or 1 as $n$ tends to infinity. Let $\mathcal{A}$ be the set containing all indices $k\in[N]$ for which the conditional entropy $H(Y_k|Y^{k-1})$ is close to 1, then the compressed signal is given by $\mathbf{z} = y_\mathcal{A}$. The original signal is recovered using an SC decoding scheme that sequentially reconstructs $y_k$. If $k\in\mathcal{A}$, the true value of $y_k$ is known, and thus we set $\hat{y}_k = y_k$. When $k\in\mathcal{A}^c$, an MAP estimator based on $\hat{y}^{k-1}$ is utilized to recover $y_k$ . Specifically, we set
\begin{equation}\label{MAP_in_binary_SC}
\hat{y}_k = \mathop{\arg\max}\limits_{y\in\{0,1\}} \P(Y_k = y|Y^{k-1} = \hat{y}^{k-1}),\text{ if }k\in\mathcal{A}^c.
\end{equation}
Define the likelihood ratio (LR) of $Y_k$ given $Y^{k-1} = y^{k-1}$ by
\begin{equation}
L_n^{(k)}(y^{k-1}) = \frac{\P(Y_k=0|Y^{k-1} = y^{k-1})}{\P(Y_k=1|Y^{k-1} = y^{k-1})},
\end{equation}
then \eqref{MAP_in_binary_SC} is equivalent to $\hat{y}_k = \mathbf{1}_{\{L_n^{(k)}\{\hat{y}^{k-1})<1\}}$. According to the recursive structure of the polar transform, $L_n^{(k)}(y^{k-1})$ satisfies the following formulas:
\begin{align}
&L_n^{(2i-1)}(y^{2i-2}) = \frac{L_{n-1}^{(i)}(y^{2i-2}_o\oplus y^{2i-2}_e)L_{n-1}^{(i)}( y^{2i-2}_e)+1}{ L_{n-1}^{(i)}(y^{2i-2}_o\oplus y^{2i-2}_e)+L_{n-1}^{(i)}( y^{2i-2}_e)},  \label{f_SC_binary}\\
&L_n^{(2i)}(y^{2i-1}) = L_{n-1}^{(i)}(y^{2i-2}_o\oplus y^{2i-2}_e)^{(-1)^{y_{2i-1}}}L_{n-1}^{(i)}( y^{2i-2}_e),  \label{g_SC_binary}
\end{align}
where $y^{2i-2}_o$ and $y^{2i-2}_e$ stand for the subvectors of $y^{2i-2}$ with odd and even indices, respectively. The initial condition is given by $L_0^{(1)} = \P(X=0)/\P(X=1)$. The recursive formulas \eqref{f_SC_binary} and \eqref{g_SC_binary}, which comprise the basic operations of binary SC decoder, characterize the evolution of LR under the basic polar transform $\mathsf{G}_1$ (refer to \cite{Arikan2009} for the details). Since a probability distribution over $\mathbb{F}_2$ can be represented by a single parameter, \eqref{f_SC_binary} and \eqref{g_SC_binary} are sufficient to track the evolution of the conditional distribution $\langle Y_k|Y^{k-1}\rangle$ under the polar transform. Thanks to the recursive nature of $\mathsf{G}_n$, the complexity of SC decoding scheme is $O(N\log N)$.

Due to the entropy polarization, $\mathcal{A}^c$ consists of the indices $k$ such that $H(Y_k|Y^{k-1})$ is close to 0. This property guarantees that the error probability of the binary SC decoder can be reduced to an arbitrarily small value. Furthermore, as $n$ tends to infinity, the fraction of high-entropy indices approaches $H(X)$. Consequently, this polarization-based scheme achieves the information-theoretical limit for lossless source coding.

\subsection{Nonsingular Distribution}\label{NRV}
Let $\mu$ be a probability measure over $\R$. By Lebesgue decomposition theorem \cite{Halmos2013}, $\mu$ can be expressed as
\begin{equation}\label{Lebesgue-decomposition}
\mu = \alpha_c \mu_c + \alpha_d\mu_d + \alpha_s\mu_s,
\end{equation}
where $\mu_c$ is an absolutely continuous measure with respect to (w.r.t.) the Lebesgue measure, $\mu_d$ is a discrete measure, $\mu_s$ is a singular measure, $\alpha_c,\alpha_d,\alpha_s\geq0$ and $\alpha_c+\alpha_d+\alpha_s=1$. We say $\mu$ is \textit{nonsingular} if it has no singular component, \ie, $\alpha_s = 0$. For example, the Bernoulli-Gaussian distribution $(1-\rho) \delta_0 + \rho\mathcal{N}(0,1)$ is nonsingular, which has been widely exploited to model the sparse signal in compressed sensing \cite{WV2012,KMSS+2012,VS2011}. We say a random variable $X$ is nonsingular if its distribution $P_X$ is nonsingular. In addition, we say a conditional distribution $\langle U|V\rangle$ is nonsingular if $\langle U|V=v\rangle$ is nonsingular $P_V$-$a.s.$ Similarly, we say $\langle U|V\rangle$ is discrete (continuous) if $\langle U|V=v \rangle$ is discrete (continuous) $P_V$-$a.s.$

Obviously, a nonsingular distribution $\mu$ is continuous-discrete-mixed, and vice versa. Along this paper, the discrete and continuous component of distributions are indicated by the subscript $d$ and $c$, respectively, such as $\mu_d$ and $\mu_c$. In particular, for a nonsingular conditional distribution $\langle U|V\rangle$, we denote by $\langle U|V=v\rangle_c$ and $\langle U|V=v\rangle_d$ the continuous and discrete component of $\langle U|V=v\rangle$, respectively. We also define the mixed representation of $\langle U|V\rangle$ as follows.
\begin{definition}[Mixed Representation]\label{Mixed---representation}
Let $\langle U|V\rangle$ be a nonsingular conditional distribution with
\begin{equation}
\langle U|V=v\rangle = \alpha_v\langle U|V=v\rangle_c + (1-\alpha_v)\langle U|V=v\rangle_d,P_V\text{-}a.s.,
\end{equation}
where $\alpha_v\in[0,1]$. The mixed representation of $\langle U|V\rangle$ is defined to be a random triple $(\Gamma,C,D)$ such that $\Gamma,C,D$ are conditionally independent given $V$ and
\begin{equation}
\begin{aligned}
\langle\Gamma|V=v\rangle \overset{d}{=} \text{Bernoulli}(\alpha_v),\ \langle C|V=v\rangle = \langle U|V=v\rangle_c,\ \langle D|V=v\rangle = \langle U|V=v\rangle_d,
\end{aligned}
\end{equation}
where ``$\overset{d}{=}$'' means ``equals in distribution''.
\end{definition}
\textbf{Remark: }If $(\Gamma,C,D)$ is the mixed representation of $\langle U|V\rangle$, then $\langle U|V=v\rangle = \langle \Gamma C+(1-\Gamma)D|V=v\rangle$, $P_V$-$a.s.$



\subsection{R\'{e}nyi Information Dimension}\label{Sec RID}
\begin{definition}[RID \cite{Renyi1959}]\label{def-RID}
Let $X$ be a real-valued random variable, the R\'{e}nyi information dimension (RID) of $X$ is defined to be
\begin{equation}\label{RID}
d(X) := \lim\limits_{n\rightarrow\infty} \frac{H\left(\lfloor nX\rfloor/n\right)}{\log n},
\end{equation}
provided the limit exists, where $\lfloor x \rfloor$ stands for the floor function of $x$.
\end{definition}

Note that $\lfloor nX\rfloor/n$ is the quantization of $X$ with resolution $1/n$, thus RID characterizes the growth rate of discrete entropy w.r.t. ever finer quantization.  For a nonsingular $X$ with distribution $P_X = \alpha\langle X\rangle_c+(1-\alpha)\langle X\rangle_d$, it was proved in \cite{Renyi1959} that $d(X) = \alpha$ if $H(\lfloor X\rfloor)<\infty$. This provides another interpretation of $d(X)$ as the weight of the continuous component of $X$. For more properties of RID, please refer to \cite{WV2010}.

For a conditional distribution $\langle U|V\rangle$, its conditional RID is defined in \cite{HA2013} as
\begin{equation}
d(U|V) := \lim\limits_{n\rightarrow\infty} \frac{H(\lfloor nU\rfloor/n | V)}{\log n},
\end{equation}
provided the limit exists. The following proposition shows that for nonsingular $\langle U|V\rangle$ satisfying mild conditions, $d(U|V)$ is equal to the average of $d(U|V=v)$.
\begin{proposition}\label{condiRID=average}
Let $\langle U|V\rangle$ be a nonsingular conditional distribution with $\E U^2 < \infty$, then
\begin{equation}
d(U|V) = \E_{V}\left[d(U|V=v)\right].
\end{equation}
\end{proposition}
\begin{IEEEproof}
See Appendix \ref{appendix A-A}.
\end{IEEEproof}
\textbf{Remark: }Let $(\Gamma,C,D)$ be the mixed representation of $\langle U|V\rangle$, then $d(U|V=v) = \P(\Gamma=1|V=v)$. If we further have $\E U^2<\infty$, then Proposition \ref{condiRID=average} implies $d(U|V) = \P(\Gamma=1)$.
\subsection{Lossless Analog Compression}
Let $X$ be a real-valued random variable and $\{X_i\}_{i=1}^\infty\overset{\iid}{\sim}X$. Define $\mathbf{X} = X^N\in\R^N$ to be the $N$-dimensional random vector representing the signal to be compressed. In linear compression, $\mathbf{X}$ is encoded by a matrix $\mathsf{A}_N\in\R^{M\times N}$ with $M<N$, then it is recovered through a decoder represented by a measurable map $\varphi_N: \R^M\rightarrow\R^N$. The aim is to design an efficient encoder-decoder pair $(\mathsf{A}_N,\varphi_N)$ with the goal of minimizing the distortion between the original signal $\mathbf{X}$ and the reconstructed signal $\varphi_N(\mathsf{A}_N\mathbf{X})$.


In \cite{WV2010}, Wu and Verd\'{u} established the fundamental limit of lossless analog compression. Let $R = M/N$ be the measurement rate and define the error probability to be
\begin{equation}
P_e(\mathsf{A}_N,\varphi_N) := \P(\varphi_N(\mathsf{A}_N\mathbf{X})\neq \mathbf{X}).
\end{equation}
For any $\epsilon>0$, define the $\epsilon$-achievable rate $R^*(\epsilon)$ to be the lowest measurement rate $R$ such that there exists a sequence of encoder-decoder pairs $(\mathsf{A}_N,\varphi_N)$ (might rely on $P_X$) with rate $R$ and $P_e(\mathsf{A}_N,\varphi_N) < \epsilon$ for sufficiently large $N$. The breakthrough work by Wu and Verd\'{u} showed that if $X$ is nonsingular, then $R^*(\epsilon) = d(X)$ for all $\epsilon >0$. In other words, RID is the fundamental limit of lossless analog compression. However, the existence of $(\mathsf{A}_N,\varphi_N)$ is guaranteed by the random projection argument without explicit construction, which leads to random measurement matrices and high-complexity decoder. Therefore, it is still necessary to design deterministic encoders with effective decoding schemes.

\subsection{Maximum a Posteriori Estimation}\label{MAPE}
\begin{definition}[MAP Estimate and Error Probability]\label{MAP-estimate}
Let $( U,V)$ be a random pair. The maximum a posteriori (MAP) estimate of $U$ given $V=v$ is defined to be
\begin{equation}\label{MAP-eq-eq-1}
U^*(v) := \mathop{\arg\max}\limits_{u\in\R} \P(U=u|V=v).
\end{equation}
The error probability of the MAP estimation for $U$ given $V=v$ is defined as
\begin{equation}
P^{\text{MAP}}_e(U|V=v) := \P(U\neq U^*(v)|V=v).
\end{equation}
The average error probability is defined to be
\begin{equation}
P^{\text{MAP}}_e(U|V) := \E_V[P^{\text{MAP}}_e(U|V=v)].
\end{equation}
\end{definition}
\textbf{Remark 1: }The error probability $P^{\text{MAP}}_e(U|V)$ only relies on the conditional distribution $\langle U|V\rangle$, hence we interpret $P^{\text{MAP}}_e(\cdot)$ as a functional of conditional distributions.\\
\textbf{Remark 2: }For continuous $\langle U|V\rangle$, $P(U=u|V=v) \equiv 0$ since continuous distribution assigns 0 probability at any single point. As a result, our MAP estimation is not well-defined for continuous conditional distributions. In such cases, we consider the MAP estimation fails as it is impossible to precisely reconstruct a continuous random variable. Note that this is different from the Bayesian statistics, in which the MAP estimation for continuous random variables is well-defined as the maximizer of probability density function.

Let $\langle U|V\rangle$ be a nonsingular conditional distribution with mixed representation $(\Gamma,C,D)$. According to the remark below Definition \ref{Mixed---representation}, for any $u\in\R$ and realization $v$ we have
\begin{equation}
\begin{aligned}
\P(U=u|V=v) &= \P(\Gamma=1|V=v)\P(C=u|V=v)+\P(\Gamma=0|V=v)\P(D=u|V=v)\\
&=\P(\Gamma=0|V=v)\P(D=u|V=v).
\end{aligned}
\end{equation}
This implies $U^*(v) = D^*(v),\forall v$. Consequently, we can write the error probability $P^{\text{MAP}}_e(U|V=v)$ as
\begin{equation}
\begin{aligned}
P^{\text{MAP}}_e(U|V=v) = \P(\Gamma=1|V=v)\P(C\neq U^*(v)|V=v)+\P(\Gamma=0|V=v)\P(D\neq D^*(v)|V=v).
\end{aligned}
\end{equation}
Since $\P(C\neq U^*(v)|V=v)=1$ and $\P(D\neq D^*(v)|V=v)=P^{\text{MAP}}_e(\langle U|V=v\rangle_d)$, we obtain
\begin{equation}
\begin{aligned}
P^{\text{MAP}}_e(U|V=v)=\P(\Gamma=1|V=v)+\P(\Gamma=0|V=v)P^{\text{MAP}}_e(\langle U|V=v\rangle_d).
\end{aligned}
\end{equation}
Note that $\P(\Gamma=1|V=v) = d(U|V=v)$. If we further assume $\E U^2<\infty$, then Proposition \ref{condiRID=average} implies that
\begin{equation}\label{Pe-MAP-upper-bound}
\begin{aligned}
P^{\text{MAP}}_e(U|V) =  d(U|V) + \E_V[(1-d(U|V=v))P^{\text{MAP}}_e(\langle U|V=v\rangle_d)].
\end{aligned}
\end{equation}
This equation suggests that $P^{\text{MAP}}_e(U|V)$ can be written as a combination of the error probabilities associated with its discrete and continuous components. Since it is impossible to precisely reconstruct a continuous random variable, the error probability associated with $\langle U|V\rangle_c$ is given by its average weight $d(U|V)$. The error probability contributed by $\langle U|V\rangle_d$ is equal to the average of $P_e^{\text{MAP}}(\langle U|V=v\rangle_d)$ weighted by $1 - d(U|V=v)$, which is represented by the second term on the right side of \eqref{Pe-MAP-upper-bound}.
\subsection{Weighted Discrete Entropy}
The Shannon's entropy $H(\cdot)$ is only specified for discrete random variables. As a generalization, we extend this concept to define the weighted discrete entropy for general nonsingular random variables.
\begin{definition}[Weighted Discrete Entropy]\label{weighted-discrete-entropy}
Let $X$ be a nonsingular random variable. The weighted discrete entropy of $X$ is defined to be
\begin{equation}
\widehat{H}(X) := (1-d(X))H(\langle X\rangle_d).
\end{equation}
The conditional weighted discrete entropy of nonsingular $\langle U|V\rangle$ is defined to be
\begin{equation}
\widehat{H}(U|V) := \E_{V}[\widehat{H}(U|V=v)].
\end{equation}
\end{definition}
\textbf{Remark: }$\widehat{H}(X)$ is equal to the entropy of $\langle X\rangle_d$ weighted by $1-d(X)$, which explains the name ``weighted discrete entropy". If $X$ is purely discrete, then $\widehat{H}(X) = H(X)$. Note that a small value of $\widehat{H}(X)$ indicates either $d(X)$ is close to 1 or $H(\langle X\rangle_d)$ is low. In both cases, the uncertainty of $X$ is barely influenced by its discrete component. Therefore, we can interpret $\widehat{H}(X)$ as a measure that quantifies the uncertainty of $X$ contributed by its discrete component.

The weighted discrete entropy has a close connection to the error probability of MAP estimation. We first focus on the purely discrete case. The subsequent proposition shows that for discrete random variables, the error probability can be bounded by its entropy.
\begin{proposition}\label{upper bound of Pe MAX}
Let $X$ be a discrete random variable. Denote $x^* = \arg\max_{x\in\R}\P(X=x)$. If $H(X)\leq 1$, then
\begin{equation}
\P(X\neq x^*) \leq h_2^{-1}(H(X)) \leq  H(X).
\end{equation}
Therefore, $P_e^{\text{MAP}}(X)\leq H(X)$ for any discrete random variable $X$.
\end{proposition}
\begin{IEEEproof}
See Appendix \ref{appendix A-B}.
\end{IEEEproof}
For nonsingular conditional distribution $\langle U|V\rangle$, from Proposition \ref{upper bound of Pe MAX} we know that
\begin{equation}\label{discrete-bounded-by-Hhat}
\begin{aligned}
\E_V[(1-d(U|V=v))P^{\text{MAP}}_e(\langle U|V=v\rangle_d)]\leq\E_V[(1-d(U|V=v))H(\langle U|V=v\rangle_d)]=\widehat{H}(U|V).
\end{aligned}
\end{equation}
Combining \eqref{Pe-MAP-upper-bound} and \eqref{discrete-bounded-by-Hhat}, we can easily obtain the following bounds on the error probability of MAP estimation.
\begin{proposition}\label{bounds on error probability}
Let $\langle U|V\rangle$ be a nonsingular conditional distribution with $\E U^2<\infty$, then
\begin{equation}
d(U|V) \leq P_e^{\text{MAP}}(U|V) \leq d(U|V) + \widehat{H}(U|V).
\end{equation}
\end{proposition}
\section{Polarization of Error Probability for MAP Estimation}\label{Probabilistic Model for Polarization under Hadamard Transform}
Let $X\in\R$ be a nonsingular random variable, and $\{X_i\}_{i=1}^\infty$ be a sequence of \iid\ random variables with distribution $P_X$. Denote by $\mathbf{X}= X^N$ the $N$-dimensional random vector. In the rest of this paper, we always assume $N=2^n$ for some integer $n$. The Hadamard matrix of order $n$ is defined as
\begin{equation}
\mathsf{H}_n := \mathsf{B}_n\left(\frac{1}{\sqrt{2}}\begin{bmatrix}1 & 1 \\ 1 & -1  \end{bmatrix}\right)^{\otimes n}\in\R^{N \times N},
\end{equation}
where $\mathsf{B}_n$ denotes the bit-reversal permutation matrix of order $n$. Let $\mathbf{Y} = \mathsf{H}_n \mathbf{X}$. The aim of this section is to show the polarization of $P^{\text{MAP}}_e(Y_k|Y^{k-1})$ as in the following theorem.
\begin{theorem}[Polarization of Error Probability]\label{polarization-of-Qn}
Suppose the source $X$ is nonsingular and satisfies
\begin{enumerate}
\item $\E X^2 < \infty$.
\item $|\spt(\langle X\rangle_d)|<\infty$.
\item $h(\langle X\rangle_c)<\infty$, $J(\langle X\rangle_c) < \infty$, where $J(\cdot)$ denotes the Fisher information (see Section \ref{PAWDE-5} for the definition).
\end{enumerate}
For any $\lambda>0$ and $\beta \in(0,1/2)$, define
\begin{equation}
\mathcal{D}_n = \{k\in[N]: P^{\text{MAP}}_e(Y_k|Y^{k-1}) \leq 2^{-\lambda n}\},\ \ \mathcal{C}_n = \{k\in[N]: P^{\text{MAP}}_e(Y_k|Y^{k-1}) \geq 1-2^{-2^{\beta n}}\}.
\end{equation}
Then
\begin{align}
\lim\limits_{n\rightarrow\infty} \frac{|\mathcal{D}_n|}{2^n}= 1 - d(X),\ \lim\limits_{n\rightarrow\infty} \frac{|\mathcal{C}_n|}{2^n}=  d(X).
\end{align}
\end{theorem}

\begin{figure}[h]
\centerline{\includegraphics[width=0.5\textwidth,trim = 105 10 105 40, clip]{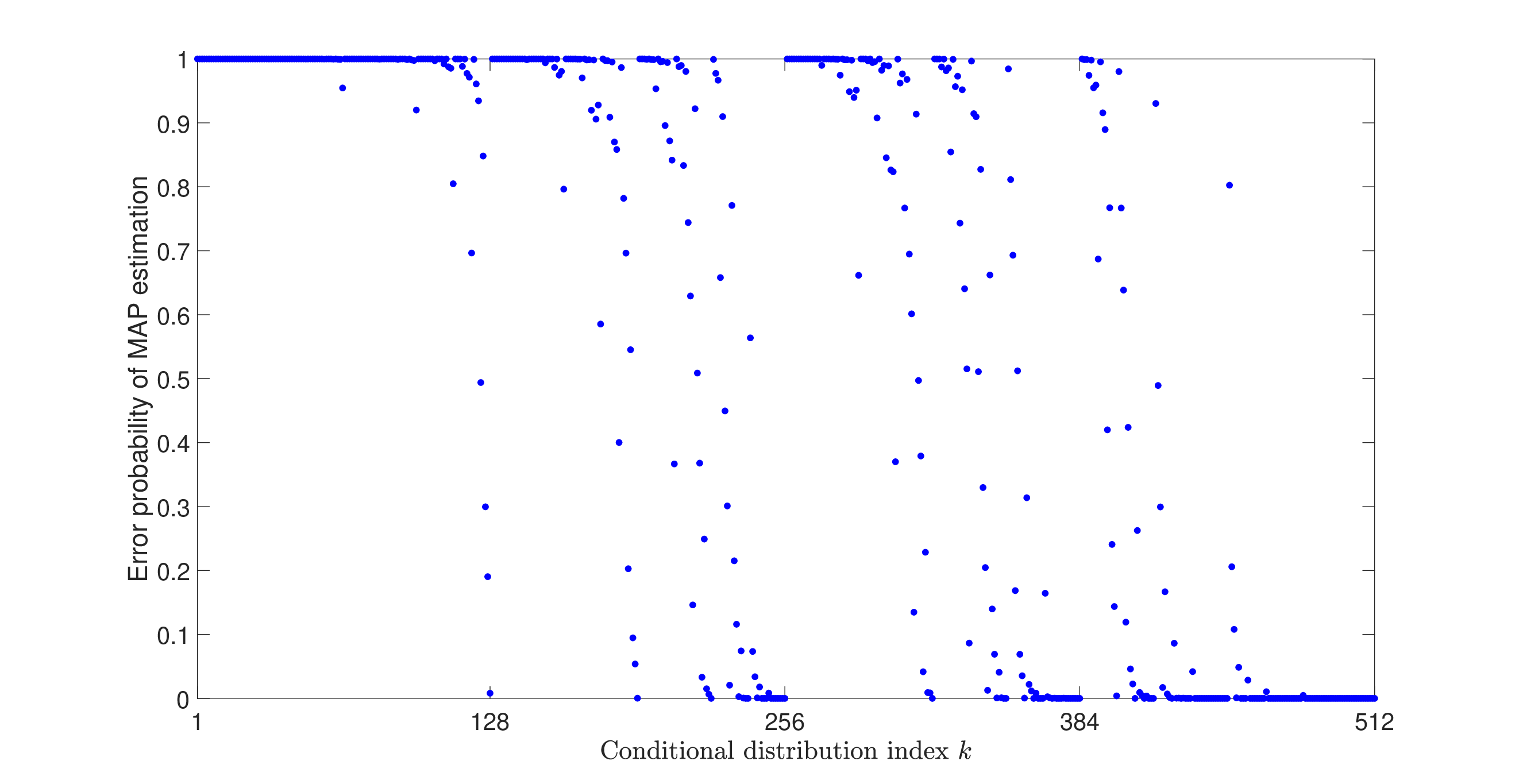}}
\caption{Plot of $P^{\text{MAP}}_e(Y_k|Y^{k-1})$ versus $k=1,2,\dots,2^9$ with $P_X = 0.5\delta_0+0.5\mathcal{N}(0,1)$.}
\label{polar-Pe-fig}
\end{figure}

Theorem \ref{polarization-of-Qn} sheds light on the compressibility of $\mathbf{Y}$. Specifically, it implies that the conditional distribution $\langle Y_k|Y^{k-1}\rangle $ becomes either completely deterministic or unpredictable. As a result, not much information is lost if we discard those $Y_k$ with $k\in\mathcal{D}_n$. Similar principles also exist in the polar codes used for source coding, where the high-entropy positions are retained to preserve information, while the low-entropy positions are discarded. The polarization of error probability with $P_X = 0.5\delta_0 + 0.5\mathcal{N}(0,1)$ for $n=9$ is demonstrated in Fig. \ref{polar-Pe-fig}.

In the upcoming subsections, we first introduce the stochastic process of conditional distribution in Section \ref{PMAPHT-1} to depict the evolution of $\langle Y_k|Y^{k-1}\rangle$. Based on this, we show the polarization of RID and the absorption of weighted discrete entropy in Section \ref{PMAPHT-2} and \ref{PMAPHT-3}, respectively. In Section \ref{PMAPHT-4}, we present the proof of Theorem \ref{polarization-of-Qn}.
\subsection{Tree-like Evolution of Conditional Distributions}\label{PMAPHT-1}
Similar to the binary polar codes \cite{Arikan2009}, we define the tree-like process to track the evolution of conditional distributions under the Hadamard transform. We first define the upper and lower Hadamard transform of conditional distributions as follows.
\begin{definition}\label{upper-and-lower-Hadamard-conditional-distribution}
Given a conditional distribution $\langle U|V\rangle $, let $(U',V')$ be an independent copy of $(U,V) $. The upper Hadamard transform of $\langle U|V\rangle $ is defined to be
\begin{equation}
\langle U|V\rangle ^{0} := \left\langle \frac{U+U'}{\sqrt{2}}\bigg|V,V'\right\rangle ,
\end{equation}
and the lower Hadamard transform of $\langle U|V\rangle$ is defined to be
\begin{equation}
\langle U|V\rangle ^{1} := \left\langle \frac{U-U'}{\sqrt{2}}\bigg|\frac{U+U'}{\sqrt{2}},V,V'\right\rangle .
\end{equation}
\end{definition}

In Definition \ref{upper-and-lower-Hadamard-conditional-distribution}, we use superscript $0$ and $1$ to represent the upper and lower Hadamard transform, respectively. Given a binary sequence $b_1b_2\cdots b_n$ and a conditional distribution $\langle U|V\rangle $, we recursively define
\begin{equation}
\langle U|V\rangle ^{b_1\cdots b_n}:=\left(\langle U|V\rangle ^{b_1\cdots b_{n-1}}\right)^{b_n}.
\end{equation}
For example, $\langle U|V\rangle ^{010}$ is obtained by successively applying upper, lower and upper Hadamard transform on $\langle U|V\rangle $.

\begin{figure}[htbp]
\centerline{\includegraphics[width=0.3\textwidth, trim = 109 52 91 31 ,clip]{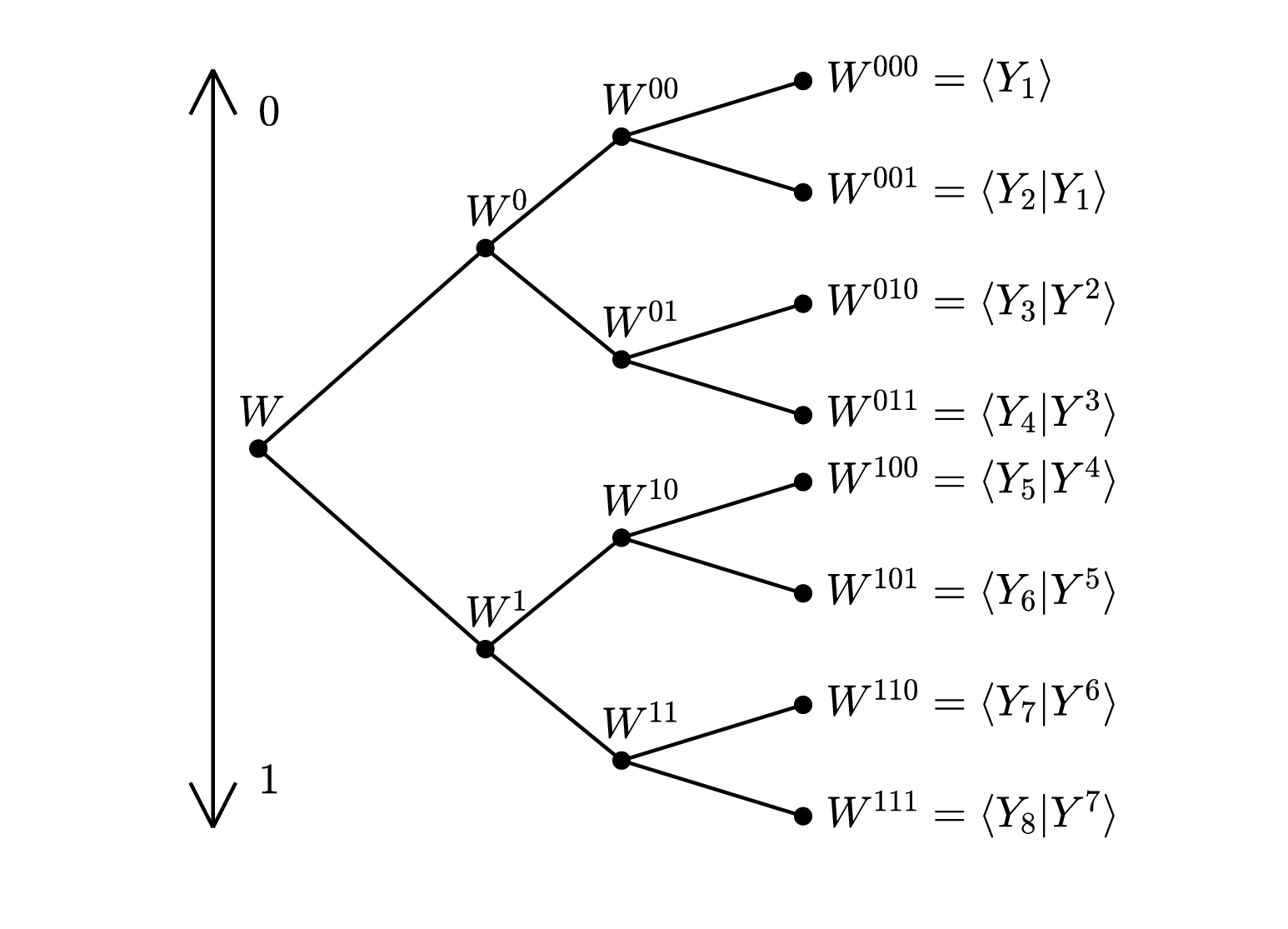}}
\caption{The tree-like evolution of conditional distributions for $N = 8$.}
\label{TP}
\end{figure}

The upper and lower Hadamard transform represents the evolution of conditional distributions under the basic transform $\mathsf{G}_1$. In fact, let $[Y_1,Y_2]^\top = \mathsf{G}_1[X_1,X_2]^\top$ then $\langle Y_1\rangle = \langle X\rangle^0$ and $\langle Y_2|Y_1\rangle = \langle X\rangle^1$. This can be easily extended using the recursive structure of Hadamard matrices. For each $k\in[N]$, let $\theta_n(k) = b_1b_2\cdots b_n$ be the binary expansion of $k-1$, \ie,\ $k = 1+\sum_{i=1}^n b_i2^{n-i}$. We have
\begin{equation}\label{distribution evolution}
\langle Y_k|Y^{k-1}\rangle  = \langle X\rangle ^{\theta_n(k)},\  \forall k \in[N].
\end{equation}
It is more intuitive to represent \eqref{distribution evolution} as a binary tree presented in Fig. \ref{TP}, where each node stands for a conditional distribution. The root node is the source distribution $W = \langle X\rangle $. Each node has two sub-nodes that represent its upper and lower Hadamard transform, respectively. The distribution $\langle Y_k|Y^{k-1}\rangle $ is obtained by the leaf nodes $W^{\theta_n(k)}, k\in[N]$.

We define a stochastic process to represent the evolution of conditional distributions. Let $\{B_i\}_{i=1}^\infty$ be a sequence of \iid\ Bernoulli(1/2) random variables that are independent of $\{X_i\}_{i=1}^\infty$. Define the conditional distribution process $\{W_n\}_{n=0}^\infty$ as $W_0 = \langle X\rangle $ and
\begin{equation}
W_n = \langle X\rangle ^{B_1B_2\cdots B_n},n\geq1.
\end{equation}
In other words, given $W_n$, $W_{n+1}$ is equal to $W_n^0$ or $W_n^1$ each with probability 1/2, which implies $\{W_n\}_{n=1}^\infty$ is a Markov process. According to \eqref{distribution evolution}, the distribution of $W_n$ is given by $\P(W_n = \langle X\rangle ^{\theta_n(k)}) = 2^{-n},\  \forall k\in[N]$.

For a functional $F(\cdot)$ that takes conditional distributions as input, if $ W = \langle U|V\rangle$ represents a conditional distribution, we denote $F(W) = F(U|V)$ for convenience. For example, we have $d(W_0) = d(X)$ and $d(\langle X\rangle ^{\theta_n(k)}) = d(Y_k|Y^{k-1})$. In the following subsections, we define the stochastic processes of RID, weighted discrete entropy and error probability by applying the corresponding functionals on $\{W_n\}_{n=1}^\infty$.
\subsection{Polarization of R\'{e}nyi Information Dimension}\label{PMAPHT-2}
\begin{definition}[RID Process \cite{HA2013}]
The RID process is defined to be
\begin{equation}
d_n := d(W_n),\ \forall n\geq 0.
\end{equation}
\end{definition}

It was shown in \cite{HA2013} that $d_n$ resembles the polarization of binary erasure channel (BEC). Formally, it was proved that
\begin{equation}\label{recursion-dn}
d_{n+1} = \left\{\begin{aligned}   &2d_n - d_n^2,& &\ \text{if }B_{n+1} = 0,\\  &d_n^2,& &\ \text{if }B_{n+1} = 1.\end{aligned}\right.
\end{equation}
This implies $d_n$ has the same behaviour as the Bhattacharyya parameter process beginning with BEC$(d_0)$ \cite{Arikan2009}. Consequently, $d_n$ polarizes in the sense that $d_n \xrightarrow{a.s.} d_\infty \in\{0,1\}$ with $\P(d_\infty=1) = 1-\P(d_\infty=0) = d(X)$. In addition, according to the rate of polarization \cite{AT2009}, for any $\beta\in(0,1/2)$ we have
\begin{align}
&\lim\limits_{n\rightarrow\infty}\P(d_n \leq 2^{-2^{\beta n}}) = 1 - d(X), \label{convgerence-rate-dn-0}\\
&\lim\limits_{n\rightarrow\infty}\P(d_n \geq 1 - 2^{-2^{\beta n}})=d(X).\label{convgerence-rate-dn-1}
\end{align}

Recall that the RID of nonsingular distribution is equal to the mass of its continuous component. Therefore, after applying Hadamard transform on $\mathbf{X}$, the resulting $\langle Y_k|Y^{k-1}\rangle$ become either purely discrete or purely continuous, and the fraction of purely continuous distributions approaches $d(X)$. This leads to the initial step of the polarization of error probability, since we can never precisely reconstruct a continuous random variable.

\subsection{Absorbtion of Weighted Discrete Entropy}\label{PMAPHT-3}
\begin{definition}[Weighted Discrete Entropy Process]
Suppose $H(\langle X\rangle_d) < \infty$. The weighted discrete entropy process is defined to be
\begin{equation}
\widehat{H}_n := \widehat{H}(W_n), \ \forall n\geq0.
\end{equation}
\end{definition}

In \cite{HAT2012}, the authors studied the entropy process initiated from purely discrete source, which is the special case of $\widehat{H}_n$ when $X$ is restricted to discrete random variable. It was proved in \cite{HAT2012} that $ \lim\limits_{n\rightarrow\infty}\widehat{H}_n \overset{a.s.}{=} 0$ if $X$ is discrete with finite support, which was named the absorption phenomenon to distinguish from that over finite fields where the discrete entropy polarizes \cite{Arikan2010}.

In this paper, we prove a stronger result on the convergence of $\widehat{H}_n$. First, we weaken the assumptions on the source $X$ by showing $\widehat{H}_n\xrightarrow{Pr.}0$ for any nonsingular source $X$ satisfying the regular conditions given in Theorem \ref{polarization-of-Qn}. Second, we further analyze the convergence rate of $\widehat{H}_n$, which is not provided in \cite{HAT2012}. The formal statement is presented in Theorem \ref{absorption-of-hat-Hn}.
\begin{theorem}[Absorption of Weighted Discrete Entropy]\label{absorption-of-hat-Hn}
Suppose $X$ satisfies the conditions given in Theorem \ref{polarization-of-Qn}. Then for any $\lambda>0$, we have
\begin{equation}\label{absorption-of-hat-Hn-eq}
\lim\limits_{n\rightarrow\infty} \P(\widehat{H}_n \leq 2^{-\lambda n}) = 1.
\end{equation}
\end{theorem}
\begin{IEEEproof}
See Section \ref{Proof of absorption of hat Hn}.
\end{IEEEproof}

As $n$ approaches infinity, the polarization of RID results in $W_n$ becoming either highly discrete or highly continuous. Theorem \ref{absorption-of-hat-Hn} further demonstrates that when $W_n$ is highly discrete, the entropy of its discrete component also becomes negligible. This contributes to the second step of error probability polarization, because a discrete random variable with low entropy can be accurately reconstructed with high probability.

\begin{figure}[htbp]
\centerline{\includegraphics[width=0.25\textwidth,trim = 158 85 685 46,clip]{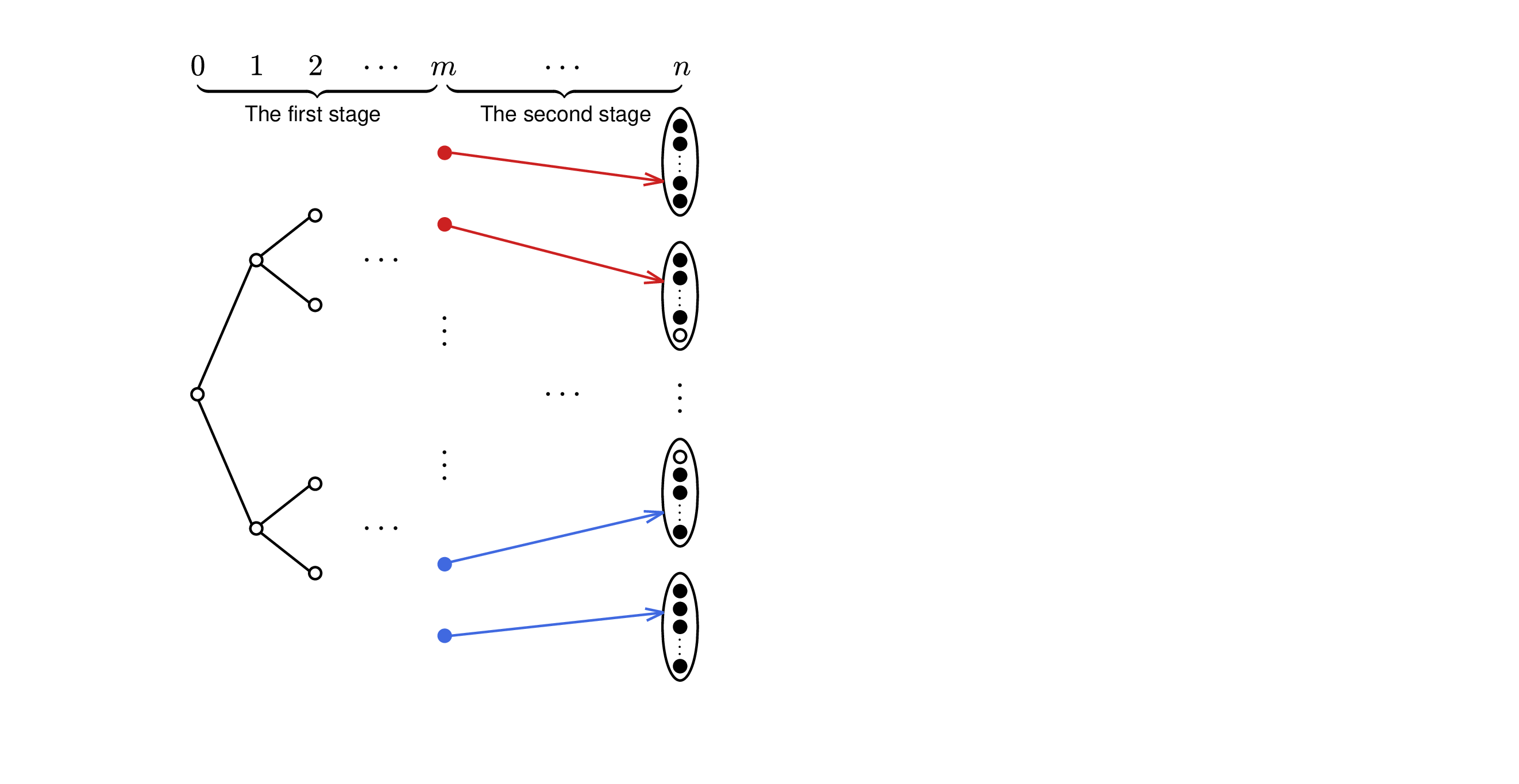}}
\caption{Illustration of the two-stages absorption of weighted discrete entropy. The red nodes and blue nodes at the $m$-th layer stand for the $W_m$ with high RID and low RID, respectively. The black nodes at the $n$-th layer represent the $W_n$ with low weighted discrete entropy.}
\label{TPF}
\end{figure}

To prove Theorem \ref{absorption-of-hat-Hn}, we divide the Hadamard transform into two stages. Let $m<n$ such that $m\rightarrow\infty$ as $n\rightarrow\infty$. In the first stage, $m$ transforms are performed, while the remaining $n-m$ transforms make up the second stage. Fig. \ref{TPF} shows the absorption of weighted discrete entropy during these two stages, where each node represents a conditional distribution (as shown in Fig. \ref{TP}). The black nodes at the $n$-th layer denote the $W_n$ with low weighted discrete entropy.

The idea behind proving Theorem \ref{absorption-of-hat-Hn} can be briefly encapsulated as follows. At the $m$-th layer, the RID of $W_m$ is close to either 0 or 1 because of polarization. In Fig. \ref{TPF}, we represent the high-RID and low-RID $W_m$ with red and blue nodes, respectively. The sub-nodes of high-RID $W_m$ (indicated by red arrows in Fig. \ref{TPF}) is expected to have small weighted discrete entropy due to the negligible mass of their discrete component. Meanwhile, for the $W_m$ with low RID, the fast polarization rate of the RID process allows us to treat them as purely discrete conditional distributions. According to the absorption of entropy for discrete source \cite{HAT2012}, it is reasonable to expect that the sub-nodes of low-RID $W_m$ (pointed to by blue arrows in Fig. \ref{TPF}) also have a vanishing weighted discrete entropy. The technical challenge is to guarantee a uniform convergence rate for all entropy process initiated from the low-RID $W_m$. We accomplish this by carefully analyzing the convergence rate of entropy process beginning with discrete source. The detailed proof can be found in Section \ref{Proof of absorption of hat Hn}.


\subsection{Proof of Theorem \ref{polarization-of-Qn}}\label{PMAPHT-4}
Fix $\lambda >0$ and $\beta\in(0,1/2)$. Define the error probability process to be
\begin{equation}
Q_n := P_e^{\text{MAP}}(W_n), \ \forall n\geq0.
\end{equation}
To prove our statement, it is equivalent to show
\begin{align}
&\lim\limits_{n\rightarrow\infty} \P(Q_n \leq 2^{-\lambda n}) = 1-d(X), \label{convgerence-rate-Qn-0}\\
&\lim\limits_{n\rightarrow\infty} \P(Q_n \geq 1 - 2^{-2^{\beta n}}) = d(X).\label{convgerence-rate-Qn-1}
\end{align}
Since $\E X^2<\infty$, we conclude that $\E Y_k^2<\infty$ for all $n\geq0$ and $k\in[N]$. It follows from Proposition \ref{bounds on error probability} that
\begin{equation}
d_n\leq Q_n\leq d_n + \widehat{H}_n.
\end{equation}
Consequently, for any $\lambda > 0$ we have $\P(Q_n\leq 2^{-\lambda n}) \leq \P(d_n\leq 2^{-\lambda n})$, and
\begin{equation}\label{small Qn}
\begin{aligned}
\P(Q_n \leq 2^{-\lambda n}) \geq \P(d_n + \widehat{H}_n \leq 2^{-\lambda n}) &\geq\P(d_n\leq 2^{-\lambda n-1},\widehat{H}_n \leq 2^{-\lambda n-1})\\
&  \geq\P(d_n\leq 2^{-\lambda n-1}) + \P(\widehat{H}_n \leq 2^{-\lambda n-1}) -1.
\end{aligned}
\end{equation}
Now \eqref{convgerence-rate-Qn-0} follows from \eqref{small Qn}, \eqref{convgerence-rate-dn-0} and Theorem \ref{absorption-of-hat-Hn}. Similarly, considering that
\begin{equation}
\P(d_n \geq 1 - 2^{-2^{\beta n}})\leq\P(Q_n \geq 1 - 2^{-2^{\beta n}})\leq1 - \P(Q_n \leq 2^{-\lambda n}) ,
\end{equation}
we can deduce \eqref{convgerence-rate-Qn-1} from \eqref{convgerence-rate-dn-1} and \eqref{convgerence-rate-Qn-0}.

\section{Partial Hadamard Compression and SC Decoding} \label{The Proposed Encoding-Decoding Scheme}
In this section, we propose the polarization-based scheme for analog compression. Let $\mathbf{x}\in\R^N$ be the realization of $\mathbf{X}$, representing the signal to be compressed. The compressed signal, denoted by $\mathbf{z} \in\R^M$, is obtained by applying a linear operation on $\mathbf{x}$. The measurement rate is given by $R = M/N$. In Section \ref{PHCSD-1} we introduce our design of the partial Hadamard matrices for linear compression. Then the analog SC decoder for signal reconstruction is presented in Section \ref{PHCSD-2}. In Section \ref{PHCSD-3} we show that the proposed scheme achieves the information-theoretical limit of lossless analog compression for nonsingular source. Lastly, the connections between the proposed scheme and binary polar codes are presented in Section \ref{PHCSD-4}

\subsection{Partial Hadamard Compression}\label{PHCSD-1}
In our compression scheme, the measurement matrix is a submatrix of $\mathsf{H}_n$, denoted by $\mathsf{H}_\mathcal{A}$, which contains the rows of $\mathsf{H}_n$ with indices in $\mathcal{A}$. The submatrix $\mathsf{H}_\mathcal{A}$ is also called partial Hadamard matrix. Let $\mathbf{y} = \mathsf{H}_n\mathbf{x}$, the compressed signal $\mathbf{z}$ is given by
\begin{equation}
\mathbf{z} = \mathsf{H}_\mathcal{A}\mathbf{x} = y_\mathcal{A}.
\end{equation}
We call $\mathcal{A}$ the reserved set and its complement $\mathcal{A}^c$ the discarded set.

To guarantee the efficiency of $\mathsf{H}_\mathcal{A}$ for compression, the reserved set $\mathcal{A}$ should be selected such that $Y_\mathcal{A}$ preserves as much information of $\mathbf{Y}$ as possible. Thanks to the polarization of error probability, we propose to reserve the components $Y_k$ such that $P^{\text{MAP}}_e(Y_k|Y^{k-1})$ is close to 1. Specifically, let
\begin{equation}\label{Q_k}
Q_n(k) := P^{\text{MAP}}_e(Y_k|Y^{k-1}),\  k\in[N].
\end{equation}
Sort the sequence $\{Q_n(k)\}_{k=1}^{N}$ with $Q_n(k_1)\geq Q_n(k_2) \geq \cdots \geq Q_n(k_N)$. Given the measurement rate $R$, let $M = \lceil RN\rceil$, where $\lceil\cdot\rceil$ denotes the ceil function. Take the reserved set $\mathcal{A}= \left\{k_{1},k_{2},\dots,k_M\right\}$. In other words, the reserved set $\mathcal{A}$ contains the indices of the $M$ largest $Q_n(k)$. Such design of $\mathcal{A}$ ensures that we can precisely recover $y_k$ given the previous $y^{k-1}$ if $k\in\mathcal{A}^c$, because Theorem \ref{polarization-of-Qn} implies that $\mathcal{A}^c$ contains the indices for which $P_e^\text{MAP}(Y_k|Y^{k-1})$ is close to 0. In practice, $\mathcal{A}$ can be determined through Monte Carlo simulation.

\subsection{Analog Successive Cancellation Decoding}\label{PHCSD-2}
Instead of directly recovering $\mathbf{x}$, the SC decoder first estimates $\hat{\mathbf{y}}$ and then set $\hat{\mathbf{x}} = \mathsf{H}_n^{-1}\hat{\mathbf{y}}$. Given the reserved set $\mathcal{A}$ and the compressed signal $\mathbf{z}=y_\mathcal{A}$, the analog SC decoder outputs the estimate $\hat{y}_k$ sequentially in the rule that
\begin{equation} \label{hard decision}
\hat{y}_k = \left\{\begin{aligned}
&y_k,&\text{ if }k\in\mathcal{A},\\
&Y_k^*(\hat{y}^{k-1}),&\text{ if }k\in\mathcal{A}^c. \end{aligned}\right.
\end{equation}
If $k\in\mathcal{A}$, the true value of $y_k$ is known thus we set $\hat{y}_k = y_k$. If $k\in\mathcal{A}^c$, the analog SC decoder outputs the MAP estimate of $Y_k$ given $Y^{k-1} = \hat{y}^{k-1}$. If $\langle Y_{k}|Y^{k-1}=\hat{y}^{k-1}\rangle$ is continuous, or equivalently, $d(Y_k|Y^{k-1}=\hat{y}^{k-1})=1$, the decoder announces failure since it is impossible to precisely reconstruct a continuous random variable. Note that the selection of $\mathcal{A}$ ensures a vanishing $P_e^\text{MAP}(Y_k|Y^{k-1})$ for $k\in\mathcal{A}^c$, which indicates that $y_k$ can be precisely reconstructed with high probability for each $k\in\mathcal{A}^c$. Starting from $k=1$, the SC decoder recovers $y_k$ sequentially until $k=N$. The reconstructed signal is given by $\hat{\mathbf{x}} = \mathsf{H}_n^{-1} \hat{\mathbf{y}} = \mathsf{H}_n \hat{\mathbf{y}}$.

The conditional distribution $\langle Y_k|Y^{k-1} = \hat{y}^{k-1}\rangle $ can be calculated recursively using the structure of Hadamard matrices. In the following, we define the analog $f$ and $g$ operations to characterize the evolution of conditional distributions under the upper and lower Hadamard transform, respectively.
\begin{definition}[f and g operations over analog domain]\label{f-g-operation}
Let $\mathcal{P}$ denote the collection of all nonsingular probability distributions over $\R$. For any $P_1,P_2\in\mathcal{P}$, let $X_1,X_2$ be independent random variables with distributions $X_1\sim P_1$, $X_2\sim P_2$. Denote $Y_1 = (X_1+X_2)/\sqrt{2}$ and $Y_2 = (X_1-X_2)/\sqrt{2}$. The map $f:\mathcal{P}\times \mathcal{P}\rightarrow \mathcal{P}$ is defined to be
\begin{equation} \label{f}
f(P_1,P_2) := \langle Y_1\rangle .
\end{equation}
The map $g:\mathcal{P}\times \mathcal{P}\times\R\rightarrow \mathcal{P}$ is defined as
\begin{equation} \label{g}
g(P_1,P_2,y) := \langle Y_2|Y_1 = y\rangle .
\end{equation}
\end{definition}
$f$ is to calculate the convolution of probability distributions, and $g$ is to calculate the conditional distribution. We derive the closed form of $f$ and $g$ in Section \ref{Basic Hadamard Transform of Nonsingular Distributions}.

Denote $\lambda^{(k)}_n(y^{k-1}) := \langle Y_k|Y^{k-1} = y^{k-1}\rangle ,\ k\in[N]$. According to the recursive structure of Hadamard matrices, the distribution $\lambda^{(k)}_n(y^{k-1})$ can be obtained by recursively applying $f$ and $g$ in the way that
\begin{equation}\label{SC recursive}
\begin{aligned}
&\lambda^{(2i-1)}_n(y^{2i-2}) = f(\lambda^{(i)}_{n-1}(\bar{y}^{i-1}),\lambda^{(i)}_{n-1}(\tilde{y}^{i-1})),\ \lambda^{(2i)}_n(y^{2i-1}) =g(\lambda^{(i)}_{n-1}(\bar{y}^{i-1}),\lambda^{(i)}_{n-1}(\tilde{y}^{i-1}),y_{2i-1}),\\
&\bar{y}^{i-1} = \frac{y^{2i-2}_o + y^{2i-2}_e}{\sqrt{2}},\ \tilde{y}^{i-1} = \frac{y^{2i-2}_o - y^{2i-2}_e}{\sqrt{2}},
\end{aligned}
\end{equation}
where $y^{2i-2}_e$ and $y^{2i-2}_o$ are subvectors of $y^{2i-2}$ with even and odd indices, respectively. This recursion can be continued down to $n=0$, at which the distribution is equal to the source $X$, \ie, $\lambda^{(1)}_0 = P_X$. The proposed analog SC decoder is summarized in Algorithm 1. Note that this decoding scheme is almost the same as the SC decoder for binary polar codes except that the basic operations are replaced by \eqref{f} and \eqref{g}. Take the complexity of calculating convolution and conditional distribution as 1, the total number of operations in the SC decoding scheme is $N \log N$.

\begin{algorithm}[t]
\caption{Analog SC decoder}
\begin{algorithmic}[1]
\REQUIRE
Reserved set $\mathcal{A}$, compressed signal $\mathbf{z}=y_\mathcal{A}$, source distribution $P_X$.
\ENSURE
The reconstructed signal $\hat{\mathbf{x}}$.
\FOR{$k= 1$ to  $N$}
\IF{$k \in \mathcal{A}$}
\STATE Set $\hat{y}_k = y_k$.
\ELSE
\STATE Recursively calculate $\langle Y_k|Y^{k-1}=\hat{y}^{k-1}\rangle$ by \eqref{SC recursive}. The initial condition is given by $P_X$.
\IF{$d(Y_k|Y^{k-1}=\hat{y}^{k-1})=1$}
\RETURN Failure
\ELSE
\STATE Set $\hat{y}_k = \mathop{\arg\max}\limits_{y\in\R}\P(Y_k=y|Y^{k-1}=\hat{y}^{k-1})$.
\ENDIF
\ENDIF
\ENDFOR
\RETURN $\hat{\mathbf{x}} = \mathsf{H}_n \hat{\mathbf{y}}$.
\end{algorithmic}
\end{algorithm}
\subsection{Achieving the Limit of Lossless Analog Compression}\label{PHCSD-3}
Using Theorem \ref{polarization-of-Qn}, we prove that the proposed partial Hadamard compression with analog SC decoder achieves the fundamental limit of lossless analog compression established in \cite{WV2010}.
\begin{theorem}\label{main theorem}
Let $X$ be a nonsingular source satisfying the conditions given in Theorem \ref{polarization-of-Qn}. If the measurement rate $R > d(X)$, then for any $p>0$ we have $P_e(\mathsf{H}_\mathcal{A},\text{SC}) = O(N^{-p})$, where $P_e(\mathsf{H}_\mathcal{A},\text{SC})$ is the error probability under the partial Hadamard matrices $\mathsf{H}_\mathcal{A}$ and analog SC decoder with measurement rate $R$ .
\end{theorem}
\begin{IEEEproof}
Let $\widehat{\mathbf{X}}_{\text{SC}}$ denote the the reconstructed signal obtained by analog SC decoder and $\widehat{\mathbf{Y}} = \mathsf{H}_n\widehat{\mathbf{X}}_{\text{SC}}$. Clearly $P_e(\mathsf{H}_\mathcal{A},\text{SC}) = \P(\widehat{\mathbf{Y}} \neq \mathbf{Y})$. Decomposing the error event $\{\widehat{\mathbf{Y}} \neq \mathbf{Y}\}$ according to the first error location, we obtain
\begin{equation}
\begin{aligned}
\P(\widehat{\mathbf{Y}} \neq \mathbf{Y})=\sum\limits_{k\in\mathcal{A}^c}\P(Y^{k-1} = \widehat{Y}^{k-1}, Y_k \neq \widehat{Y}_k)&=\sum\limits_{k\in\mathcal{A}^c}\P(Y^{k-1} = \widehat{Y}^{k-1}, Y_k \neq Y_k^*(\widehat{Y}^{k-1}))\\
&=\sum\limits_{k\in\mathcal{A}^c}\P(Y^{k-1} = \widehat{Y}^{k-1}, Y_k \neq Y_k^*(Y^{k-1}))\\
&\leq\sum\limits_{k\in\mathcal{A}^c}\P(Y_k \neq Y_k^*(Y^{k-1}))\\
&= \sum\limits_{k\in\mathcal{A}^c}P_e^{\text{MAP}}(Y_k|Y^{k-1}).
\end{aligned}
\end{equation}
For any $p>0$, let $\mathcal{I}_n = \left\{k\in[N]: Q_n(k)\leq 2^{-(p+1)n}\right\}$, where $Q_n(k)$ is given by \eqref{Q_k}. By Theorem \ref{polarization-of-Qn} we obtain
\begin{equation}
\lim\limits_{n\rightarrow\infty} \frac{|\mathcal{I}_n|}{2^n} = 1 - d(X).
\end{equation}
Since $R > d(X)$ and $\mathcal{A}^c$ contains the indices of the $(1-R)N$ smallest $Q_n(k)$, for sufficiently large $n$ we have $\mathcal{A}^c\subset \mathcal{I}_n$. This implies
\begin{equation}
\sum\limits_{k\in\mathcal{A}^c} P^{\text{MAP}}_e(Y_k|Y^{k-1}) \leq 2^{-(p+1)n}2^n=N^{-p}.
\end{equation}
\end{IEEEproof}

\subsection{Connections to Polar Codes}\label{PHCSD-4}

\renewcommand\arraystretch{1.5}
\begin{table*}[t]
\caption{The connections between the analog Hadamard compression and binary polar codes}
\label{comparion_of_analog_and_binar}
\centering
\begin{tabular}{c|c|c|c|c}
\hline
\hline
\multicolumn{3}{c|}{}& Analog Hadamard compression & Binary polar codes\\
\hline
\multicolumn{3}{c|}{Theoretical basis} & Polarization over $\R$ & Polarization over $\mathbb{F}_2$\\
\hline
& \multicolumn{2}{c|}{Commonness} & \multicolumn{2}{c}{Selecting rows from the base matrix using polarization-based principle}\\ \cline{2-5}
Encoding& \multirow{2}{*}{Difference}& Base Matrix &$\small \mathsf{H}_n = \mathsf{B}_n\left(\frac{1}{\sqrt{2}}\begin{bmatrix}1 & 1 \\ 1 & -1  \end{bmatrix}\right)^{\otimes n}$& $\small \mathsf{G}_n =  \mathsf{B}_n\begin{bmatrix}1 & 1\\0 & 1\end{bmatrix}^{\otimes n} $ \\ \cline{3-5}
& &Construction & Rows of $ \mathsf{H}_n$ with high error probability & Rows of $\mathsf{G}_n$ with high discrete entropy\\
\hline
\multirow{2}{*}{Decoding} & \multicolumn{2}{c|}{Commonness}  & \multicolumn{2}{c}{Sequential reconstruction with MAP estimation for discarded entries} \\ \cline{2-5}
& Difference & Basic operations & \makecell[c]{Calculating the convolution as \eqref{f} \\and the conditional distribution as \eqref{g}} & Calculating LR as \eqref{f_SC_binary} and \eqref{g_SC_binary} \\
\hline
\hline
\end{tabular}
\end{table*}

The proposed scheme has substantial similarities to binary polar codes for source coding, while there are also notable differences. Regarding the encoding process, the Hadamard matrices, employed as the base matrix for analog compression, possess a recursive structure similar to the polar transform. Furthermore, a similar polarization-based principle is utilized to select rows from the Hadamard matrices for constructing the encoding matrices. On the decoding side, the analog SC decoder closely resembles the binary SC decoder, with the exception that the basic operations are replaced by their counterparts over analog domain. Specifically, since the probability distributions over $\mathbb{F}_2$ can be represented by a single parameter, it is sufficient to calculate likelihood ratio for the MAP estimation in binary SC decoder. However, the probability distributions over $\R$ cannot be parameterized in general. Therefore, the analog SC decoder needs to calculate the convolution and conditional distribution over $\R$. The connections between the analog Hadamard compression and binary polar codes are summarized in Table \ref{comparion_of_analog_and_binar}.

\section{Basic Hadamard Transform of Nonsingular Distributions}\label{Basic Hadamard Transform of Nonsingular Distributions}
In this section, we focus on the basic Hadamard transform of nonsingular distributions, \ie, we provide the closed form of the operations $f$ and $g$ defined in Definition \ref{f-g-operation}. Throughout this section, $X_1$ and $X_2$ are assumed to be two independent nonsingular random variables with mixed representation $(\Gamma_1,C_1,D_1)$ and $(\Gamma_2,C_2,D_2)$, respectively. Without loss of generality, assume
\begin{equation}\label{equal-general-initial-distribution-mixed-representation}
X_1 = \Gamma_1C_1+(1-\Gamma_1)D_1,\ X_2 = \Gamma_2C_2+(1-\Gamma_2)D_2.
\end{equation}
Suppose the distributions of $X_1$ and $X_2$ are given by
\begin{equation}\label{general-initial-distribution-mixed-representation}
\begin{aligned}
D_1 \sim \sum\limits_i p_i\delta_{x_i},\ C_1 \sim \varphi_1(t),\  \Gamma_1 \sim \text{Bernoulli}(\rho_1), \ \ D_2 \sim \sum\limits_j q_j\delta_{y_j},\ C_2 \sim \varphi_2(t),\  \Gamma_2 \sim \text{Bernoulli}(\rho_2),
\end{aligned}
\end{equation}
where $C_i \sim \varphi_i(t)$ means $C_i$ has the density $\varphi_i(t)$, $i=1,2$. Let $Y_1 = (X_1+X_2)/\sqrt{2}$ and $Y_2 = (X_1-X_2)/\sqrt{2}$. Clearly $\langle Y_1\rangle $ and $\langle Y_2|Y_1\rangle $ are nonsingular. The aim of this section is to find the mixed representations of $\langle Y_1\rangle $ and $\langle Y_2|Y_1\rangle $. For convenience, denote $\bar{C}_i = C_i/\sqrt{2}$ and $\bar{D}_i = D_i/\sqrt{2}$, $i=1,2$.
\subsection{Distribution of $Y_1$}
Let * denote the convolution of probability measures. Then
\begin{equation}
\begin{aligned}
\langle Y_1\rangle & = (\rho_1 \langle \bar{C}_1\rangle  + (1-\rho_1)\langle \bar{D}_1\rangle )*(\rho_2 \langle \bar{C}_2\rangle  + (1-\rho_2)\langle \bar{D}_2\rangle )\\
&=\rho_1\rho_2 \langle \bar{C}_1\rangle *\langle \bar{C}_2\rangle  + \rho_1(1-\rho_2)\langle \bar{C}_1\rangle *\langle \bar{D}_2\rangle +(1-\rho_1)\rho_2\langle \bar{D}_1\rangle *\langle \bar{C}_2\rangle +(1-\rho_1)(1-\rho_2)\langle \bar{D}_1\rangle *\langle \bar{D}_2\rangle .
\end{aligned}
\end{equation}
Among the four components of $\langle Y_1\rangle $, $\langle \bar{D}_1\rangle *\langle \bar{D}_2\rangle $ is discrete and the other three are continuous. As a result, let $\Gamma^0,C^0,D^0$ be independent and
\begin{equation}\label{f operation}
\begin{aligned}
\rho^0 &= 1 - (1-\rho_1)(1-\rho_2),\ \Gamma^0 \sim \text{Bernoulli}(\rho^0),\\
C^0 &\sim \frac{\rho_1\rho_2}{\rho^0} \langle \bar{C}_1+\bar{C}_2\rangle  + \frac{\rho_1(1-\rho_2)}{\rho^0}\langle \bar{C}_1+\bar{D}_2\rangle +\frac{(1-\rho_1)\rho_2}{\rho^0}\langle \bar{D}_1+\bar{C}_2\rangle ,\\
D^0 &= \bar{D}_1+\bar{D}_2,
\end{aligned}
\end{equation}
then $(\Gamma^0,C^0,D^0)$ is the mixed representation of $\langle Y_1\rangle$. In other words,
\begin{equation}\label{f operation-concrete}
f(\langle X_1\rangle, \langle X_2\rangle)= \langle\Gamma^0C^0 + (1-\Gamma^0)D^0\rangle.
\end{equation}
To find the density of $C^0$, denote $\bar{D}_1+\bar{C}_2\sim L_1(y)$, $\bar{C}_1+\bar{D}_2\sim L_2(y)$ and $\bar{C}_1+\bar{C}_2\sim L_3(y)$. Let
\begin{equation}\label{Fz}
\begin{aligned}
F_1(y) &= (1-\rho_1)\rho_2L_1(y)= (1-\rho_1)\rho_2\sum\limits_ip_i\sqrt{2}\varphi_2(\sqrt{2}y-x_i),\\
F_2(y) &= \rho_1(1-\rho_2)L_2(y)= \rho_1(1-\rho_2)\sum\limits_jq_j\sqrt{2}\varphi_1(\sqrt{2}y-y_j),\\
F_3(y) &= \rho_1\rho_2L_3(y) =\rho_1\rho_2\int_\R\sqrt{2}\varphi_1(s)\varphi_2(\sqrt{2}y-s)ds,\\
F(y) &= F_1(y) + F_2(y) + F_3(y),
\end{aligned}
\end{equation}
then the density of $C^0$ is given by $F(y) / \rho^0$.
\subsection{Distribution of $Y_2$ conditioned on $Y_1$}
We first introduce the concept of regular conditional distribution \cite[Chapter 5.1.3]{Durrett2010}.
\begin{definition}[Regular conditional distribution \cite{Durrett2010}]\label{RCD}
Let $(\Omega,\mathcal{F},\P)$ be a probability space, $X: (\Omega,\mathcal{F})\rightarrow (S,\mathcal{S})$ a measurable map, and $\mathcal{G}\subset\mathcal{F}$ a sub $\sigma$-algebra. A two-variable function $Q(\omega,A): \Omega\times\mathcal{S}\rightarrow [0,1]$ is said to be a regular conditional distribution for $X$ given $\mathcal{G}$ if
\begin{enumerate}
\item For each $A\in\mathcal{S}$, $Q(\cdot,A)\overset{a.s.}{=}\P(X\in A|\mathcal{G})$.
\item For a.s. $\omega$, $Q(\omega,\cdot)$ is a probability measure on $(S,\mathcal{S})$.
\end{enumerate}
\end{definition}

We need to find a function $Q(y,A)$ such that $Q(Y_1,A)\overset{a.s.}{=}\P(Y_2\in A|Y_1)$ for any Borel set $A$, and $Q(y,\cdot)$ is a probability measure over $\R$ almost surely. Once such function $Q(y,A)$ is found, the conditional distribution $\langle Y_2|Y_1=y\rangle $ is given by $\P(Y_2\in A|Y_1=y) = Q(y,A)$.

\begin{proposition}\label{prop-g-operation}
For $y\in\R$, define $\Gamma^1_y,C^1_y$ and $D^1_y$ to be independent random variables such that
\begin{equation}\label{Cy1Dy1}
\begin{aligned}
\rho^1_y &= F_3(y)/F(y), \ \Gamma^1_y\sim \text{Bernoulli}(\rho^1_y),\\
C^1_y &\sim \langle \bar{C}_1-\bar{C}_2|\bar{C}_1+\bar{C}_2=y\rangle ,\\
D^1_y &\sim \frac{F_1(y)}{F_1(y)+F_2(y)}\langle \bar{D}_1-\bar{C}_2|\bar{D}_1+\bar{C}_2=y\rangle +\frac{F_2(y)}{F_1(y)+F_2(y)}\langle \bar{C}_1-\bar{D}_2|\bar{C}_1+\bar{D}_2=y\rangle .
\end{aligned}
\end{equation}
For any $y\in\R$ and Borel set $A$, define
\begin{equation}\label{g operation}
\begin{aligned}
Q(y,A) = \left\{\begin{aligned}&\P(\bar{D}_1-\bar{D}_2\in A|\bar{D}_1+\bar{D}_2=y),&\text{if}\ y\in \spt(D^0),\\
&\P(\Gamma_y^1C_y^1+(1-\Gamma_y^1)D_y^1\in A),&\text{if}\ y\notin \spt(D^0).\\
\end{aligned}\right.
\end{aligned}
\end{equation}
Then $Q(y,A)$ is the regular conditional distribution for $Y_2$ given $\sigma(Y_1)$.
\end{proposition}
\begin{IEEEproof}
See Appendix \ref{appendix B}.
\end{IEEEproof}
\textbf{Remark: } According to Proposition \ref{prop-g-operation}, if $y\in \spt(D^0)$, then $\langle Y_2|Y_1=y\rangle $ is purely discrete and has the same distribution as $\langle \bar{D}_1-\bar{D}_2|\bar{D}_1+\bar{D}_2=y\rangle $. If $y\notin \spt(D^0)$, then $( \Gamma^1_y,C^1_y,D^1_y) $ is the mixed representation of $\langle Y_2|Y_1=y\rangle $. In summary, we have
\begin{equation}\label{g operation-concrete}
\begin{aligned}
&g(\langle X_1\rangle, \langle X_2\rangle, y) = \left\{\begin{aligned}&\langle \bar{D}_1-\bar{D}_2|\bar{D}_1+\bar{D}_2=y\rangle,&\text{if }y\in\spt(D^0),\\
&\langle \Gamma^1_yC^1_y + (1-\Gamma^1_y)D^1_y\rangle,&\text{if }y\notin\spt(D^0).
\end{aligned}\right.
\end{aligned}
\end{equation}

The detailed proof of Proposition \ref{prop-g-operation} is given in Appendix \ref{appendix B}. Here we provide some heuristical explanations. Let $P_{ab}(y) =\P(\Gamma_1=a,\Gamma_2=b|Y_1=y),\ a,b\in\{0,1\}$. Then $\langle Y_2|Y_1=y\rangle$ can be decomposed as
\begin{equation}\label{P(T=t|Z=z)}
\begin{aligned}
\langle Y_2|Y_1=y\rangle  &= P_{00}(y)\langle \bar{D}_1-\bar{D}_2|\bar{D}_1+\bar{D}_2=y\rangle +P_{01}(y)\langle \bar{D}_1-\bar{C}_2|\bar{D}_1+\bar{C}_2=y\rangle \\
&\ \ \ +P_{10}(y)\langle \bar{C}_1-\bar{D}_2|\bar{C}_1+\bar{D}_2=y\rangle +P_{11}(y)\langle \bar{C}_1-\bar{C}_2|\bar{C}_1+\bar{C}_2=y\rangle .
\end{aligned}
\end{equation}
If $y\in \spt(D^0)$, we conclude that $P_{00}(y) = 1$. This is because continuous measure assigns 0 probability to any countable set. Therefore, in this case we have $\langle Y_2|Y_1=y\rangle =\langle \bar{D}_1-\bar{D}_2|\bar{D}_1+\bar{D}_2=y\rangle$. If $y\notin \spt(D^0)$, then clearly $P_{00}(y) = 0$. The remaining three terms in the right side of \eqref{P(T=t|Z=z)} correspond to the three components of $C^0$ as in \eqref{f operation}, thus their weights are given by $P_{01}(y) = F_1(y)/F(y)$, $P_{10}(y) = F_2(y)/F(y)$ and $P_{11}(y) = F_3(y)/F(y)$, respectively. As a result, we have $\langle Y_2|Y_1=y\rangle_c=\langle \bar{C}_1-\bar{C}_2|\bar{C}_1+\bar{C}_2=y\rangle $, and $\langle Y_2|Y_1=y\rangle_d$ is equal to the combination of $\langle \bar{D}_1-\bar{C}_2|\bar{D}_1+\bar{C}_2=y\rangle $ and $\langle \bar{C}_1-\bar{D}_2|\bar{C}_1+\bar{D}_2=y\rangle $.
\subsection{Reproduce the Polarization of RID}
The key step to show the RID polarization is the recursive formulas of the RID process $d_n$, which was proved under linear algebra setting in \cite{HA2017}. We show that the recursive formulas can be obtained by a straightforward calculation using \eqref{f operation-concrete} and \eqref{g operation-concrete}. Specifically, in the following we prove \eqref{recursion-dn}. Suppose $W_n = \langle U|V\rangle$, then $d_n = d(W_n) = d(U|V)$. Let $(U',V')$ be the independent copy of $(U,V)$. By \eqref{f operation} and \eqref{f operation-concrete},
\begin{equation}
\begin{aligned}
d\left(\frac{U+U'}{\sqrt{2}}\bigg|V=v,V'=v'\right)& = d(f(\langle U|V=v\rangle,\langle U'|V'=v'\rangle))\\
&= 1 - (1-d(U|V=v))(1-d(U'|V'=v')).
\end{aligned}
\end{equation}
Taking expectation in both side we obtain
\begin{equation}
\begin{aligned}
d\left(\frac{U+U'}{\sqrt{2}}\bigg|V,V'\right) &= \E_{V,V'}[1 - (1-d(U|V=v))(1-d(U'|V'=v'))] \\
&= 1 - (1-\E_V[d(U|V=v)])(1-\E_{V'}[d(U'|V'=v')])\\
&=2d_n - d_n^2.
\end{aligned}
\end{equation}
As a result, $d_{n+1} = d(W_n^0) = 2d_n - d_n^2$ if $B_{n+1} = 0$. Now we consider the case $B_{n+1} = 1$. Fix $v,v'$, denote $\mu_1 = \langle U|V=v\rangle, \mu_2 = \langle U'|V'=v'\rangle$ and $\mu^0=f(\mu_1,\mu_2)$. We have
\begin{equation}\label{dn recursive lower}
\begin{aligned}
d\left(\frac{U-U'}{\sqrt{2}}\bigg|\frac{U+U'}{\sqrt{2}},V=v,V'=v'\right) &= \E_{u\sim\mu^0}\left[d\left(\frac{U-U'}{\sqrt{2}}\bigg|\frac{U+U'}{\sqrt{2}} = u,V=v,V'=v'\right)\right]\\
& = \E_{u\sim\mu^0}[d(g(\mu_1,\mu_2,u))].
\end{aligned}
\end{equation}
The next proposition gives the evolution of RID under the lower Hadamard transform.
\begin{proposition}\label{RID of g operation}
Let $X_1,X_2$ be two independent nonsingular random variables with $\E X_1^2,\E X_2^2<\infty$, and $Y_1 = (X_1+X_2)/\sqrt{2}, Y_2 = (X_1-X_2)/\sqrt{2}$. Then
\begin{equation}\label{RID of g operation-eq}
d(Y_2|Y_1) = \E_{y\sim Y_1}[d(g(\langle X_1\rangle,\langle X_2\rangle,y))] = d(X_1)d(X_2).
\end{equation}
\end{proposition}
\begin{IEEEproof}
The first equality in \eqref{RID of g operation-eq} follows from Proposition \ref{condiRID=average}. Suppose the distributions of $X_1$ and $X_2$ are given by \eqref{equal-general-initial-distribution-mixed-representation} and \eqref{general-initial-distribution-mixed-representation}, then using \eqref{g operation-concrete} we have
\begin{equation}\label{d(Y_2|Y_1=y)}
d(g(\langle X_1\rangle,\langle X_2\rangle,y)) = \left\{\begin{aligned} & 0,&\text{ if }y\in \spt(D^0),\\& \rho^1_y,&\text{ if }y\notin \spt(D^0), \end{aligned}\right.
\end{equation}
where $D^0$ is given by \eqref{f operation}, and $\rho_y^1$ is defined in \eqref{Cy1Dy1}. Since $Y_1$ has mixed representation $(\Gamma^0,C^0,D^0)$, it follows that
\begin{equation}
\begin{aligned}
\E_{y\sim Y_1}[d(g(\langle X_1\rangle,\langle X_2\rangle,y))]&=\rho^0\E[d(g(\langle X_1\rangle,\langle X_2\rangle,C^0))] + (1-\rho^0)\E[d(g(\langle X_1\rangle,\langle X_2\rangle,D^0))]\\
&=\rho^0\int_{\spt(D^0)^c} \rho_y^1\frac{F(y)}{\rho^0}dy=\int_{\R}F_3(y)dy = d(X_1)d(X_2),
\end{aligned}
\end{equation}
where $F(y)$ and $F_3(y)$ are given by \eqref{Fz}.
\end{IEEEproof}

From Proposition \ref{RID of g operation} we know that
\begin{equation}
\E_{u\sim\mu^0}[d(g(\mu_1,\mu_2,u))] = d(\mu_1)d(\mu_2) = d(U|V=v)d(U'|V'=v').
\end{equation}
Consequently, using \eqref{dn recursive lower} we have
\begin{equation}
d\left(\frac{U-U'}{\sqrt{2}}\bigg|\frac{U+U'}{\sqrt{2}},V,V'\right)= \E_{V,V'}[d(U|V=v)d(U'|V'=v')]=d_n^2.
\end{equation}
Therefore, $d_{n+1} = d(W_n^1) = d_n^2$ if $B_{n+1}=1$. This completes the proof of \eqref{recursion-dn}.

\section{Proofs of the absorption of weighted discrete entropy}\label{Proof of absorption of hat Hn}
In this section we provide the proof of Theorem \ref{absorption-of-hat-Hn}. First, we establish some preliminaries.

For a nonsingular $\langle U|V\rangle $, we define the conditional distribution process $W_n^{\langle U|V\rangle }$, with the input source specified in the superscript, to be $W_0^{\langle U|V\rangle}= \langle U|V\rangle $ and
\begin{equation}
W_n^{\langle U|V\rangle } := \langle U|V\rangle ^{B_1B_2\cdots B_n},\ \forall n\geq1,
\end{equation}
where $\{B_i\}_{i=1}^\infty$ are the Bernoulli(1/2) random variables defined in Section \ref{PMAPHT-1}. This generalizes the tree-like process $\{W_n\}_{n=0}^\infty$ by allowing arbitrary conditional distribution to be the root. For convenience, if the input source is $X$, we still denote $W_n = W^{\langle X\rangle}_n$.

For a nonsingular $\langle U|V\rangle$, we define
\begin{equation}
\begin{aligned}
&H_d(U|V=v) := H(\langle U|V=v\rangle_d),\ H_d(U|V):=\E_V[H_d(U|V=v)],\\
&K(\langle U|V\rangle) := \sup\limits_{v} |\spt(\langle U|V=v\rangle_d)|.
\end{aligned}
\end{equation}
$H_d(U|V)$ stands for the entropy of the discrete component of $\langle U|V\rangle$, and $K(\langle U|V\rangle)$ represents the largest support size of $\langle U|V=v\rangle_d$ across all possible realizations $v$. Clearly we have
\begin{equation}\label{H_hat-lower-1}
\widehat{H}(U|V) = \E_V[(1-d(U|V=v))H_d(U|V=v)]
\end{equation}
and
\begin{equation}\label{H_hat-lower-2}
H_d(U|V=v) \leq \log K(\langle U|V\rangle),\ \forall v.
\end{equation}
If we further assume $\E U^2<\infty$, then \eqref{H_hat-lower-1}, \eqref{H_hat-lower-2} and Proposition \ref{condiRID=average} implies
\begin{equation}\label{H_hat-and-d-and-K}
\widehat{H}(U|V)\leq (1 - d(U|V))\log K(\langle U|V\rangle).
\end{equation}

The next proposition provides an upper bound on the support size of discrete component generated by the Hadamard transform, which will be extensively utilized in our proof.
\begin{proposition}\label{expontial-rate-of-H}
Let $\langle U|V\rangle$ be a nonsingular conditional distribution with $K(\langle U|V\rangle)<\infty$. Then
\begin{equation}\label{upper bound on K}
K(W_n^{\langle U|V\rangle}) \leq (K(\langle U|V\rangle)+1)^{2^n}, \forall n\geq0.
\end{equation}
\end{proposition}
\begin{IEEEproof}
If $K(\langle U|V\rangle)=0$, then $\langle U|V\rangle$ is continuous hence $K(W_n^{\langle U|V\rangle})\equiv0$. Otherwise we have $K(\langle U|V\rangle)\geq 1$. The statement is proved by induction on $n$. The case $n=0$ is obvious. Suppose \eqref{upper bound on K} holds for $n=k$. Let $W_{k}^{\langle U|V\rangle} = \langle U_k|V_k\rangle$ and denote $ P_{v_k}= \langle U_k|V_k=v_k\rangle$. We have
\begin{equation}
\begin{aligned}
K(W_{k+1}^{\langle U|V\rangle}) = \left\{\begin{aligned}&\sup\limits_{v_k,v_k'} |\spt(f(P_{v_k},P_{v_k'})_d)|,&\text{ if }B_{k+1} = 0,\\
&\sup\limits_{v_k,v_k',y} |\spt(g(P_{v_k},P_{v_k'},y)_d)|,&\text{ if }B_{k+1} = 1.\end{aligned}\right.
\end{aligned}
\end{equation}
For any nonsingular distributions $\mu$ and $\nu$, from \eqref{f operation-concrete} and \eqref{g operation-concrete} we know that
\begin{equation}\label{spt-size-recursive-bound}
\begin{aligned}
|\spt(f(\mu,\nu)_d)| &\leq |\spt(\mu_d)||\spt(\nu_d)|,\\
|\spt(g(\mu,\nu,y)_d)| &\leq |\spt(\mu_d)|+|\spt(\nu_d)|,\ \forall y\in\R.
\end{aligned}
\end{equation}
It follows that
\begin{equation}\label{recursion-K}
\begin{aligned}
K(W_{k+1}^{\langle U|V\rangle})&\leq \left\{\begin{aligned}&K(W_{k}^{\langle U|V\rangle})^2,\text{ if }B_{k+1} = 0,\\
&2K(W_{k}^{\langle U|V\rangle}),\text{ if }B_{k+1} = 1.\end{aligned}\right.\\
\end{aligned}
\end{equation}
The inductive assumption implies $K(W_{k}^{\langle U|V\rangle})\leq(K(\langle U|V\rangle)+1)^{2^k}$. Since $K(\langle U|V\rangle)\geq 1$, we have
\begin{equation}
\begin{aligned}
K(W_{k+1}^{\langle U|V\rangle})&\leq 2K(W_{k}^{\langle U|V\rangle}) \vee K(W_{k}^{\langle U|V\rangle})^2\\
& \leq2(K(\langle U|V\rangle)+1)^{2^k}\vee (K(\langle U|V\rangle)+1)^{2^{k+1}} \\
&=(K(\langle U|V\rangle)+1)^{2^{k+1}},
\end{aligned}
\end{equation}
which implies that \eqref{upper bound on K} also holds for $n=k+1$.
\end{IEEEproof}

Now we present the proof of Theorem \ref{absorption-of-hat-Hn}. Before diving into the details, we first outline the proof structure as follows. We divide the absorption of weighted discrete entropy into two stages as in Fig. \ref{TPF}. The first stage consists of $m$ transforms, where the value of $m$ will be specified in \eqref{m-formula}, and the second stage contains the remaining $n-m$ transforms. Due to the Markov property of $\{W_i\}_{i=0}^\infty$, we can consider $W_n$ as the conditional distribution process initiated from $W_m$, \ie,\ $W_n \overset{d}{=} W^{W_m}_{n-m}$. In the first stage, the RID polarizes, leading to a highly continuous or highly discrete $W_m$. For the highly continuous $W_m$, we know its RID is closed to 1. Therefore, in this case we can show that $\widehat{H}_n$ approaches 0 using \eqref{H_hat-and-d-and-K} and Proposition \ref{expontial-rate-of-H}. For the highly discrete $W_m$, the proof is split into three lemmas (namely, Lemma \ref{hat-Hn-controled-by-Hn}, \ref{convergence-rate-of-Hn} and \ref{regular-condition-on-upper-bound-of-Hdn} that will be presented in the proof). Firstly, Lemma \ref{hat-Hn-controled-by-Hn} implies that we can further treat the highly discrete $W_m$ as purely discrete, which enables us to focus on $\widehat{H}(W^{\widetilde{W}_m}_{n-m})$, where $\widetilde{W}_m$ is the discrete component of $W_m$. Secondly, using martingale methods and a novel variant of EPI, we establish the convergence rate of entropy process initiated from purely discrete source in Lemma \ref{convergence-rate-of-Hn}. Since $\widetilde{W}_m$ is purely discrete, Lemma \ref{convergence-rate-of-Hn} provides a convergence rate of $\widehat{H}(W^{\widetilde{W}_m}_{n-m})$. Lastly, to apply Lemma \ref{convergence-rate-of-Hn} with various $W_m$, we show in Lemma \ref{regular-condition-on-upper-bound-of-Hdn} that $H_d(W_n)$ can be uniformly bounded with high probability by tracking the evolution of mixed entropy and Fisher information (see Section \ref{PAWDE-5} for the definitions). These three steps allow us to conclude the absorption of $\widehat{H}_n$ when $W_m$ is highly discrete.

\begin{IEEEproof}[Proof of Theorem \ref{absorption-of-hat-Hn}]
Fix $\lambda > 0$. Choose $\alpha$ and $\beta$ such that $\alpha > \lambda+2$ and $\beta \in (0,1/2)$. Let
\begin{equation}\label{m-formula}
m = \left\lceil\frac{1}{\beta}\log(\alpha n)\right\rceil.
\end{equation}
Since $m = \Theta(\log n)$, by \eqref{convgerence-rate-dn-0} and \eqref{convgerence-rate-dn-1} we know that
\begin{equation}\label{dm-convergence-rate}
\begin{aligned}
&\lim\limits_{n\rightarrow\infty}\P(d_m \leq 2^{-\alpha n}) = 1 - d(X),\ \ \  \lim\limits_{n\rightarrow\infty}\P(d_m \geq 1 - 2^{-\alpha n}) = d(X).
\end{aligned}
\end{equation}
According to the value of $d_m$, we decompose $\P(\widehat{H}_n>2^{-\lambda n})=P_{1,n}+P_{2,n}+P_{3,n}$, where
\begin{equation}
\begin{aligned}
&P_{1,n}:=\P(\widehat{H}_n>2^{-\lambda n},d_m \geq 1 - 2^{-\alpha n}),\\
&P_{2,n}:=\P(\widehat{H}_n>2^{-\lambda n},d_m \in (2^{-\alpha n} , 1 - 2^{-\alpha n})),\\
&P_{3,n}:=\P(\widehat{H}_n>2^{-\lambda n},d_m \leq 2^{-\alpha n}).
\end{aligned}
\end{equation}
In the following three parts we show $P_{i,n}\xrightarrow{n\rightarrow\infty}0$ for $i=1,2,3$, respectively.

\textbf{Part 1: }Since $\E X^2,K(\langle X\rangle) < \infty$, using \eqref{H_hat-and-d-and-K} and Proposition \ref{expontial-rate-of-H} we have
\begin{equation}
\begin{aligned}
\widehat{H}_n \leq2^n(1-d_n)\log(K(\langle X\rangle)+1).
\end{aligned}
\end{equation}
Denoting $c = (\log(K(\langle X\rangle)+1))^{-1}$, we obtain
\begin{equation}
P_{1,n} \leq \P(d_n < 1 - c2^{-(\lambda+1)n},d_m\geq1-2^{-\alpha n}).
\end{equation}
Let $S_{m,n} = \sum_{i=m+1}^n B_i$. By \eqref{recursion-dn} we know $d_n \geq d_m^{2^{S_{m,n}}}$, which is followed by
\begin{equation}
\begin{aligned}
\P(d_n < 1 - c2^{-(\lambda+1)n},d_m\geq1-2^{-\alpha n})&\leq\P((1 - c2^{-(\lambda+1)n})^{2^{-S_{m,n}}}>1-2^{-\alpha n})\\
&\overset{(a)}{\leq}\P(1 - c2^{-S_{m,n}-(\lambda+1)n}>1-2^{-\alpha n})\\
&=\P(S_{m,n} > (\alpha-\lambda-1)n+\log c),
\end{aligned}
\end{equation}
where $(a)$ holds for sufficiently large $n$ because of Bernoulli's inequality with general exponent, which states that $(1+x)^r \leq 1+rx$ for any $x>-1$ and $r\in[0,1]$. Note that $\alpha -\lambda - 1> 1 $, since we have chosen $\alpha$ to satisfy $\alpha > \lambda + 2$. Therefore, for $n$ large enough we have
\begin{equation}
\begin{aligned}
P_{1,n} &\leq \P\left(S_{m,n} > (\alpha-\lambda-1)n+\log c\right)=0
\end{aligned}
\end{equation}

\textbf{Part 2:} It follows from \eqref{dm-convergence-rate} that $P_{2,n} \leq \P(d_m\in(2^{-\alpha n} , 1 - 2^{-\alpha n}))\xrightarrow{n\rightarrow\infty}0$.

\textbf{Part 3:} In this part we show $P_{3,n}\xrightarrow{n\rightarrow\infty}0$. The following lemma allows us to further treat the low-RID $W_m$ as purely discrete due to the fast polarization rate of $d_m$.
\begin{lemma}\label{hat-Hn-controled-by-Hn}
For nonsingular $\langle U|V\rangle $ with mixed representation $(\Gamma,C,D)$, let $W^{\langle U|V\rangle }_n$ and $W_n^{\langle D|V\rangle }$ be the conditional distribution processes beginning with $\langle U|V\rangle $ and $\langle D|V\rangle $, respectively. If $\E U^2<\infty$, then for all $n\geq 0$,
\begin{equation}\label{hat-Hn-controled-by-Hn-eq}
\begin{aligned}
\widehat{H}^{\langle U|V\rangle }_n \leq \widehat{H}^{\langle D|V\rangle}_n +2^{2n}d(U|V) \log(K(\langle U|V\rangle)+1).
\end{aligned}
\end{equation}
\end{lemma}
\begin{IEEEproof}
See Section \ref{PAWDE-3}.
\end{IEEEproof}

Let $\widetilde{W}_m$ be the discrete component of $W_m$, \ie,\ if $W_m = \langle U|V\rangle$ with mixed representation $(\Gamma, C, D)$, then $\widetilde{W}_m = \langle D|V\rangle$. For all $l\geq 0$, define
\begin{equation}
\begin{aligned}
\widehat{H}^{W_m}_l := \widehat{H}(W_m^{B_{m+1}B_{m+2}\cdots B_{m+l}}),\ \ \widehat{H}^{\widetilde{W}_m}_l := \widehat{H}(\widetilde{W}_m^{B_{m+1}B_{m+2}\cdots B_{m+l}}),
\end{aligned}
\end{equation}
then we have $\widehat{H}_n = \widehat{H}^{W_m}_{n-m}$. According to Lemma \ref{hat-Hn-controled-by-Hn} and Proposition \ref{expontial-rate-of-H},
\begin{equation}\label{H_hat_n-and_H-hat_l}
\begin{aligned}
\widehat{H}_{n-m}^{W_m}\leq \widehat{H}^{\widetilde{W}_m}_{n-m} + 2^{2(n-m)}d_m\log(K(W_m)+1)\leq \widehat{H}^{\widetilde{W}_m}_{n-m} + 2^{2n-m}d_m\log(K(\langle X\rangle)+2).
\end{aligned}
\end{equation}
If $d_m \leq 2^{-\alpha n}$, then $2^{2n-m}d_m\log(K(\langle X\rangle)+2) = o(2^{-\lambda n})$ because $\alpha -2 > \lambda$ and $m = \Theta(\log n)$. Therefore, we can find $\bar{\lambda}\in(0,\lambda)$ such that
\begin{equation}\label{upper-Pen-step1}
P_{3,n} =\P(\widehat{H}^{W_m}_{n-m}>2^{-\lambda n},d_m \leq 2^{-\alpha n})\leq \P(\widehat{H}^{\widetilde{W}_m}_{n-m}>2^{-\bar{\lambda} n},d_m \leq 2^{-\alpha n})
\end{equation}
for $n$ large enough. This implies it is sufficient to focus on the entropy process initiated from $\widetilde{W}_m$.

The next lemma provides the convergence rate of entropy process initiated from discrete conditional distributions.
\begin{lemma}\label{convergence-rate-of-Hn}
Let $\langle D|V\rangle$ be a discrete conditional distribution with $H(D|V) < \infty$.
Then for any $\lambda,\epsilon>0$, there exists constants $c_1,c_2$ (only relying on $\lambda$ and $\epsilon$) such that
\begin{equation}\label{convergence-rate-of-Hn-eq}
\P(\widehat{H}_n^{\langle D|V\rangle} >2^{-\lambda n}) \leq \frac{c_1 H(D|V)}{\sqrt{n}} + \epsilon,
\end{equation}
provided by $n \geq 2H(D|V)^2 \vee c_2$.
\end{lemma}
\begin{IEEEproof}
See Section \ref{PAWDE-4}.
\end{IEEEproof}
\textbf{Remark: }According to Lemma \ref{convergence-rate-of-Hn}, the convergence rate of $\widehat{H}_n^{\langle D|V\rangle}$ is influenced by the source $\langle D|V\rangle$ only through the entropy $H(D|V)$. Consequently, the entropy processes initiated from a family of discrete conditional distributions with bounded entropy will exhibit a uniform convergence rate.

To utilize Lemma \ref{convergence-rate-of-Hn}, it is essential to ensure that $H(\widetilde{W}_m)$ can be uniformly bounded (w.r.t. the random binary sequence $B_1\dots B_m$) by a term of order $o(\sqrt{n})$. To achieve this objective, we define
\begin{equation}
H_{d,m} := H_d(W_m) = H(\widetilde{W}_m),
\end{equation}
In the subsequent lemma, we show that with high probability $H_{d,m}$ cannot increase with a super-linear rate when $d_m$ approaches 0.
\begin{lemma}\label{regular-condition-on-upper-bound-of-Hdn}
For any $\beta \in(0,1/2)$ and a sequence $\{\xi_n\}_{n=0}^\infty$ such that $\xi_n = \omega(n)$,
\begin{equation}\label{regular-condition-on-upper-bound-of-Hdn-eq}
\lim_{n\rightarrow\infty} \P(d_n\leq2^{-2^{\beta n}},H_{d,n}>\xi_n)= 0.
\end{equation}
\end{lemma}
\begin{IEEEproof}
See Section \ref{PAWDE-5}.
\end{IEEEproof}

Now choose a sequence $\{\xi_n\}_{n=0}^\infty$ such that $\xi_n = o\left(\sqrt{n}\right)$ and $\xi_n = \omega(\log n)$. We further decompose the right side of \eqref{upper-Pen-step1} into
\begin{equation}
\begin{aligned}
P_{3,n}^{(1)} &:= \P(\widehat{H}^{\widetilde{W}_m}_{n-m}>2^{-\bar{\lambda} n},d_m \leq 2^{-\alpha n},H_{d,m}\leq\xi_n),\\
P_{3,n}^{(2)} &:= \P(\widehat{H}^{\widetilde{W}_m}_{n-m}>2^{-\bar{\lambda} n},d_m \leq 2^{-\alpha n},H_{d,m}>\xi_n).
\end{aligned}
\end{equation}
Since $\xi_n = \omega(\log n) = \omega(m)$, by Lemma \ref{regular-condition-on-upper-bound-of-Hdn} we have
\begin{equation}\label{P3n-2}
P_{3,n}^{(2)}\leq\P(d_m \leq 2^{-\alpha n},H_{d,m}>\xi_n)\xrightarrow{n\rightarrow\infty}0.
\end{equation}

Concerning $P_{3,n}^{(1)}$, we can observe that $\xi_n$ acts as a uniform upper bound for $H_{d,m}$. By utilizing Lemma \ref{convergence-rate-of-Hn} and the Markov property of $\{W_n\}_{n=0}^\infty$, we deduce that for any $\epsilon >0$, there exist constants $c_1$ and $c_2$ that solely depend on $\bar{\lambda}$ and $\epsilon$, such that
\begin{equation}
\begin{aligned}
P_{3,n}^{(1)} \leq \P(\widehat{H}^{\widetilde{W}_m}_{n-m}>2^{-\bar{\lambda} n}|H_{d,m}\leq\xi_n)\leq\frac{c_1 \xi_n}{\sqrt{n-m}} + \epsilon, \ \ \text{if } n\geq m+ (\xi_n^2\vee c_2).
\end{aligned}
\end{equation}
Using $\xi_n = o(\sqrt{n})$ and $m = \Theta(\log n)$, we conclude that $\limsup\limits_{n\rightarrow\infty}P_{3,n}^{(1)}\leq \epsilon$. Since $\epsilon$ is arbitrary, we obtain $P_{3,n}^{(1)}\xrightarrow{n\rightarrow\infty}0$, which completes the proof of Theorem \ref{absorption-of-hat-Hn}.
\end{IEEEproof}

\subsection{Proof of Lemma \ref{hat-Hn-controled-by-Hn}}\label{PAWDE-3}
Our proof is based on the following observation.
\begin{proposition}\label{fd-and-gd}
Let $\mu$ and $\nu$ be two nonsingular distributions, then
\begin{align}
&f(\mu,\nu)_d = f(\mu_d,\nu_d),\label{fd}\\
&g(\mu,\nu,y)_d = g(\mu_d,\nu_d,y),\text{ if }y\in\spt(f(\mu_d,\nu_d)).\label{gd}
\end{align}
\end{proposition}
\begin{IEEEproof}
\eqref{fd} and \eqref{gd} are clearly indicated by \eqref{f operation-concrete} and \eqref{g operation-concrete}, respectively.
\end{IEEEproof}

According to \eqref{fd}, the upper Hadamard transform yields a discrete component that is equivalent to directly applying the transform to the discrete component of input distributions. Similarly, as shown in \eqref{gd}, the same rules holds for the lower Hadamard transform when the distribution generated by the upper Hadamard transform takes value from its discrete support. Therefore, if the input distribution demonstrates a high discreteness (\ie,\ is of low RID), the discrete component of the distributions generated by the Hadamard transform, will closely resemble those obtained by directly applying the transform to the discrete component of input distribution. In the following we prove Lemma \ref{hat-Hn-controled-by-Hn} based on this observation.

\begin{IEEEproof}[Proof of Lemma \ref{hat-Hn-controled-by-Hn}]
For any $n\geq 0$, let
\begin{equation}
\begin{aligned}
\{(U_i,V_i)\}_{i=1}^N\overset{\iid}{\sim}(U,V),\ \ \{(\Gamma_i, C_i, D_i)\}_{i=1}^N\overset{\iid}{\sim}(\Gamma, C, D),
\end{aligned}
\end{equation}
and $\mathbf{L} = \mathsf{H}_n U^N,\widetilde{\mathbf{L}} = \mathsf{H}_n D^N$. To prove the statement, it is equivalent to show that for all $k\in[N]$,
\begin{equation}\label{hat-Hn-controled-by-Hn-eq-equi}
\begin{aligned}
\widehat{H}(L_k|L^{k-1},\mathbf{V})\leq H(\widetilde{L}_k|\widetilde{L}^{k-1},\mathbf{V})+2^{2n}d(U|V)\log(K(\langle U|V\rangle)+1).\\
\end{aligned}
\end{equation}
For convenience, define the functions $\underline{d}(\cdot),\underline{H_d}(\cdot)$ and $\underline{\widehat{H}}(\cdot)$ as
\begin{equation}\label{functionlize}
\begin{aligned}
&\underline{d}(l^{k-1},\mathbf{v}) := d(L_k|L^{k-1}=l^{k-1},\mathbf{V}=\mathbf{v}),\\
&\underline{H_d}(l^{k-1},\mathbf{v}) := H_d( L_k|L^{k-1}=l^{k-1},\mathbf{V}=\mathbf{v}),\\
&\underline{\widehat{H}}(l^{k-1},\mathbf{v}) := \widehat{H}(L_k|L^{k-1}=l^{k-1},\mathbf{V}=\mathbf{v}) = [1-\underline{d}(l^{k-1},\mathbf{v})]\underline{H_d}(l^{k-1},\mathbf{v}).
\end{aligned}
\end{equation}
With these notations, we interpret $\underline{d}(L^{k-1},\mathbf{V})$, $ \underline{H_d}(L^{k-1},\mathbf{V})$ and $\underline{\widehat{H}}(L^{k-1},\mathbf{V})$ as random variables obtained by applying $\underline{d}(\cdot),\underline{H_d}(\cdot)$ and $\underline{\widehat{H}}(\cdot)$ to $(L^{k-1},\mathbf{V})$, respectively. Without loss of generality, let us assume
\begin{equation}
U_i = \Gamma_i C_i + (1-\Gamma_i)D_i,\ \forall i\in[N].
\end{equation}
Define the event $A = \{\Gamma_i=0,\forall i\in[N]\}$. We have
\begin{equation}
\begin{aligned}
\widehat{H}(L_k|L^{k-1},\mathbf{V}) =\E\underline{\widehat{H}}(L^{k-1},\mathbf{V})\leq\E\underline{H_d}(L^{k-1},\mathbf{V})=\E[\underline{H_d}(L^{k-1},\mathbf{V})\mathbf{1}_A]+\E[ \underline{H_d}(L^{k-1},\mathbf{V})\mathbf{1}_{A^c}].\\
\end{aligned}
\end{equation}

On the event $A$, we have $U_i = D_i,\forall i\in[N]$ and hence $L^{k-1}=\widetilde{L}^{k-1}$. To deal with this case, we extend Proposition \ref{fd-and-gd} to the Hadamard transform with arbitrary order.
\begin{proposition}\label{equi-distribution-given-all-discrete}
For any realizations $\mathbf{v},\mathbf{l}$ and any $k\in[N]$, if $l^{k-1} \in \spt(\langle\widetilde{L}^{k-1}|\mathbf{V}=\mathbf{v}\rangle)$, then
\begin{equation}\label{equi-distribution-given-all-discrete-eq}
\langle L_k|L^{k-1}= l^{k-1},\mathbf{V}=\mathbf{v}\rangle_d= \langle\widetilde{L}_k | \widetilde{L}^{k-1} = l^{k-1},\mathbf{V}=\mathbf{v}\rangle.
\end{equation}
\end{proposition}
\begin{IEEEproof}
See Appendix \ref{appendix C}.
\end{IEEEproof}

Similar to \eqref{functionlize}, define
\begin{equation}
\underline{H}(\tilde{l}^{k-1},\mathbf{v}) = \underline{H}(\widetilde{L}_k|\widetilde{L}^{k-1}=\tilde{l}^{k-1},\mathbf{V}=\mathbf{v}).
\end{equation}
If follows from Proposition \ref{equi-distribution-given-all-discrete} that
\begin{equation}\label{two-parts-by-A-1}
\begin{aligned}
\E[\underline{H_d}(L^{k-1},\mathbf{V})\mathbf{1}_A] = \E[\underline{H}(\widetilde{L}^{k-1},\mathbf{V})\mathbf{1}_A]\le\E[\underline{H}(\widetilde{L}^{k-1},\mathbf{V})] = H(\widetilde{L}_k\big|\widetilde{L}^{k-1},\mathbf{V}).
\end{aligned}
\end{equation}
By \eqref{H_hat-lower-2} and Proposition \ref{expontial-rate-of-H} we know that
\begin{equation}\label{two-parts-by-A-2}
\E[\underline{H_d}(L^{k-1},\mathbf{V})\mathbf{1}_{A^c}]\leq 2^n\log(K(\langle U|V\rangle+1)\P(A^c).
\end{equation}
Consequently, the proof is completed by
\begin{equation}
\begin{aligned}
\P(A^c) = 1 - \prod\limits_{i=1}^{N}\P(\Gamma_i=0)=1 - \prod\limits_{i=1}^{N}\E[\P(\Gamma_i=0|V_i)] = 1 - (1-d(U|V))^{2^n}\leq 2^nd(U|V),
\end{aligned}
\end{equation}
where the last inequality follows from Bernoulli's inequality, which states that $(1+x)^r\geq 1+rx$ for any $x\geq-1$ and $r\geq1$.
\end{IEEEproof}
\subsection{Proof of Lemma \ref{convergence-rate-of-Hn}}\label{PAWDE-4}
For convenience, denote $H_n = \widehat{H}_n^{\langle D|V\rangle}$. We first show that $H_n$ converges to 0 almost surely, and then establish the convergence rate as in \eqref{convergence-rate-of-Hn-eq}.

We use similar arguments as in \cite{HAT2012} to show $H_n\xrightarrow{a.s.}0$. Let $\mathcal{F}_n = \sigma(\{B_i\}_{i=1}^n)$ be the $\sigma$-algebra generated by $\{B_i\}_{i=1}^n$. Denote $W_n^{\langle D|V\rangle} = \langle D_n|V_n\rangle$. Let $(D'_n,V'_n)$ be an independent copy of $(D_n,V_n)$. Then we have $H_n = H(D_n|V_n)$ and
\begin{equation}\label{recursion-of-Hn}
H_{n+1}=\left\{
\begin{aligned}
&H(D_n + D'_n|V_n,V'_n),&\text{ if }B_{n+1} = 0,\\
&H(D_n - D'_n|D_n + D'_n,V_n,V'_n),&\text{ if }B_{n+1} = 1.
\end{aligned}
\right.
\end{equation}
By the chain rule of entropy,
\begin{equation}
\begin{aligned}
\E[H_{n+1}|\mathcal{F}_n] &=\frac{1}{2}[H(D_n + D'_n|V_n,V'_n)+H(D_n - D'_n|D_n + D'_n,V_n,V'_n)]\\
&=\frac{1}{2}H(D_n, D'_n|V_n,V'_n)=H(D_n|V_n) = H_n,
\end{aligned}
\end{equation}
which implies $\{H_n,\mathcal{F}_n\}_{n=0}^\infty$ is a positive martingale. The martingale convergence theorem \cite[Theorem 5.2.8]{Durrett2010} implies that $H_n$ converges almost surely to a limit $H_\infty$. To determine $H_\infty$, we examine the difference
\begin{equation}\label{martingale-difference}
|H_{n+1}-H_n|=H(D_n + D'_n|V_n,V'_n) - H(D_n|V_n),
\end{equation}
To deal with \eqref{martingale-difference}, we prove the following lemma, which generalizes the result of \cite[Theorem 3]{HAT2014}.
\begin{lemma}\label{conditional EPI}
There exists an increasing continuous function $L(x)$ with $L(x) = 0 \Leftrightarrow x = 0$ such that for all discrete $\langle D|V\rangle$ with $H(D|V) < \infty$,
\begin{equation}\label{conditional EPI a1}
H(D+D'|V,V') - H(D|V) \geq L(H(D|V)),
\end{equation}
where $(D',V')$ is an independent copy of $(D,V)$. In addition,
\begin{equation}\label{conditional EPI a2}
L(x) \geq C x^4,\ \ \forall x< 4,
\end{equation}
where $C$ is an absolute constant.
\end{lemma}
\begin{IEEEproof}
See Appendix \ref{appendix D}.
\end{IEEEproof}
\textbf{Remark:} Compared with \cite[Theorem 3]{HAT2014}, our contribution lies in two aspects. On the one hand, we weaken the condition in \cite[Theorem 3]{HAT2014} that both $D$ and $V$ are required to be discrete with finite support, while we only need the conditional distribution $\langle D|V\rangle$ to be discrete. On the other hand, we provide a polynomial lower bound on the function $L(x)$ when $x$ is small. We prove \eqref{conditional EPI a1} by decomposing $H(D|V) = \E_V[H(D|V=v)]$ according to the value of $H(D|V=v)$, and prove \eqref{conditional EPI a2} by a careful estimation based on \cite[Theorem 2]{HAT2014}. The detailed proof are shown in Appendix \ref{appendix D}.

By \eqref{martingale-difference} and Lemma \ref{conditional EPI} we have
\begin{equation}\label{lower-bound-on-martingale-difference}
|H_{n+1} - H_n| \geq L(H_n).
\end{equation}
Using the continuity of $L(x)$ we obtain
\begin{equation*}
0 \leq L(H_\infty) \overset{a.s.}{=} \lim\limits_{n\rightarrow\infty} L(H_n) \leq \lim\limits_{n\rightarrow\infty}|H_{n+1} - H_n| \overset{a.s.}{=} 0.
\end{equation*}
This implies $H_\infty \overset{a.s.}{=}0$ since $L(x) = 0$ if and only if $x=0$.

Next we prove \eqref{convergence-rate-of-Hn-eq} to establish the convergence rate of $H_n$. To accomplish this, we present two lemmas that capture the fundamental aspects of the proof. The first lemma gives the decay rate of the probability that $H_n$ has not reached a small value within $n$ steps.
\begin{lemma}\label{random walk}
For any $\kappa \in (0,1)$, define
\begin{equation}
\tau_\kappa := \inf\{n\geq0:H_n\leq \kappa\}
\end{equation}
to be the first time $H_n$ hits $(0,\kappa]$. Then there exist absolute constants $\tilde{c}_1$ and $\tilde{c}_2$ (independent of $\langle D|V\rangle$) such that
\begin{equation}\label{large entropy}
\P(\tau_\kappa > n)\leq \kappa^{-8}(\tilde{c}_1+\tilde{c}_2H(D|V))n^{-1/2},
\end{equation}
provided that $n\geq H(D|V)^2$.
\end{lemma}
\begin{IEEEproof}
See Appendix \ref{appendix E}.
\end{IEEEproof}
\textbf{Remark:} Since $|H_{n+1} - H_n| \geq L(H_n) \geq L(\kappa)$ when $n < \tau_\kappa$, $H_n$ behaves like a \textit{random walk} with lower-bounded step length during this period. Therefore, we can consider $\tau_\kappa$ as the first hitting time of a random walk (hence $\tau_\kappa$ is a stopping time). This enables us to apply martingale methods with stopping time to derive \eqref{large entropy}. The proof of Lemma \ref{random walk} is presented in Appendix \ref{appendix E}.

The second lemma is a novel variant of EPI for discrete random variables, which is used to establish the dynamics of $H_n$.
\begin{lemma}\label{new EPI}
Let $X,Y$ be independent discrete random variables over $\R$. If $H(X),H(Y) \leq 1$, then
\begin{align}\label{new EPI 1}
H(X+Y) \geq (1 -\delta)(H(X) +H(Y))- 6\delta,
\end{align}
where $\delta = h_2^{-1}(H(X))+h_2^{-1}(H(Y))$.
\end{lemma}
\begin{IEEEproof}
See Appendix \ref{appendix F-A}.
\end{IEEEproof}
\textbf{Remark:} It is readily apparent that the entropy of the sum of two independent random variables $H(X+Y)$, satisfies the relationship $H(X) + H(Y) \geq H(X+Y) \geq (H(X)+H(Y))/2$. The problem at the core of EPI concerns the gap $H(X+Y) - (H(X)+H(Y))/2$, an area of research that has been extensively pursued (\eg, \cite{HAT2014},\cite{TAO2010}). Lemma \ref{new EPI} introduces a novel variant of EPI, which provides an estimate for the difference $H(X)+H(Y) - H(X+Y)$. This estimate demonstrates that when $H(X)$ and $H(Y)$ are sufficiently small, the difference $H(X)+H(Y) - H(X+Y)$ is no greater than $O(h_2^{-1}(H(X)) + h_2^{-1}(H(Y)))$. This provides valuable insights into the dynamics of $H_n$.

Based on Lemma \ref{new EPI}, we can derive the following corollary, which provides an upper bound on the evolution of $H_n$ when $H_n$ is small.
\begin{corollary}\label{single step upper bound on Hn when it is small}
For any $\gamma > 0$, there exists $s =s(\gamma)> 0$ such that if $H_n \leq s$, then
\begin{equation}\label{small entropy}
H_{n+1} \leq \left\{\begin{aligned}&2H_n,\ \ \text{if}\ B_{n+1} = 0,\\&\gamma H_n,\ \ \text{if}\ B_{n+1} = 1.  \end{aligned}\right.
\end{equation}
\end{corollary}
\begin{IEEEproof}
See Appendix \ref{appendix F-B}.
\end{IEEEproof}
\textbf{Remark:} \eqref{small entropy} provides the dynamics of $H_n$, which plays an essential role in the analysis of convergence rate. By Corollary \ref{single step upper bound on Hn when it is small}, we can conclude that the effect of lower Hadamard transform is more significant than the upper Hadamard transform when $H_n$ is small.

Before presenting the detailed proof of \eqref{convergence-rate-of-Hn-eq}, we first explain the main idea behind. The convergence of $H_n$ can be divided into two phases. In the first phase $H_n> \kappa$, where $\kappa$ is a small constant specified in the following proofs. Since $H_n$ is a bounded-below martingale, it behaves like a random walk with lower-bounded step length, which implies $H_n$ hits $(0,\kappa]$ eventually. We use Lemma \ref{random walk} to estimate the tail probability of the first hitting time. Once $H_n$ hits $(0,\kappa]$, the second phase begins and $H_n$ is absorbed to 0 exponentially fast due to Corollary \ref{single step upper bound on Hn when it is small}.
\begin{IEEEproof}[Proof of \eqref{convergence-rate-of-Hn-eq}] Fix $\lambda,\epsilon>0$. Let $\gamma = 2^{-(4\lambda +2)}$ and $s = s(\gamma)$ such that \eqref{small entropy} holds. For any $\kappa\in(0,s\wedge 1)$, define $\tau_\kappa = \inf\{n\geq 0:\ H_n\leq \kappa\}$. We have
\begin{equation}\label{Thm3.1 a1}
\begin{aligned}
\P(H_n \leq 2^{-\lambda n}) \geq \sum\limits_{j\leq n/2}\P(H_n \leq 2^{-\lambda n}, \tau_\kappa = j).
\end{aligned}
\end{equation}
Fix integer $j\in[0,n/2]$. Define $\widetilde{H}_{0} = \kappa$ and
\begin{equation}
\widetilde{H}_{n} := \left\{\begin{aligned}& 2 \widetilde{H}_{n-1}, \ \ \text{if }B_{j+n} = 0 \\
& \gamma \widetilde{H}_{n-1},\ \ \text{if }B_{j+n} = 1\end{aligned}\right. ,\forall n\geq1.
\end{equation}
Let $E = \{\widetilde{H}_{n} \leq s,\ \forall n\geq0\}$. By Corollary \ref{single step upper bound on Hn when it is small} we know that
\begin{equation}\label{decomposed-by-tau-control}
E\cap\{\tau_\kappa = j\}\subset \{\widetilde{H}_{n} \geq H_{n+j},\ \forall n\geq0\}.
\end{equation}
As a result,
\begin{equation}\label{rate-Hn-a2}
\begin{aligned}
\P(H_n \leq 2^{-\lambda n}, \tau_\kappa = j) &\geq \P(\{H_{n}\leq 2^{-\lambda n}\}\cap E\cap\{\tau_\kappa = j\})\\
&\overset{(a)}{\geq}\P(\{\widetilde{H}_{n-j}\leq 2^{-\lambda n}\}\cap E\cap\{\tau_\kappa = j\})\\
&\overset{(b)}{=}\P(\{\widetilde{H}_{n-j} \leq 2^{-\lambda n} \}\cap E)\P(\tau_\kappa = j)\\
&\geq(\P(\widetilde{H}_{n-j} \leq 2^{-\lambda n}) + \P(E) - 1)\P(\tau_\kappa = j),
\end{aligned}
\end{equation}
where $(a)$ follows from \eqref{decomposed-by-tau-control}, and $(b)$ holds because $\{\widetilde{H}_{n-j} \leq \epsilon 2^{-\lambda n} \}\cap E \in \sigma(\{B_k\}_{k\geq j+1})$, which is independent of $\{\tau_\kappa = j\}\in\mathcal{F}_j$. Let $S_{n} = \sum_{i=j+1}^{j+n} B_i$, then
\begin{equation}
\widetilde{H}_{n} = \kappa\gamma^{S_{n}}2^{n-S_{n}},\ \forall n\geq0.
\end{equation}
Choose $\alpha$ such that $\alpha\in(\frac{2\lambda+1}{4\lambda+3},\frac{1}{2})$, define $A_\alpha = \{S_{n-j} \geq \alpha (n-j)\}$. By the law of large numbers, there exists $c_0 = c_0(\alpha,\epsilon)$ such that $\P(A_\alpha) \geq 1 - \epsilon/4$ for $n \geq c_0$. On the event $A_\alpha$ we have
\begin{equation}
\widetilde{H}_{n-j} = \kappa \gamma^{S_{n-j}}2^{n-j - S_{n-j}} \leq (\gamma^\alpha2^{1-\alpha})^{n-j}\overset{(a)}{\leq}2^{-2\lambda(n-j)}\overset{(b)}{\leq}2^{-\lambda n},
\end{equation}
where $(a)$ holds because $\alpha>\frac{2\lambda+1}{4\lambda+3}$ and $\gamma = 2^{-4\lambda+2}$, and $(b)$ follows from $j\leq n/2$. As a result, we have $A_\alpha \subset \{\widetilde{H}_{n-j}\leq2^{-\lambda n}\}$ and hence
\begin{equation}\label{rate-Hn-a3}
\P(\widetilde{H}_{n-j} \leq 2^{-\lambda n}) \geq \P(A_\alpha) \geq 1 - \epsilon/4,\forall n\geq c_0.
\end{equation}
Next we consider the lower bound on $\P(E)$. We have
\begin{equation}
\begin{aligned}
\P(E) = 1 - \P(E^c)&\geq 1 - \sum_{n\geq0}\P(\widetilde{H}_n > s)\\
&= 1 - \sum_{n\geq0}\P(\kappa\gamma^{S_{n}}2^{n-S_{n}} > s) \\
&=  1 - \sum_{n\geq0} \P\left(S_n < \frac{n-\log (s/\kappa)}{1 - \log\gamma}\right)\\
& =  1 - \sum_{n > \log( s/\kappa)} \P\left(S_n < \frac{n-\log (s/\kappa)}{1 - \log\gamma}\right)\\
&\geq  1 - \sum_{n > \log( s/\kappa)} \P(S_n < n/3),
\end{aligned}
\end{equation}
where the last inequality follows from $1/(1 - \log\gamma) < 1/3$ and $s > \kappa$. Using the Chernoff's bound \cite[p.531]{Gallager1968}, we obtain
\begin{equation}\label{Thm3.1 a4}
\begin{aligned}
\P(E)\geq1- \sum_{n > \log (s/\kappa)} 2^{-n(1 - h_2(1/3))}.
\end{aligned}
\end{equation}
This implies we can take $\kappa = \kappa(s,\epsilon)$ small enough such that $\P(E) \geq 1 - \epsilon /4$. Now by \eqref{rate-Hn-a2}, \eqref{rate-Hn-a3} and \eqref{Thm3.1 a4},
\begin{equation}\label{Thm3.1 a5}
\begin{aligned}
\P(H_n \leq 2^{-\lambda n}, \tau_\kappa = j)\geq \left(1 - \epsilon/2\right)\P(\tau_\kappa = j),\ \forall n\geq c_0,j\leq n/2.
\end{aligned}
\end{equation}
Consequently, from \eqref{Thm3.1 a1} we obtain
\begin{equation}
\P(H_n \leq 2^{-\lambda n})\geq (1-\epsilon/2)\P(\tau_\kappa\leq n/2),\forall n\geq c_0.
\end{equation}
By Lemma \ref{random walk}, for $n\geq 2H(D|V)^2$ we have
\begin{equation}
\P(\tau_\kappa \leq n/2) \geq 1-\sqrt{2}\kappa^{-8}(\tilde{c}_1+\tilde{c}_2H(D|V))n^{-1/2}.
\end{equation}
Let $c_1 = \sqrt{2}\tilde{c}_2\kappa^{-8}$ and take $c_2=c_2(\epsilon,\kappa)>c_0$ such that $\sqrt{2}\tilde{c}_1n^{-\frac{1}{2}}\kappa^{-8} < \epsilon/2$ for $n\geq c_2$. Then
\begin{equation}
\begin{aligned}
\P(H_n \leq 2^{-\lambda n})\geq(1-\epsilon/2)(1-\epsilon/2-c_1H(U|V)n^{-1/2})\geq\ 1 - c_1H(D|V)n^{-1/2} - \epsilon,
\end{aligned}
\end{equation}
provided that $n \geq 2H(D|V)^2 \vee c_2$.
\end{IEEEproof}

\subsection{Proof of Lemma \ref{regular-condition-on-upper-bound-of-Hdn}}\label{PAWDE-5}
We aim to show that $H_{d,n}$ is uniformly bounded by any sequence of order $\omega(n)$ with high probability when $d_n$ is small. To this end, we define the mixed entropy (see Definition \ref{mixed entropy}) for nonsingular distributions. We prove that the mixed entropy exhibits supermartingale properties under the Hadamard transform, thereby providing an upper bound for the combination of the entropy of discrete and continuous components. Furthermore, we analyze the evolution of Fisher information under the Hadamard transform, which enables us to bound the entropy of the continuous component from below by a linear function. These two steps allow us to prove the desired result. In the following, we first establish the preliminaries of mixed entropy and Fisher information, and then we prove Lemma \ref{regular-condition-on-upper-bound-of-Hdn}.

\textbf{Mixed Entropy:}
The concept of entropy is well-defined for discrete distributions using the discrete entropy $H(\cdot)$, and for continuous distributions using the differential entropy $h(\cdot)$. However, the definition of entropy for general probability distributions remains unclear. Extensive researches have been conducted in this area \cite{Renyi1959,CAR2022,KPRH2016,NPS2006}. Building upon these existing studies, we propose the mixed entropy for nonsingular distributions.
\begin{definition} [Mixed Entropy] \label{mixed entropy}
Let $X$ be a nonsingular random variable with mixed representation $(\Gamma, C,D)$. Denote $\rho = d(X)$. The mixed entropy of $X$ is defined to be
\begin{equation}\label{mixed entropy a1}
\mathcal{H}(X)  := \rho h(C) + (1-\rho)H(D) + h_2(\rho).
\end{equation}
The conditional mixed entropy of nonsingular $\langle U|V\rangle$ is defined to be $\mathcal{H}(U|V) := \E_V[\mathcal{H}(U|V=v)]$.
\end{definition}
\textbf{Remark: } Definition \ref{mixed entropy} aligns with several entropy definitions proposed in previous studies for general probability distributions. Specifically, the mixed entropy $\mathcal{H}(\cdot)$ corresponds to (i) the $\rho$-dimensional entropy defined in \cite{Renyi1959}; (ii) the dimensional rate bias (DRB) defined in \cite[Definition 9]{CAR2022}, and (iii) the entropy defined for mixed-pairs in \cite[Definition 2.3]{NPS2006}.

The lemma presented below shows that the mixed entropy satisfies a form of ``chain rule'' under the Hadamard transform.
\begin{lemma}\label{chain rule of mixed entropy}
Let $X_1$ and $X_2$ be independent nonsingular random variables such that $|\mathcal{H}(X_1)|,|\mathcal{H}(X_2)|<\infty$. Denote $Y_1 = (X_1+X_2)/\sqrt{2}$, $Y_2 = (X_1-X_2)/\sqrt{2}$ and $\rho_1=d(X_1),\rho_2=d(X_2)$, then
\begin{equation}\label{chain rule a1}
\begin{aligned}
\mathcal{H}(Y_1) + \mathcal{H}(Y_2|Y_1) = \mathcal{H}(X_1) + \mathcal{H}(X_2)- \frac{\rho_1(1-\rho_2)+\rho_2(1-\rho_1)}{2}.
\end{aligned}
\end{equation}
\end{lemma}
\begin{IEEEproof}
See Appendix \ref{appendix G}.
\end{IEEEproof}
\textbf{Remark:} By setting $\rho_1=\rho_2=0$ and $\rho_1=\rho_2=1$ in Lemma \ref{chain rule of mixed entropy}, the results are compatible with the chain rule of discrete and differential entropy.

Define the mixed entropy process to be
\begin{equation}
\mathcal{H}_n := \mathcal{H}(W_n),\ \forall n\geq 0,
\end{equation}
where $W_n = \langle X\rangle ^{B_1\cdots B_n}$. Utilizing Lemma \ref{chain rule of mixed entropy}, it is easy to show that for any $n\geq 0$,
\begin{equation}\label{supermartingale-mixed-entropy-1}
\E[\mathcal{H}_{n+1}|\mathcal{F}_n] = \mathcal{H}_n - d_n(1-d_n)/2\leq \mathcal{H}_n,
\end{equation}
which indicates that $\{\mathcal{H}_n,\mathcal{F}_n\}_{n=0}^\infty$ is a supermartingale. As a result, we conclude that $\E \mathcal{H}_n \leq \E\mathcal{H}_0 = \mathcal{H}(X)<\infty$. This provides an upper bound for the average mixed entropy under the Hadamard transform.

\textbf{Fisher Information: }For a continuous random variable $X$ with density $\varphi(x)$, the Fisher information of $X$ is defined as \cite[Chapter 17.7]{Cover2006}
\begin{equation}
J(X) := \int_{\R} \frac{\varphi'(x)^2}{\varphi(x)}dx.
\end{equation}
Since $J(\cdot)$ is a functional of the density $\varphi$, we refer to $J(\varphi)$ and $J(X)$ interchangeably. The conditional Fisher information of continuous $\langle U|V\rangle$ is defined to be
\begin{equation}
J(U|V) := \E_V[J(U|V=v)].
\end{equation}
For nonsingular distributions, we define the weighted Fisher information as follows.
\begin{definition}[Weighted Fisher Information]
Let $X$ be a nonsingular random variable. The weighted Fisher information of $X$ is defined to be
\begin{equation}
\hat{J}(X) := d(X)J(\langle X\rangle_c).
\end{equation}
The conditional weighted Fisher information of nonsingular $\langle U|V\rangle$ is defined to be $\hat{J}(U|V) := \E_V[\hat{J}(U|V=v)]$.
\end{definition}

The following lemma establishes upper bounds on the evolution of weighted Fisher information under the upper and lower Hadamard transform.
\begin{lemma}\label{Fisher-information-basic-Hadamard-transform}
Let $X_1$ and $X_2$ be independent nonsingular random variables with $\hat{J}(X_1),\hat{J}(X_2)<\infty$. Let $Y_1 = (X_1+X_2)/\sqrt{2}$ and $Y_2 = (X_1-X_2)/\sqrt{2}$, then
\begin{equation}
\begin{aligned}
\hat{J}(Y_1)\leq \frac{5}{2}(\hat{J}(X_1) + \hat{J}(X_2)),\ \ \hat{J}(Y_2|Y_1)\leq\frac{1}{2}(\hat{J}(X_1)+\hat{J}(X_2)).
\end{aligned}
\end{equation}
\end{lemma}
\begin{IEEEproof}
See Appendix \ref{appendix H}.
\end{IEEEproof}

Define the weighted Fisher information process to be
\begin{equation}
\hat{J}_n := \hat{J}(W_n),\ \forall n\geq 0.
\end{equation}
Using Lemma \ref{Fisher-information-basic-Hadamard-transform}, it is not hard to see that $\hat{J}_{n+1}\leq 5\hat{J}_n$. Consequently, we have
\begin{equation}\label{rate-of-Fisher-process}
\hat{J}_n \leq \hat{J}(X) 5^n,\ \forall n\geq0.
\end{equation}
This indicates that the weighted Fisher information process increases at most exponentially fast.

For a nonsingular $\langle U|V\rangle$, define the weighted differential entropy of $\langle U|V\rangle$ to be
\begin{equation}
\hat{h}(U|V) := \E_V[d(U|V=v)h(\langle U|V=v\rangle_c)].
\end{equation}
The next lemma reveals the connection between the weighted Fisher information and weighted differential entropy.
\begin{lemma}\label{Fisher-information-EPI}
Let $\langle U|V\rangle$ be a nonsingular conditional distribution with $\E U^2<\infty$ and $\hat{J}(U|V)<\infty$, then
\begin{equation}
\hat{h}(U|V) \geq  \frac{d(U|V)}{2}\log\left(\frac{2\pi\e\,d(U|V)}{\hat{J}(U|V)}\right).
\end{equation}
\end{lemma}
\begin{IEEEproof}
See Appendix \ref{appendix I}.
\end{IEEEproof}

Define the weighted differential entropy process to be
\begin{equation}
\hat{h}_n := \hat{h}(W_n),\ \forall n\geq0.
\end{equation}
Since $\E X^2<\infty$ and $J(\langle X\rangle_c)<\infty$, using \eqref{rate-of-Fisher-process} and Lemma \ref{Fisher-information-EPI} we obtain
\begin{equation}
\begin{aligned}
\hat{h}_n&\geq\frac{d_n}{2}\log(2\pi\e\,d_n\hat{J}_n^{-1})\geq\frac{d_n}{2}\log(2\pi\e\,d_n\hat{J}(X)^{-1}5^{-n})\geq-\frac{\log5}{2}n+\frac{d_n}{2}\log(2\pi\e\hat{J}(X)^{-1}\,d_n),\ \forall n\geq0.
\end{aligned}
\end{equation}
Note that $d_n\log(d_n)$ is bounded since $d_n\in[0,1]$. Therefore, we can find a positive constant $T$ only depending on $\hat{J}(X)$ such that
\begin{equation}\label{lower-h-hat-n}
\hat{h}_n \geq -T n ,\ \forall n \geq 1.
\end{equation}
This provides a lower bound for the entropy of continuous component under the Hadamard transform.

\textbf{Proof of Lemma \ref{regular-condition-on-upper-bound-of-Hdn}: }
By \eqref{H_hat-lower-1}, \eqref{H_hat-lower-2} and Proposition \ref{expontial-rate-of-H} we obtain
\begin{equation}
\widehat{H}_n \geq H_{d,n} - 2^nd_n\log(K(\langle X\rangle)+1).
\end{equation}
If $d_n \leq 2^{-2^{\beta n}}$ and $H_{d,n}> \xi_n$, then
\begin{equation}\label{lower-H-hat-n}
\widehat{H}_n \geq \xi_n - 2^{n-2^{\beta n}}\log(K(\langle X\rangle)+1) =\xi_n - o(1).
\end{equation}
As a result, on the event $\{d_n\leq 2^{-2^{\beta n}},H_{d,n}> \xi_n\}$ we have
\begin{equation}
\mathcal{H}_{n} \overset{(a)}{\geq} \widehat{H}_n+\hat{h}_n \overset{(b)}{\geq} \xi_n - o(1)- Tn,
\end{equation}
where $(a)$ follows from the definition of mixed entropy, and $(b)$ holds because of \eqref{lower-h-hat-n} and \eqref{lower-H-hat-n}. Since $\xi_n = \omega(n)$, it follows that
\begin{equation}
\{d_n\leq 2^{-2^{\beta n}},H_{d,n}> \xi_n\}\subset\{\mathcal{H}_n>\xi_n/2\}
\end{equation}
for $n$ large enough. Consequently, it is sufficient to show $\P(\mathcal{H}_n>\xi_n/2)\xrightarrow{n\rightarrow\infty}0$. Note that
\begin{equation}
\mathcal{H}_n +Tn \geq \hat{h}_n +Tn \geq 0,\ \forall n\geq1.
\end{equation}
By Markov's inequality,
\begin{equation*}
\P(\mathcal{H}_{n} > \xi_n/2) \leq \P(\mathcal{H}_{n} +T n> \xi_n/2) \leq \frac{\E\mathcal{H}_{n} + Tn}{\xi_n /2}.
\end{equation*}
From \eqref{supermartingale-mixed-entropy-1} we know that $\E\mathcal{H}_{n} \leq \mathcal{H}(X)< \infty$. It follows from $\xi_n = \omega(n)$ that
\begin{equation}
\lim\limits_{n\rightarrow\infty}\P(\mathcal{H}_{n} > \xi_n/2)=0,
\end{equation}
which completes the proof.

\section{Numerical Experiments} \label{Numerical Experiments}
\begin{figure}[!t]
\centerline{\includegraphics[width=0.55\textwidth,trim = 104 26 109 27,clip]{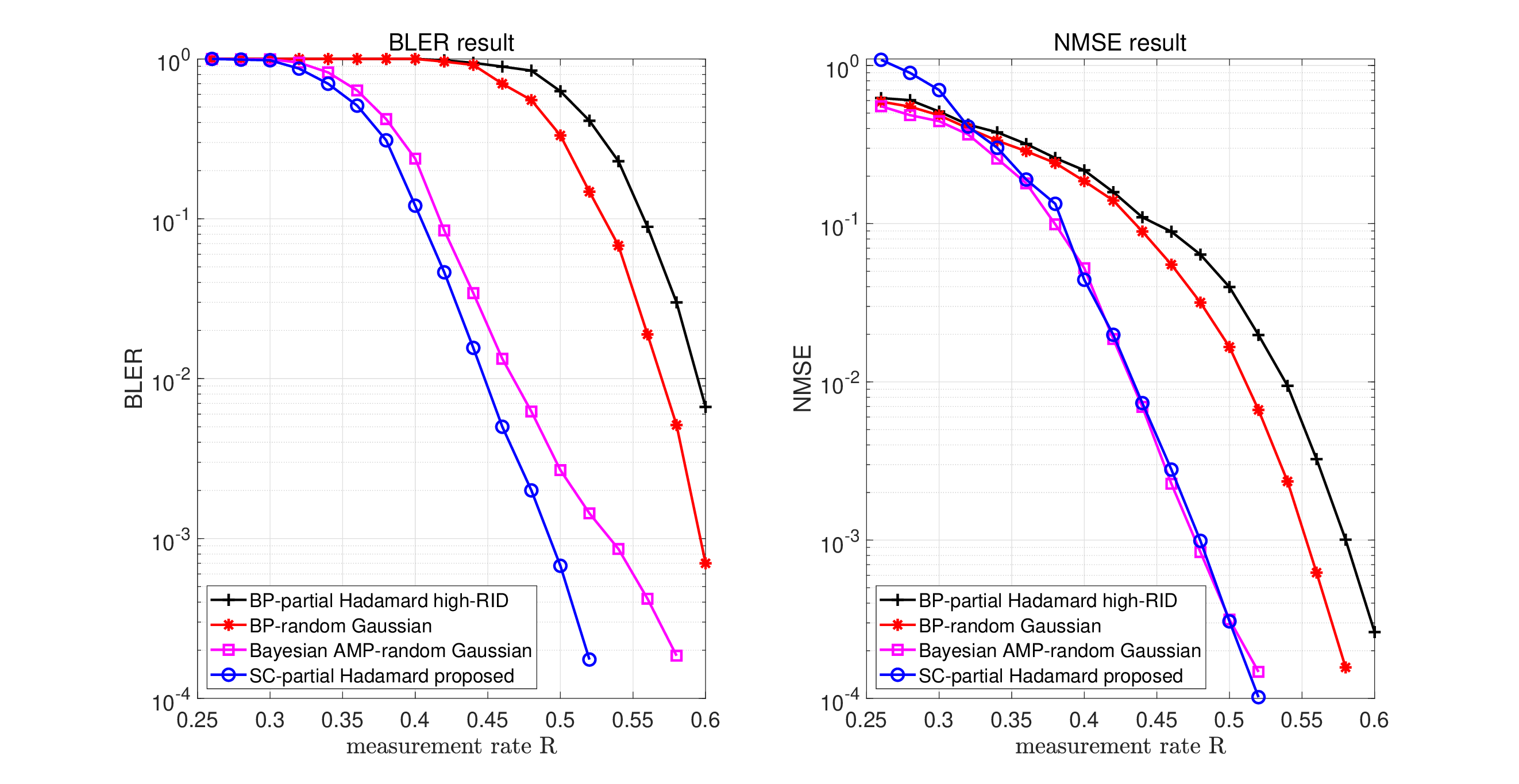}}
\caption{The BLER (left) and NMSE (right) under different measurement rate. Signal length $N =512$, source distribution $P_X = 0.8\delta_0 + 0.2\mathcal{N}(0,1)$. }
\label{BLER-NMSE}
\end{figure}

We further evaluate the performance of the proposed Hadamard compression and analog SC decoder on noiseless compressed sensing. Let the signal length $N=512$ and the source distribution $P_X = 0.8\delta_0 + 0.2\mathcal{N}(0,1)$. That is, $X=0$ with probability 0.8 and $X$ distributes as standard Gaussian with probability 0.2. As a result, $d(X) = 0.2$ and around $80\%$ components of $\mathbf{X}$ are exactly 0. The performance is gauged by the normalized mean square error (NMSE) given by
\begin{equation}
\text{NMSE} = \frac{\E \|\widehat{\mathbf{X}}-\mathbf{X}\|^2}{\E\|\mathbf{X}\|^2},
\end{equation}
and the ``block error rate (BLER)'' which is defined as
\begin{equation}
\text{BLER} = \P(\|\widehat{\mathbf{X}}-\mathbf{X}\|^2 > \eta\E\|\mathbf{X}\|^2 ),
\end{equation}
that is, the recovery fails if the reconstruction error larger than the tolerance $\eta$ which is set to $10^{-2}$. The proposed analog SC decoder might fail without any output, in this case we set the output to the least square estimate $\hat{\mathbf{x}} = \mathsf{H}_\mathcal{A}^\top (\mathsf{H}_\mathcal{A}\mathsf{H}_\mathcal{A}^\top)^{-1}\mathbf{z}$. The simulation results of BLER and NMSE under different measurement rate are presented in Fig. \ref{BLER-NMSE}. The proposed scheme is compared with the classic Basis Pursuit (BP) algorithm and the Bayesian AMP algorithm \cite{DMM2010}.  Both BP and AMP employ random Gaussian measurements for recovery. Furthermore, the partial Hadamard matrix chosen from high-RID rows, as proposed in \cite{HA2017}, is also taken into account for the BP decoding. To ensure the convergence of AMP, we initialize with 10 random values and select the optimal one as the finial output. Different from both BP and AMP, the SC decoding is non-iterative and involves only $N\log N$ operations.

Due to the incorporation of prior information, the performance of the SC and Bayesian AMP decoder is better than that of the BP algorithm. While the SC decoder exhibits only a marginal improvement in BLER over the AMP decoder under low measurement rate, its superiority becomes larger as the measurement rate increases. Notably, the SC decoder requires much lower measurement rate to achieve the same BLER compared to BP. In fact, the BLER curve of SC decoder starts decreasing at $R = 0.32$, while the curve of BP starts at $R = 0.44$. The reason is that for an $N$-dimensional signal with $k$ non-zero components, BP requires $O(2k\log (N/k))$ measurements for precise reconstruction \cite{Donoho2006}, while $O(k)$ measurements are enough for the SC decoder as proved in Theorem \ref{main theorem}. It is observed from the NMSE result that, under moderate measurement rate the SC decoder outperforms BP and maintains comparable performance to AMP. However, under more stringent low measurement conditions ($R \leq 0.3$), the NMSE of SC decoder is comparatively higher. This issue may be attributed to the error propagation of SC decoder, which leads to severely degraded reconstruction if the recovery fails. This observation also suggests a lack of robustness in the analog SC decoder, which is an important challenge to be addressed in future research.

\section{Conclusion and Future Works} \label{Conclusion}
In this paper, we study the lossless analog compression via the polarization-based framework. We prove that for nonsingular source, the error probability of MAP estimation polarizes under the Hadamard transform. Based on the analog polarization, we propose the partial Hadamard matrices and the corresponding analog SC decoder. The measurement matrix is deterministically constructed by selecting rows from the Hadamard matrix, and the SC decoder for binary polar codes is generalized for the reconstruction of analog signal. Thanks to the polarization of error probability, we prove that the proposed scheme achieves the information-theoretical limit for lossless analog compression. We define the weighted discrete entropy to quantify the uncertainty of general random variable, and show that the weighted discrete entropy vanishes under the Hadamard transform, which generalizes the absorption phenomenon of discrete entropy. As the key step of the proof, we develop a novel variant of entropy power inequality and use martingale methods with stopping time to obtain the convergence rate of the discrete entropy process. The performance of the proposed approach is numerically evaluated on the noiseless compressed sensing. The simulation result shows that the proposed method yields superior performance than the Basis Pursuit reconstruction, and maintains comparable performance to the Bayesian AMP algorithm.

In future investigations, it is important to develop computationally efficient methods to approximate the analog $f$ and $g$ operations. Despite the fact that only $N\log N$ operations are required in the analog SC decoder, each operation involves computing a convolution of probability measure over the real line or a conditional distribution, which remains a computationally intensive task. Additionally, enhancing the robustness of the analog SC decoder is another critical issue, particularly in view of its potential application in practical scenarios.

\appendices
\section{Proofs of Proposition \ref{condiRID=average} and Proposition \ref{upper bound of Pe MAX}}\label{appendix A}
\subsection{Proof of Proposition \ref{condiRID=average}}\label{appendix A-A}
Since $\E U^2<\infty$, we have $\E[\log(1+|U|)]<\infty$. From \cite[Proposition 1]{WV2010} we obtain $H(\lfloor U\rfloor|V)\leq H(\lfloor U\rfloor)<\infty$. It was shown in \cite[Eq.(11)]{Renyi1959} that
$H(\lfloor nX\rfloor/n)\leq H(\lfloor X\rfloor) + \log n$ for any random variable $X$. Thus for $n\geq2$,
\begin{equation}
\frac{H(\lfloor nU\rfloor/n|V=v)}{\log n} \leq H(\lfloor U\rfloor|V=v) + 1,\  \forall v.
\end{equation}
Since $H(\lfloor U\rfloor|V)<\infty$, by dominated convergence theorem we obtain
\begin{equation}
\begin{aligned}
d(U|V) = \lim\limits_{n\rightarrow\infty}\E_{V}\left[\frac{H(\lfloor nU\rfloor/n|V=v)}{\log n}\right]= \E_{V}\left[\lim\limits_{n\rightarrow\infty}\frac{H(\lfloor nU\rfloor/n|V=v)}{\log n}\right]= \E_{V}[d(U|V=v)].
\end{aligned}
\end{equation}

\subsection{Proof of Proposition \ref{upper bound of Pe MAX}}\label{appendix A-B}
Suppose $P_X = \sum_{i\geq1} p_i\delta_{x_i}$ with $p_1\geq p_2\geq\cdots$. Then $x^* = x_1$ and $\P(X= x^*) = p_1$. It is followed by
\begin{equation}\label{HDh2p1}
H(X) \geq H(\mathbf{1}_{\{X=x_1\}})  = h_2(p_1).
\end{equation}
On the other hand,
\begin{equation}
1 \geq H(X) = \sum_i -p_i\log p_i \geq -\log p_1\sum_i p_i = -\log p_1,
\end{equation}
which implies $p_1 \geq 1/2$. Combining with \eqref{HDh2p1} we obtain
\begin{equation}
p_1 = 1 - h_2^{-1}(h_2(p_1)) \geq 1 - h_2^{-1}(H(X)).
\end{equation}
Note that $h_2^{-1}(x)\leq x$ for any $x\in[0,1]$. The proof is completed by $p_1 \geq 1 - h_2^{-1}(H(X))\geq 1 - H(X)$.

\section{Proof of Proposition \ref{prop-g-operation}}\label{appendix B}
We show that $Q(y,A)$ meets the two conditions in Definition \ref{RCD}. Clearly condition $2)$ is satisfied. To verify condition $1)$, it is enough to show that
\begin{equation}\label{RCDaim}
\E[\phi(Y_1)Q(Y_1,A)] = \E[\phi(Y_1)\mathbf{1}_{\{Y_2\in A\}}]
\end{equation}
holds for any Boreal set $A$ and measurable function $\phi$. We prove it by thoroughly calculating the left side of \eqref{RCDaim}.

Using the distribution of $Y_1$, we have
\begin{equation}\label{S1+S2}
\begin{aligned}
\E[\phi(Y_1)Q(Y_1,A)] &=(1-\rho^0)\E[\phi(D^0)Q(D^0,A)] + \rho^0\E[\phi(C^0)Q(C^0,A)].\\
\end{aligned}
\end{equation}
On the one hand,
\begin{equation}\label{D a1}
\begin{aligned}
(1-\rho^0)\E[\phi(D^0)Q(D^0,A)] &=\P(\Gamma_1=0,\Gamma_2=0)\E[\phi(D^0)\P(\bar{D}_1-\bar{D}_2\in A|D^0)]\\
&=\P(\Gamma_1=0,\Gamma_2=0)\E[\phi(D^0)\mathbf{1}_{\{\bar{D}_1-\bar{D}_2\in A\}}]\\
&\overset{(a)}{=}\E[\phi(D^0)\mathbf{1}_{\{\bar{D}_1-\bar{D}_2\in A\}}\mathbf{1}_{\{\Gamma_1=0,\Gamma_2=0\}}]\\
&=\E[\phi(Y_1)\mathbf{1}_{\{Y_2\in A\}}\mathbf{1}_{\{\Gamma_1=0,\Gamma_2=0\}}],
\end{aligned}
\end{equation}
where $(a)$ follows from the independence between $\Gamma_1,\Gamma_2$ and $D_1,D_2$. On the other hand, since $C^0\sim F(y)/\rho^0$, where $F(y)$ is given by \eqref{Fz}, we obtain
\begin{equation}\label{I1+I2}
\begin{aligned}
\rho^0\E[\phi(C^0)Q(C^0,A)]&=\rho^0\int_\R \phi(y) Q(y,A) \frac{F(y)}{\rho^0} dy\\
&=\underbrace{\int_\R  \phi(y)\P(D^1_y\in A)(F_1(y) +F_2(y))dy}_{I_1} + \underbrace{\int_\R  \phi(y)\P(C^1_y\in A)F_3(y)dy}_{I_2}.\\
\end{aligned}
\end{equation}
Similar to \eqref{D a1} we can deduce that
\begin{equation}\label{I2}
\begin{aligned}
I_2 = \rho_1\rho_2\E[\phi(\bar{C}_1+\bar{C}_2)\P(\bar{C}_1-\bar{C}_2\in A|\bar{C}_1+\bar{C}_2)]=\E[\phi(Y_1)\mathbf{1}_{\{Y_2\in A\}} \mathbf{1}_{\{\Gamma_1 = 1,\Gamma_2 = 1\}}].
\end{aligned}
\end{equation}
For the term $I_1$, using \eqref{Cy1Dy1} we have
\begin{equation}\label{I1ss}
\begin{aligned}
I_1 &= \int_\R \phi(y)F_1(y)\P(\bar{D}_1-\bar{C}_2\in A|\bar{D}_1+\bar{C}_2=y)dy+\int_\R\phi(y)F_2(y)\P(\bar{C}_1-\bar{D}_2\in A|\bar{C}_1+\bar{D}_2=y)dy\\
& = (1-\rho_1)\rho_2\E[\phi(\bar{D}_1+\bar{C}_2)\P(\bar{D}_1-\bar{C}_2\in A|\bar{D}_1+\bar{C}_2)]+\rho_1(1-\rho_2)\E[\phi(\bar{C}_1+\bar{D}_2)\P(\bar{C}_1-\bar{D}_2\in A|\bar{C}_1+\bar{D}_2)]\\
&=\E[\phi(Y_1)\mathbf{1}_{\{Y_2\in A\}} (\mathbf{1}_{\{\Gamma_1 = 0,\Gamma_2 = 1\}}+\mathbf{1}_{\{\Gamma_1 = 1,\Gamma_2 = 0\}})].
\end{aligned}
\end{equation}
Now combing \eqref{S1+S2}--\eqref{I1ss} we obtain
\begin{equation}
\begin{aligned}
\E[\phi(Y_1)Q(Y_1,A)] = \sum\limits_{a,b\in\{0,1\}} \E[\phi(Y_1)\mathbf{1}_{\{Y_2\in A\}} \mathbf{1}_{\{\Gamma_1 = a,\Gamma_2 = b\}}] = \E[\phi(Y_1)\mathbf{1}_{\{Y_2\in A\}}].
\end{aligned}
\end{equation}

\section{Proof of Proposition \ref{equi-distribution-given-all-discrete}}\label{appendix C}
We prove \eqref{equi-distribution-given-all-discrete-eq} by induction on $n$. For $n = 0$, we have $L_1 = U_1$ and $\widetilde{L}_1 = D_1$, hence \eqref{equi-distribution-given-all-discrete-eq} obviously holds. Now suppose \eqref{equi-distribution-given-all-discrete-eq} holds for $ n = k $. When $n = k+1$, denote $N_k = 2^k$ and
\begin{equation}
\begin{aligned}
\mathbf{S} = \mathsf{H}_{k} U^{N_k},\ \ \mathbf{T} = \mathsf{H}_{k} U_{N_k+1}^{2N_k},\ \ \widetilde{\mathbf{S}} = \mathsf{H}_{k} D^{N_k},\ \ \widetilde{\mathbf{T}} = \mathsf{H}_{k} D_{N_k+1}^{2N_k}.
\end{aligned}
\end{equation}
According to the recursive structure of Hadmard matrices, for any $i\in[N_k]$ we have
\begin{equation}\label{Z-S-T}
\begin{aligned}
L_{2i-1} = \frac{1}{\sqrt{2}}(S_i +T_i ),\ \ L_{2i} = \frac{1}{\sqrt{2}}(S_i -T_i ),\ \
\widetilde{L}_{2i-1} = \frac{1}{\sqrt{2}}(\widetilde{S}_i +\widetilde{T}_i ),\ \ \widetilde{L}_{2i} = \frac{1}{\sqrt{2}}(\widetilde{S}_i -\widetilde{T}_i ).
\end{aligned}
\end{equation}
Let $\mathbf{s}= \frac{1}{\sqrt{2}}(l_e+l_o)$ and $\mathbf{t}= \frac{1}{\sqrt{2}}(l_e-l_o)$, where $l_e$ and $l_o$ are the sub-vectors of $l^{2N_k}$ with even and odd indices, respectively. For each $i\in[N_k]$, denote
\begin{equation}
\begin{aligned}
&\mu = \langle S_i|S^{i-1} = s^{i-1},V^{N_k}=v^{N_k}\rangle,&\nu = \langle T_i|T^{i-1} = t^{i-1},V_{N_k+1}^{2N_k}=v_{N_k+1}^{2N_k}\rangle,\\
&\tilde{\mu} = \langle \widetilde{S}_i|\widetilde{S}^{i-1} = s^{i-1},V^{N_k}=v^{N_k}\rangle, &\tilde{\nu} = \langle\widetilde{T}_i|\widetilde{T}^{i-1} = t^{i-1},V_{N_k+1}^{2N_k}=v_{N_k+1}^{2N_k}\rangle.
\end{aligned}
\end{equation}
It is followed by
\begin{equation}
\begin{aligned}
&\langle L_{2i-1}|L^{2i-2}= l^{2i-2},\mathbf{V}=\mathbf{v}\rangle= f(\mu,\nu),\ \ \quad\ \langle \widetilde{L}_{2i-1}|\widetilde{L}^{2i-2}= l^{2i-2},\mathbf{V}=\mathbf{v}\rangle= f(\tilde{\mu},\tilde{\nu}),\\
&\langle L_{2i}|L^{2i-1}= l^{2i-1},\mathbf{V}=\mathbf{v}\rangle= g(\mu,\nu,l_{2i-1}),\ \ \langle \widetilde{L}_{2i}|\widetilde{L}^{2i-1}= l^{2i-1},\mathbf{V}=\mathbf{v}\rangle= g(\tilde{\mu},\tilde{\nu},l_{2i-1}).
\end{aligned}
\end{equation}
By the inductive assumption, we have $\mu_d = \tilde{\mu}$ and $\nu_d = \tilde{\nu}$. Then Proposition \ref{fd-and-gd} implies that
\begin{equation}\label{k=2i-1}
\begin{aligned}
\langle L_{2i-1}|L^{2i-2}= l^{2i-2},\mathbf{V}=\mathbf{v}\rangle_d = f(\mu,\nu)_d = f(\mu_d,\nu_d) =f(\tilde{\mu},\tilde{\nu}) =\langle \widetilde{L}_{2i-1}|\widetilde{L}^{2i-2}= l^{2i-2},\mathbf{V}=\mathbf{v}\rangle .
\end{aligned}
\end{equation}
Since $l^{2i-1} \in \spt(\langle\widetilde{L}^{2i-1}|\mathbf{V}=\mathbf{v}\rangle)$, we have
\begin{equation}
\begin{aligned}
l_{2i-1} \in \spt(\langle\widetilde{L}_{2i-1}|\widetilde{L}^{2i-2}=l^{2i-2},\mathbf{V}=\mathbf{v}\rangle)=\spt(f(\tilde{\mu},\tilde{\nu})) = \spt(f(\mu_d,\nu_d)).
\end{aligned}
\end{equation}
Using Proposition \ref{fd-and-gd} again we obtain
\begin{equation}\label{k=2i}
\begin{aligned}
\langle L_{2i}|L^{2i-1}= l^{2i-1},\mathbf{V}=\mathbf{v}\rangle_d = g(\mu,\nu,l_{2i-1})_d = g(\mu_d,\nu_d,l_{2i-1})=g(\tilde{\mu},\tilde{\nu},l_{2i-1}) = \langle \widetilde{L}_{2i}|\widetilde{L}^{2i-1}= l^{2i-1},\mathbf{V}=\mathbf{v}\rangle .
\end{aligned}
\end{equation}
Since \eqref{k=2i-1} and \eqref{k=2i} holds for all $i\in[N_k]$, we conclude that \eqref{equi-distribution-given-all-discrete-eq} also holds for $n= k+1$.

\section{Proof of Lemma \ref{conditional EPI}}\label{appendix D}
Our proof is based on \cite[Theorem 2]{HAT2014}, which proves an EPI for integer-valued random variables. It is not hard to extend it to all discrete random variables. For completeness we restate their result in the next lemma.
\begin{lemma}[\cite{HAT2014}, Theorem 2]\label{g(c,d)}
For any independent discrete random variables $X,Y$ with $H(X),H(Y)<\infty$,
\begin{equation}
H(X+Y) - \frac{H(X)+H(Y)}{2} \geq q(H(X),H(Y)),
\end{equation}
where $q: \R^+\times \R^+ \rightarrow \R^+$ is given by
\begin{equation*}
\begin{aligned}
q(c,d) &= \frac{1}{2}\min\limits_{x,y\in[0,1]}\{\left(dx-h_2(x)+ cy-h_2(y)\right)\vee l(x,y)\},\\
l(x,y)& = \min\limits_{(a,b)\in T(x,y)} \frac{\log e}{8}\left((1-x)^2a^2+(1-y)^2b^2\right),\\
T(x,y) &= \{a,b\geq 0:\ a\geq(4y-2)^+,b\geq(4x-2)^+,a+b\geq 2 -x -y\}.
\end{aligned}
\end{equation*}
In addition, $q(c,d)$ is continuous, doubly-increasing (\textit{i.e.}, fix one of $c$ or $d$, $q(c,d)$ is an increasing function w.r.t. the other variable), and $q(c,d) = 0 \Leftrightarrow (c,d) = (0,0)$.
\end{lemma}

In the following we first prove \eqref{conditional EPI a1}, and then we show the polynomial lower bound \eqref{conditional EPI a2}.
\subsection{Proof of \eqref{conditional EPI a1}}
Let $m(z) = \inf_{x+y\geq z/4} q(x,y)$, where $q(x,y)$ is given in Lemma \ref{g(c,d)}. Define $L(0) = 0$ and
\begin{equation}
L(x) = \inf\left\{z\in[0,m(x)]:\ 2z \geq \frac{x}{2}\sqrt{1 - \frac{z}{m(x)}}\right\}, \forall x > 0.
\end{equation}
It is easy to verify that $L(x)$ is increasing and continuous when $x>0$, and $L(x) = 0$ if and only if $x=0$. Note that $0\leq L(x)\leq m(x)$ and $m(x)\xrightarrow{x\rightarrow0}0$, which implies $L(x)$ is also continuous at $x=0$.

For convenience, let us denote
\begin{equation}\label{abbreviate-of-H(X|y)}
\begin{aligned}
&H(D|v) = H(D|V=v), H(D'|v') = H(D'|V'=v'),H(D+D'|v,v') = H(D+D'|V=v,V'=v').
\end{aligned}
\end{equation}
Let $x = H(D|V)$, $\triangle = H(D+D'|V,V') - H(D|V)$ and
\begin{equation}
\triangle(v,v') = H(D+D'|v,v') - \frac{H(D|v) + H(D'|v')}{2}.
\end{equation}
To prove \eqref{conditional EPI a1}, it is equivalent to show $\triangle \geq L(x)$. Let $A = \{v:H(D|v)\leq x/4\}$. By Lemma \ref{g(c,d)} we obtain
\begin{equation}
\triangle(v,v') \geq q\left(H(D|v),H(D'|v')\right) \geq m(x),\ \forall (v,v')\in(A\times A)^c.
\end{equation}
It follows that
\begin{equation}\label{AA-1}
\begin{aligned}
\triangle =\E[\triangle(V,V')]\geq \E[\triangle(V,V')\mathbf{1}_{\{(V,V')\in(A\times A)^c\}}]\geq m(x)(1 - \P(V\in A)^2).
\end{aligned}
\end{equation}
If $\triangle \geq m(x)$, then $\triangle\geq m(x) \geq L(x)$, the proof is done. Otherwise we have
\begin{equation}
\P(V\in A) \geq \sqrt{1 - \frac{\triangle}{m(x)}}.
\end{equation}
On the other hand, since $H(D+D'|v,v') \geq H(D|v)$ for all $v,v'$, we have
\begin{equation}
\begin{aligned}
\triangle &\geq \E[\triangle(V,V')\mathbf{1}_{\{V\in A^c,V'\in A\}}]\\
&\geq \E_{(v,v')\sim(V,V')}\left[\left(\frac{H(D|v)}{2} - \frac{x}{8}\right)\mathbf{1}_{\left\{v\in A^c,v'\in A\right\}}\right]\\
&\geq \P(V\in A)\left(\frac{1}{2}\E_{v\sim V}[H(D|v)\mathbf{1}_{\{v\in A^c\}}] - \frac{x}{8}\right),
\end{aligned}
\end{equation}
As a result,
\begin{equation}
\E_{v\sim V}[H(D|v)\mathbf{1}_{\{v\in A^c\}}] \leq \frac{2\triangle}{\P(V\in A)} + \frac{x}{4}\leq \frac{2\triangle}{\sqrt{1 - \frac{\triangle}{m(x)}}} + \frac{x}{4}.
\end{equation}
It is followed by
\begin{equation}
\begin{aligned}
x = \E_{v\sim V}[H(D|v)\mathbf{1}_{\{v\in A\}}] + \E_{v\sim V}[H(D|v)\mathbf{1}_{\{v\in A^c\}}] \leq \frac{x}{4} + \frac{2\triangle}{\sqrt{1 - \frac{\triangle}{m(x)}}} + \frac{x}{4},
\end{aligned}
\end{equation}
which implies
\begin{equation}
2\triangle \geq \frac{x}{2}\sqrt{1 - \frac{\triangle}{m(x)}}.
\end{equation}
Finally, by the definition of $L(x)$ we have $\triangle \geq L(x)$. This completes the proof of \eqref{conditional EPI a1}.
\subsection{Proof of \eqref{conditional EPI a2}}
Initially, we show two propositions that give lower bounds on $l(x,y)$ and $q(c,d)$ given in Lemma \ref{g(c,d)}.
\begin{proposition}\label{lower bound of l}
$l(x,y) \geq \frac{\log e}{256}(2-x-y)^2, \forall x,y\in[0,1]$.
\end{proposition}
\begin{IEEEproof}
Consider 4 different cases.

\textbf{Case 1:} $x\in[0,\frac{3}{4}], y\in[0,\frac{3}{4}]$. By Cauchy-Schwarz inequality and the definition of $T(x,y)$,
\begin{equation}
(1-x)^2a^2+(1-y)^2b^2 \geq \frac{(a+b)^2}{\frac{1}{(1-x)^2}+\frac{1}{(1-y)^2}}\geq \frac{(2-x-y)^2}{32}.
\end{equation}
Thus $l(x,y) \geq \frac{\log e}{256}(2-x-y)^2$.\vspace{0.3em}

\textbf{Case 2:} $x\in[0,\frac{3}{4}], y\in[\frac{3}{4},1]$. Since $a \geq (4y-2)^+ \geq 1$ and $(1-x)^2 \geq (1-y)^2$,
\begin{equation}
\begin{aligned}
l(x,y) \geq \frac{\log e}{8}(1-x)^2 \geq \frac{\log e}{16}\left((1-x)^2+(1-y)^2\right) \geq \frac{\log e}{32} (2-x-y)^2 \geq \frac{\log e}{256}(2-x-y)^2.
\end{aligned}
\end{equation}

\textbf{Case 3:} $x\in[\frac{3}{4},1], y\in[0,\frac{3}{4}]$. The proof is similar to case 2.

\textbf{Case 4:} $x\in[\frac{3}{4},1], y\in[\frac{3}{4},1]$. Since $a,b\geq 1$,
\begin{equation}
\begin{aligned}
l(x,y) \geq \frac{\log e}{8}((1-x)^2+(1-y)^2) \geq \frac{\log e}{16}(2-x-y)^2 \geq \frac{\log e}{256}(2-x-y)^2.
\end{aligned}
\end{equation}
\end{IEEEproof}
\begin{proposition}\label{polynomial lower h}
If $c+d\leq 1$, then $q(c,d) \geq C_0(c+d)^4$, where $C_0 = \frac{\log e /256}{2((\log e / 256) + 2\sqrt{2} + 1)^4}$.
\end{proposition}
\begin{IEEEproof}
It is easy to show $h_2(x) \leq  2\sqrt{1-x}$ for any $x\in[0,1]$. By Proposition \ref{lower bound of l} we have
\begin{equation}
\begin{aligned}
q(c,d) \geq \frac{1}{2} \min\limits_{x,y\in[0,1]} &\left\{t(x,y)\vee s(x,y)\right\},
\end{aligned}
\end{equation}
where $t(x,y) = dx - 2\sqrt{1-x} + cy - 2\sqrt{1-y}$ and $s(x,y) = \alpha(2-x-y)^2, \alpha = \log e / 256$. Since $t(x,y)$ is doubly-increasing and $s(x,y)$ is doubly-decreasing, and both $t$ and $s$ are continuous with $t(0,0) < s(0,0)$ and $t(1,1)>s(1,1)$, we conclude that the minimizer of $t(x,y)\vee s(x,y)$ over $[0,1]$ must satisfy $t(x,y) = s(x,y)$. As a result,
\begin{equation}
\begin{aligned}
q(c,d) \geq \frac{1}{2}\min\{s(x,y): x,y\in[0,1],t(x,y) = s(x,y)\}.
\end{aligned}
\end{equation}
Let $u = 1 - x$ and $v = 1- y$ then we obtain
\begin{equation}
\begin{aligned}
q(c,d)\geq\frac{\alpha}{2}\min\limits_{(u,v)\in A_{c,d}}(u+v)^2,
\end{aligned}
\end{equation}
where $A_{c,d} = \{u,v\in[0,1]: du + cv + \alpha(u+v)^2 + 2(\sqrt{u}+\sqrt{v}) = c + d\}$. Since $d+c\leq 1$, for any $(u,v)\in A_{c,d}$,
\begin{equation}
\begin{aligned}
c + d  = du + cv + \alpha(u+v)^2 + 2(\sqrt{u}+\sqrt{v})\leq \alpha(u+v)^2 + u+v + 2\sqrt2\sqrt{u+v}.
\end{aligned}
\end{equation}
If $u+v\leq 1$, then $\alpha(u+v)^2 + u+v + 2\sqrt2\sqrt{u+v}\leq (\alpha + 2\sqrt{2}+1)\sqrt{u+v}$, and hence
\begin{equation}\label{q_lower_1}
(u+v)^2\geq\frac{(c + d)^4}{(\alpha + 2\sqrt{2}+1)^4}.
\end{equation}
If $u+v> 1$, we have $\alpha(u+v)^2 + u+v + 2\sqrt2\sqrt{u+v}\leq(\alpha + 2\sqrt{2}+1)(u+v)^2$, which is followed by
\begin{equation}\label{q_lower_2}
(u+v)^2 \geq \frac{c + d}{\alpha + 2\sqrt{2}+1}.
\end{equation}
As a result, from \eqref{q_lower_1} and \eqref{q_lower_2} we obtain
\begin{equation}
(u+v)^2 \geq \left(\frac{(c + d)^4}{(\alpha + 2\sqrt{2}+1)^4}\wedge\frac{c + d}{\alpha + 2\sqrt{2}+1}\right) = \frac{(c + d)^4}{(\alpha + 2\sqrt{2}+1)^4},\ \forall (u,v)\in A_{c,d},
\end{equation}
which is followed by
\begin{equation}
\begin{small}
\begin{aligned}
q(c,d) &\geq \frac{\alpha}{2}\frac{(c + d)^4}{(\alpha + 2\sqrt{2}+1)^4}= C_0 (c+d)^4.\\
\end{aligned}
\end{small}
\end{equation}
\end{IEEEproof}

Now we are ready to prove \eqref{conditional EPI a2}. By Proposition \ref{polynomial lower h} we have $m(x) \geq C_0 (x/4)^4,\forall x < 4$. Let $C_1 = C_0/4^4$, then for any $x<4$,
\begin{equation}
\begin{aligned}
L(x) &\geq \inf\left\{z\geq 0:\ 2z \geq \frac{x}{2}\sqrt{1 - \frac{z}{C_1x^4}}\right\}\\
&= \inf\{z\geq 0:\ 16C_1x^2z^2 + z - C_1x^4\geq 0\}\\
&= \frac{\sqrt{1+64C_1^2x^6}-1}{32C_1x^2}.
\end{aligned}
\end{equation}
Since $\sqrt{1+t}-1 \geq t/3$ for any $t\in[0,3]$, we have
\begin{equation}
L(x) \geq \frac{64C_1^2x^6}{3\cdot 32C_1x^2}=C x^4,\ \forall x < 4\wedge(3/64C_1^2)^{1/6} =4,
\end{equation}
where $C = 2C_1/3$. This completes the proof of \eqref{conditional EPI a2}.

\section{Proof of Lemma \ref{random walk}}\label{appendix E}
If $H_0 = H(D|V) \leq \kappa$, there is nothing to prove. Otherwise, for any $a \geq H_0$ , define \begin{equation}
\tau = \inf\{n\geq0:\ H_n \leq \kappa\ \text{or}\ H_n \geq a\}.
\end{equation}
Clearly $\tau$ is a stopping time w.r.t. $\{\mathcal{F}_n\}_{n=1}^\infty$. Since $H_n\xrightarrow{a.s.}0$, we have $\tau<\infty,a.s.$ Denote $\eta = C^2\kappa^8$ where $C$ is the constant given in \eqref{conditional EPI a2}. We split the proof into the following propositions.
\begin{proposition}\label{subMG}
$\{H_{n\wedge \tau}^2 - \eta(n\wedge\tau),\mathcal{F}_n\}_{n\geq 0}$ is a submartingale.
\end{proposition}
\begin{IEEEproof}
Let $L(x)$ be the function given in Lemma \ref{conditional EPI}. Using \eqref{lower-bound-on-martingale-difference} we obtain
\begin{equation}
|H_{n\wedge\tau} - H_{(n-1)\wedge\tau}|=\mathbf{1}_{\{\tau\geq n\}}|H_n-H_{n-1}|\geq\mathbf{1}_{\{\tau\geq n\}}L(H_{n-1}).
\end{equation}
On the event $\{\tau\geq n\}$ we have $H_{n-1} > \kappa$, thus
\begin{equation}
\begin{aligned}
\mathbf{1}_{\{\tau\geq n\}}L(H_{n-1})\geq \mathbf{1}_{\{\tau\geq n\}}L(\kappa) \geq \mathbf{1}_{\{\tau\geq n\}}\sqrt{\eta}.
\end{aligned}
\end{equation}
By Doob's optimal stopping theorem (\cite[Theorem 5.7.4]{Durrett2010}), $\{H_{n\wedge\tau},\mathcal{F}_n\}_{n\geq0}$ is a martingale and hence
\begin{equation}
\begin{aligned}
\E[H_{n\wedge \tau}^2 - \eta(n\wedge\tau)|\mathcal{F}_{n-1}] &=H_{(n-1)\wedge\tau}^2 + \E[(H_{n\wedge\tau} - H_{(n-1)\wedge\tau})^2- \eta(n\wedge\tau)|\mathcal{F}_{n-1}]\\
&\geq H_{(n-1)\wedge\tau}^2 + \eta\E[\mathbf{1}_{\{\tau\geq n\}}- (n\wedge\tau)|\mathcal{F}_{n-1}].
\end{aligned}
\end{equation}
Finally, the desired result follows from $\mathbf{1}_{\{\tau\geq n\}}-(n\wedge\tau) = -[(n-1)\wedge\tau] \in \mathcal{F}_{n-1}$.
\end{IEEEproof}

\begin{proposition}\label{E tau}
$\E \tau \leq (\kappa^2+4aH_0)/\eta$.
\end{proposition}
\begin{IEEEproof}
By Proposition \ref{subMG},
\begin{equation}\label{B2 1}
\E[H_{n\wedge \tau}^2 - \eta(n\wedge\tau)] \geq \E[H_{0\wedge \tau}^2 - \eta(0\wedge\tau)] \geq 0.
\end{equation}
From \eqref{recursion-of-Hn} we know $H_n \leq 2H_{n-1}$, which implies $H_{n\wedge\tau} \leq 2a$. By monotone convergence theorem and dominated convergence theorem, we obtain the following limits:
\begin{equation}
\lim\limits_{n\rightarrow\infty}\E[n\wedge\tau] =\E\tau,\ \lim\limits_{n\rightarrow\infty}\E H_{n\wedge\tau}^2=\E H_\tau^2.
\end{equation}
Taking the limit as $n\rightarrow\infty$ in \eqref{B2 1}, we have $\eta\E\tau \leq \E H_{\tau}^2$. Finally, the proof is completed by
\begin{equation}
\begin{aligned}
\E H_{\tau}^2 = \E[H_{\tau}^2\mathbf{1}_{\{H_\tau \leq\kappa\}}]+ \E[H_{\tau}^2\mathbf{1}_{\{H_\tau \geq a\}}]\overset{(a)}{\leq} \kappa^2 + 4a^2\frac{\E H_\tau }{a} \overset{(b)}{=} \kappa^2 + 4aH_0,
\end{aligned}
\end{equation}
where (a) holds due to $H_\tau = \lim\limits_{n\rightarrow\infty} H_{n\wedge\tau} \leq 2a$ and Markov's inequality, and (b) follows from
\begin{equation}
\E H_\tau = \lim\limits_{n\rightarrow\infty} \E H_{n\wedge\tau} = \E H_{0\wedge\tau}=H_0.
\end{equation}
\end{IEEEproof}

Recall that $\tau_\kappa = \inf\{n\geq 0: H_n\leq \kappa\}$.
\begin{proposition} \label{B2 Prop3}
For $n\geq H_0^2$ we have
\begin{equation}\label{B2 Prop3 eq}
\P(\tau_\kappa \geq n) \leq \frac{1 + (4+C^2)H_0}{\kappa^8C^2\sqrt{n}}.
\end{equation}
\end{proposition}
\begin{IEEEproof}
Since $\{\tau_\kappa \geq n\}\subset \{\tau \geq n\}\cup \{H_\tau \geq a\}$, using Markov's inequality and Proposition \ref{E tau} we have
\begin{equation}\label{B2 2}
\begin{aligned}
\P(\tau_\kappa \geq n)\leq \P(\tau \geq n) + \P(H_\tau\geq a) \leq \frac{\E \tau}{n} + \frac{\E H_\tau}{a} \leq \frac{\kappa^2+4aH_0}{\eta n} + \frac{H_0}{a}.
\end{aligned}
\end{equation}
Since \eqref{B2 2} holds for all $a > H_0$, taking $a = \sqrt{n}$ and using $\kappa < 1$ we obtain \eqref{B2 Prop3 eq}.
\end{IEEEproof}

Finally, the statement of Lemma \ref{random walk} follows from Proposition \ref{B2 Prop3} by taking $\tilde{c}_1 = 1 / C^2$ and $\tilde{c}_2 = (4+C^2)/C^2$.

\section{Proofs of Lemma \ref{new EPI} and Corollary \ref{single step upper bound on Hn when it is small}} \label{appendix F}
\subsection{Proof of Lemma \ref{new EPI}}\label{appendix F-A}
Suppose $X\sim p = \sum_{i\geq1} p_i\delta_{x_i}$ and $Y\sim q = \sum_{j\geq1} q_j\delta_{y_j}$ with $p_1 = \max_ip_i$ and $q_1 = \max_jq_j$. Let $*$ denote the convolution of probability measures over $\R$, then $X+Y$ has the distribution $p*q$. Since translation does not change entropy, we assume $x_1 = y_1 = 0$ (Otherwise, consider $X' = X - x_1$ and $Y' = Y - y_1$). Assume $p_1,q_1 < 1$,  let \begin{equation}
\tilde{p} = \sum\limits_{i=2}^\infty\frac{p_i\delta_{x_i}}{1-p_1},\ \tilde{q} = \sum\limits_{j=2}^\infty \frac{q_j\delta_{y_j}}{1-q_1},
\end{equation}
then $p$ and $q$ can be decomposed into
\begin{equation}\label{decomposion of p,q}
p = p_1 \delta_0 + (1-p_1)\tilde{p},\ q = q_1\delta_0 + (1-q_1)\tilde{q}.
\end{equation}
Consequently, we have
\begin{equation}\label{p*q-decomposition}
\begin{aligned}
p*q &= (p_1 \delta_0 + (1-p_1)\tilde{p})*(q_1\delta_0 + (1-q_1)\tilde{q})\\
& = p_1q_1\delta_0 + p_1(1-q_1)\tilde{q} + q_1(1-p_1)\tilde{p} + (1-p_1)(1-q_1)(\tilde{p}*\tilde{q}).
\end{aligned}
\end{equation}
Define $\{T_i\}_{i=1}^4$ and $S$ to be independent random variables such that $T_1\sim \delta_0,\ T_2\sim\tilde{q},\ T_3\sim\tilde{p},\ T_4\sim\tilde{p}*\tilde{q}$ and $S$ has the distribution
\begin{equation}
\P(S=1)=p_1q_1,\ \P(S=2)=p_1(1-q_1),\ \P(S=3)=q_1(1-p_1),\ \P(S=4)=(1-p_1)(1-q_1).
\end{equation}
Clearly $T_S$ has the distribution $p*q$. Therefore,
\begin{equation}
H(X+Y) = H(p*q) = H(T_S) = H(T_S,S) - H(S|T_S).
\end{equation}
For the term $H(T_S,S)$,
\begin{equation}
\begin{aligned}
H(T_S,S) = H(T_S|S) + H(S) =\sum_{k=1}^4 \P(S=k)H(T_k|S=k) + H(S) =\sum_{k=1}^4 \P(S=k)H(T_k) + H(S).\\
\end{aligned}
\end{equation}
By Proposition \ref{upper bound of Pe MAX}, we have $p_1\geq1 - h_2^{-1}(H(p))\geq 1 - \delta$ and $q_1 \geq 1 - h_2^{-1}(H(q))\geq1-\delta$. As a result,
\begin{equation}\label{H(Y_J,J)}
\begin{aligned}
H(T_S,S)&\geq\sum_{k=1}^3 \P(S=k)H(T_k) + H(S)\\
&= p_1(1-q_1)H(\tilde{q}) + q_1(1-p_1)H(\tilde{p})+ h_2(p_1) + h_2(q_1)\\
&\geq(1 - \delta)[(1-p_1)H(\tilde{p})+ h_2(p_1)+(1-q_1)H(\tilde{q})+  h_2(q_1)]\\
&=(1 - \delta)[H(p) + H(q)].
\end{aligned}
\end{equation}

Now it is enough to show $H(S|T_S) \leq 6\delta$. By considering the conditional entropy expression, we have
\begin{equation}\label{H(J|Y_J)}
\begin{aligned}
&H(S|T_S) =\P(T_S=0)H(S|T_S=0) +\sum\limits_{t\neq 0} \P(T_S=t)H(S|T_S=t) \leq H(S|T_S=0) + 2\P(T_S \neq 0),
\end{aligned}
\end{equation}
where the final inequality holds because $H(S|T_S=t) \leq \log|\spt(S)|=\log 4 = 2$. The term $\P(T_S\neq0)$ can be bounded as
\begin{equation}\label{P(Y_J_neq_0)}
\begin{aligned}
\P(T_S\neq 0)  = 1 - \P(T_S=0) \leq 1 - p_1q_1\leq 1 - p_1 + 1 - q_1 \leq  h_2^{-1}(H(p))+h_2^{-1}(H(q))=\delta.
\end{aligned}
\end{equation}
Note that $S$ can only equal $1$ or $4$ conditioned on $T_S= 0$, since $\tilde{p},\tilde{q}$ has no probability mass at $0$. Then
\begin{equation}
H(S|T_S=0) = h_2(\P(S=1|T_S=0)).
\end{equation}
We have
\begin{equation}
\begin{aligned}
\P(S=1|T_S =0) &= \frac{\P(T_S=0|S=1)\P(S=1)}{\sum_{i=1}^4\P(T_i=0|S=i)\P(S=i)}\\
& = \frac{p_1q_1}{p_1q_1 + (1-p_1)(1-q_1)\P(T_4=0)} \\
&\geq \frac{p_1q_1}{p_1q_1 + (1-p_1)(1-q_1)}.
\end{aligned}
\end{equation}
Let
\begin{equation}
\lambda = \frac{p_1q_1}{p_1q_1 + (1-p_1)(1-q_1)}.
\end{equation}
Since $p_1 \geq 1 - h_2^{-1}(H(p)),q_1 \geq 1 - h_2^{-1}(H(q))$ and $H(p),H(q) \leq 1$, we have $p_1,q_1 \geq 1/2$. It follows that
\begin{equation}\label{lambda-new-EPI}
\lambda = \frac{1}{1 + (\frac{1}{p_1}-1)(\frac{1}{q_1}-1)} \geq \frac{1}{2}.
\end{equation}
Note that $h_2(x)$ is decreasing on $x\in[1/2,1]$, which implies
\begin{equation}\label{H(J|Y_J=0)}
\begin{aligned}
H(S|T_S=0) \leq h_2(\lambda) &\overset{(a)}{\leq} - 2(1-\lambda)\log(1-\lambda)\\
&=-\frac{2(1-p_1)(1-q_1)}{p_1q_1+(1-p_1)(1-q_1)}\log\frac{(1-p_1)(1-q_1)}{p_1q_1+(1-p_1)(1-q_1)}\\
&\overset{(b)}{\leq}-4(1-p_1)(1-q_1)\log\left[(1-p_1)(1-q_1)\right]\\
&\leq4h_2(p_1)(1-q_1) + 4h_2(q_1)(1-p_1)\\
&\leq4[H(p)h_2^{-1}(H(q))+H(q)h_2^{-1}(H(p))] \\
&\leq 4\delta,
\end{aligned}
\end{equation}
where $(a)$ holds because $-x\log x \leq -(1-x)\log(1 - x)$ for all $x\in[1/2,1]$, and $(b)$ follows from the fact that $1/2 \leq xy+(1-x)(1-y) \leq 1 $ for all $x,y\in [1/2,1)$. Finally, by  \eqref{H(J|Y_J)}, \eqref{P(Y_J_neq_0)} and \eqref{H(J|Y_J=0)} we obtain $H(S|T_S)\leq 6\delta$, which completes the proof of Lemma \ref{new EPI}.

\subsection{Proof of Corollary \ref{single step upper bound on Hn when it is small}}\label{appendix F-B}
By \eqref{recursion-of-Hn} we have $H_{n+1}\leq2H_n$ when $B_{n+1} = 0$.
If $B_{n+1} = 1$, then
\begin{equation}
\begin{aligned}
H_{n+1} = H(D_n - D'_n|D_n + D'_n,V_n,V'_n) = 2H(D_n|V_n) - H(D_n + D'_n|V_n,V'_n).
\end{aligned}
\end{equation}
Therefore, it is enough to show that for any discrete $\langle D|V\rangle$,
\begin{equation}\label{C 4}
\lim\limits_{H(D|V)\rightarrow 0} \frac{H(D+D'|V,V')}{2H(D|V)} = 1,
\end{equation}
where $(D',V')$ is an independent copy of $(D,V)$.

For convenience, let us denote $H(D|v),H(D'|v')$ and $H(D+D'|v,v')$ as in \eqref{abbreviate-of-H(X|y)}. Since $h_2^{-1}(x) = o(x)$ when $x\rightarrow0$, by Lemma \ref{new EPI} we conclude that $\forall \epsilon >0,\ \exists \delta_1 > 0$ such that
\begin{equation}
H(D+D'|v,v')\geq\left(1-\epsilon/2\right)(H(D|v)+H(D'|v'))
\end{equation}
for all $v,v'\in A:=\{v:H(D|v) \leq \delta_1\}$. It follows that
\begin{equation}
\begin{aligned}
&H(D+D'|V,V') \\
=\ &\E_{(v,v')\sim(V,V')}[H(D+D'|v,v')]\\
\geq\ &\E_{(v,v')\sim(V,V')}[(1 - \epsilon/2)(H(D|v)+H(D'|v'))\mathbf{1}_{\{v\in A,v'\in A \}}+H(D'|v')\mathbf{1}_{\{v\in A,v'\in A^c \}}+H(D|v)\mathbf{1}_{\{v\in A^c,v'\in A \}}] \\
=\ &2(1 - \epsilon/2) \P(V\in A)\E_{v\sim V}[H(D|v)\mathbf{1}_{\{v\in A\}}] + 2\P(V\in A)\E_{v\sim V}[H(D|v)\mathbf{1}_{\{v\in A^c\}}]\\
\geq\ &2(1 - \epsilon/2)\P(V\in A)H(D|V).
\end{aligned}
\end{equation}
By Markov's inequality,
\begin{equation}
\P(V\in A)  = 1 - \P_V(H(D|v) > \delta_1) \geq 1 - \frac{H(D|V)}{\delta_1}.
\end{equation}
Let $\delta = \epsilon\delta_1/2$, then $\P(V\in A) \geq 1 - \epsilon/2$ if $H(D|V) \leq \delta$. As a result, when $H(D|V) \leq \delta$ we have
\begin{equation}
H(D+D'|V,V') \geq 2(1 - \epsilon/2)^2H(D|V) \geq 2(1-\epsilon)H(D|V).
\end{equation}
On the other hand, we know $H(D+D'|V,V')\leq 2H(D|V)$, and this completes the proof of \eqref{C 4}.

\section{Proof of Lemma \ref{chain rule of mixed entropy}}\label{appendix G}
Suppose the distribution of $X_1$ and $X_2$ are given by \eqref{equal-general-initial-distribution-mixed-representation} and \eqref{general-initial-distribution-mixed-representation}. We prove the statement by a straightforward calculation using \eqref{f operation} and \eqref{g operation}. First we have
\begin{equation}\label{Af}
\begin{aligned}
\mathcal{H}(Y_1) &= \rho^0h(C^0) + (1-\rho^0)H(D^0) + h_2(\rho^0)\\
&= -\int_\R F(y)\log F(y)dy + (1-\rho_1)(1-\rho_2)H(D_1+D_2)- (1-\rho_1)(1-\rho_2)\log[(1-\rho_1)(1-\rho_2)],
\end{aligned}
\end{equation}
If $y\in \spt(D^0)$, by \eqref{g operation} we know $\mathcal{H}(Y_2|Y_1=y)= H(\bar{D}_1-\bar{D}_2|\bar{D}_1+\bar{D}_2=y)$. Therefore,
\begin{equation}\label{Ag1}
\begin{aligned}
& \E_{y\sim Y_1} [\mathcal{H}(Y_2|Y_1=y)\mathbf{1}_{\{y\in \spt(D^0)\}}]\\
=\ &(1-\rho^0)\E_{y\sim D^0}[\mathcal{H}(Y_2|Y_1=y)\mathbf{1}_{\{y\in \spt(D^0)\}}]+\rho^0\E_{y\sim C^0}[\mathcal{H}(Y_2|Y_1=y)\mathbf{1}_{\{y\in \spt(D^0)\}}]\\
=\ & (1-\rho_1)(1-\rho_2) H(\bar{D}_1-\bar{D}_2|\bar{D}_1+\bar{D}_2)\\
\end{aligned}
\end{equation}
From \eqref{g operation}, we can calculate the expectation of $\mathcal{H}(Y_2|Y_1=y)$  over $\spt(D^0)^c$ as
\begin{equation}\label{Ag2}
\begin{aligned}&\ \E_{y\sim Y_1} [\mathcal{H}(Y_2|Y_1=y)\mathbf{1}_{\{y\notin \spt(D^0)\}}]\\
= \ & \int_\R F(y) [\rho^1_yh(C^1_y) + (1-\rho^1_y)H(D^1_y) + h_2(\rho^1_y) ]dy\\
= \ &\underbrace{\int_\R F_3(y)h(C^1_y)dy}_{I_1} + \underbrace{\int_\R (F_1(y)+F_2(z))H(D^1_y)dy}_{I_2} +\underbrace{\int_\R F(y)h_2(\rho^1_y)dy}_{I_3}.
\end{aligned}
\end{equation}
Since $F_3(y)/(\rho_1\rho_2)$ is the density of $\bar{C}_1+\bar{C}_2$ and $C^1_y\sim \langle\bar{C}_1-\bar{C}_2|\bar{C}_1+\bar{C}_2=y\rangle$, then
\begin{equation}\label{Ag2-1}
\begin{aligned}
I_1= \rho_1\rho_2h(\bar{C}_1-\bar{C}_2|\bar{C}_1+\bar{C}_2).
\end{aligned}
\end{equation}
Let
\begin{equation}\label{tilde-p-and-tilde-q}
\begin{aligned}
\tilde{p}_i(y) = (1-\rho_1)\rho_2\sqrt{2}p_i\varphi_2(\sqrt{2}y-x_i), \ \ \tilde{q}_j(y) = \rho_1(1-\rho_2)\sqrt{2}q_j\varphi_1(\sqrt{2}y-y_j).
\end{aligned}
\end{equation}
Then the distribution of $D_y^1$ can be written as
\begin{equation}\label{D^1_z}
D^1_y \sim \frac{\sum\limits_i \tilde{p}_i(y) \delta_{\sqrt{2}x_i-y} +\sum\limits_j \tilde{q}_j(y) \delta_{y-\sqrt{2}y_j}}{F_1(y)+F_2(y)}.
\end{equation}
If $y\notin \spt(D^0)$, it is impossible that $\sqrt{2}x_i - y = y - \sqrt{2}y_j$ for some $i,j$. Using this and \eqref{D^1_z} we can calculate $I_2$ as
\begin{equation}\label{Ag2-2}
\begin{aligned}
I_2 &= -\int_\R \sum\limits_i \tilde{p}_i(y)\log\frac{\tilde{p}_i(y)}{F_1(y)+F_2(y)}dy -\int_\R \sum\limits_j \tilde{q}_j(y)\log\frac{\tilde{q}_j(y)}{F_1(y)+F_2(y)}dy\\
&=  -\int_\R \sum\limits_i  (1-\rho_1)\rho_2 p_i\sqrt{2}\varphi_2(\sqrt{2}y-x_i)\log\frac{(1-\rho_1)\rho_2 p_i\sqrt{2}\varphi_2(\sqrt{2}y-x_i)}{F_1(y)+F_2(y)}dy\\
& \quad -\int_\R \sum\limits_j \rho_1(1-\rho_2) q_j\sqrt{2}\varphi_1(\sqrt{2}y-y_j)\log\frac{\rho_1(1-\rho_2) q_j\sqrt{2}\varphi_1(\sqrt{2}y-y_j)}{F_1(y)+F_2(y)}dy\\
& = -(1-\rho_1)\rho_2 \log[(1-\rho_1)\rho_2] - \frac{(1-\rho_1)\rho_2}{2}+(1-\rho_1)\rho_2 [H(D_1)+h(C_2)]-\rho_1(1-\rho_2) \log[\rho_1(1-\rho_2)]\\
&\quad - \frac{\rho_1(1-\rho_2)}{2}+\rho_1(1-\rho_2) [H(D_2)+h(C_1)]+ \int_\R (F_1(y)+F_2(y))\log[F_1(y)+F_2(y)]dy.
\end{aligned}
\end{equation}
Using $\rho^1_y = F_3(y)/F(y)$, for the term $I_3$ we have
\begin{equation}\label{Ag2-3}
\begin{aligned}
I_3& =-\int_\R (F_1(y) + F_2(y))\log(F_1(y)+F_2(y))dy-\int_\R F_3(y)\log F_3(y) dy + \int_\R F(y)\log F(y) dy\\
&=  -\int_\R (F_1(y) + F_2(y))\log(F_1(y)+F_2(y))dy -  \rho_1\rho_2\log (\rho_1\rho_2) + \rho_1\rho_2 h(\bar{C}_1+\bar{C}_2)+ \int_\R F(y)\log F(y) dy.
\end{aligned}
\end{equation}
Combining \eqref{Af}--\eqref{Ag2-1}, \eqref{Ag2-2} and \eqref{Ag2-3},  after canceling out common terms and carefully manipulating the resulting expression, we ultimately arrive at
\begin{equation}
\begin{aligned}
&\mathcal{H}(Y_1)+\mathcal{H}(Y_2|Y_1) \\
=\ & \mathcal{H}(Y_1)+\E_{y\sim Y_1} [\mathcal{H}(Y_2|Y_1=y)\mathbf{1}_{\{y\in \spt(D^0)\}}]+\E_{y\sim Y_1} [\mathcal{H}(Y_2|Y_1=y)\mathbf{1}_{\{y\notin \spt(D^0)\}}]\\
=\ &\rho_1h(C_1)+(1-\rho_1)H(D_1) + h_2(\rho_1) + \rho_2h(C_2)+(1-\rho_2)H(D_2) + h_2(\rho_2)-\frac{\rho_1(1-\rho_2)+\rho_2(1-\rho_1)}{2}\\
=\ &\mathcal{H}(X_1) + \mathcal{H}(X_2) - -\frac{\rho_1(1-\rho_2)+\rho_2(1-\rho_1)}{2}.
\end{aligned}
\end{equation}

\section{Proof of Lemma \ref{Fisher-information-basic-Hadamard-transform}}\label{appendix H}
We begin with introducing some useful properties of Fisher information.
\begin{proposition}\label{H1}
Let $\{X_i\}_{i=1}^n$ be independent continuous random variables with $J(X_i)<\infty,\forall i$. For any $\{\lambda_i\}_{i=1}^n\in[0,1]$ such that $\sum_i \lambda_i^2=1$, we have
\begin{equation}
J\left(\sum\limits_{i=1}^n \lambda X_i\right) \leq \sum\limits_{i=1}^n \lambda_i^2 J(X_i).
\end{equation}
\end{proposition}
\begin{IEEEproof}
We refer to \cite{Stam1959} or \cite[Lemma 1.3]{CS1991}.
\end{IEEEproof}

\begin{proposition}\label{H2}
Suppose $\{\varphi_i\}_{i=1}^\infty$ is a sequence of density functions with $J(\varphi_i)<\infty,\forall i$. Let $\varphi = \sum\limits_i \alpha_i \varphi_i$ with $\alpha_i \geq 0$ and $\sum_i \alpha_i = 1$. Then
\begin{equation}
J(\varphi) \leq \sum\limits_i \alpha_i J(\varphi_i).
\end{equation}
\end{proposition}
\begin{IEEEproof}
Note that
\begin{equation}
\begin{aligned}
\varphi'(x)^2 = \left(\sum\limits_i \alpha_i \varphi_i'(x)\right)^2 \overset{(a)}{\leq} \left(\sum\limits_i\frac{\alpha_i^2 \varphi_i'(x)^2}{\alpha_i \varphi_i(x)}\right)\left(\sum\limits_i\alpha_i \varphi_i(x)\right)= \varphi(x)\left(\sum\limits_i\alpha_i\frac{ \varphi_i'(x)^2}{\varphi_i(x)}\right),
\end{aligned}
\end{equation}
where $(a)$ follows from Cauchy-Schwarz inequality. It follows that
\begin{equation}
\begin{aligned}
J(\varphi) = \int_\R \frac{\varphi'(x)^2}{\varphi(x)}dx \leq \int_\R \sum\limits_i\alpha_i\frac{ \varphi_i'(x)^2}{\varphi_i(x)}dx = \sum\limits_i\alpha_i\int_\R \frac{ \varphi_i'(x)^2}{\varphi_i(x)}dx = \sum\limits_i \alpha_i J(\varphi_i).
\end{aligned}
\end{equation}
\end{IEEEproof}

\begin{proposition}\label{H3}
Let $X_1$ be a continuous random variable, and $X_2$ be a discrete random variable that is independent of $X_1$. If $J(X_1)<\infty$, then for any $\lambda\in(0,1]$ we have
\begin{equation}
J(\lambda X_1 + \sqrt{1-\lambda^2} X_2) \leq \lambda^{-2}J(X_1).
\end{equation}
\end{proposition}
\begin{IEEEproof}
Suppose $\langle X_2\rangle = \sum\limits_j q_j \delta_{y_j}$. Denote by $\varphi_j(x)$ the density of $\lambda X_1 + \sqrt{1-\lambda^2} y_j$, then the density of $\lambda X_1 + \sqrt{1-\lambda^2} X_2$ is given by $\phi(x) = \sum_j q_j\varphi_j(x)$. Since $J(\varphi_j) = \lambda^{-2}J(X_1),\ \forall j$, it follows from Proposition \ref{H2} that
\begin{equation}
J(\lambda X_1 + \sqrt{1-\lambda^2} X_2) = J(\phi) \leq \sum\limits_j q_jJ(\varphi_j) = \lambda^{-2}J(X_1).
\end{equation}
\end{IEEEproof}
\begin{proposition}\label{H4}
Let $X_1$ and $X_2$ be independent continuous random variables with $J(X_1),J(X_2)<\infty$. For any $\lambda\in[0,1]$, let $Y_1 = \lambda X_1 + \sqrt{1-\lambda^2} X_2$ and $Y_2 = \sqrt{1-\lambda^2} X_1 - \lambda X_2$. Then
\begin{equation}
J(Y_2|Y_1) = (1-\lambda^2)J(X_1) + \lambda^2J(X_2).
\end{equation}
\end{proposition}
\begin{IEEEproof}
Let $\varphi_1(x)$ and $\varphi_2(x)$ be the density functions of $X_1$ and $X_2$, respectively. Then the density of $Y_1$ is given by
\begin{equation}
\varphi_{Y_1}(y) = \int_\R \varphi_1(\lambda y+\sqrt{1-\lambda^2}t)\varphi_{2}(\sqrt{1-\lambda^2}y-\lambda t)dt.
\end{equation}
The density of the conditional distribution $\langle Y_2|Y_1=y\rangle$ can be written as
\begin{equation}
\varphi_{Y_2|Y_1}(t|y) = \frac{\varphi_1(\lambda y+\sqrt{1-\lambda^2}t)\varphi_2(\sqrt{1-\lambda^2}y-\lambda t)}{\varphi_{Y_1}(y)}.
\end{equation}
As a result,
\begin{equation}
\begin{aligned}
J(Y_2|Y_1) &= \int_\R \varphi_{Y_1}(y)\int_\R \frac{(\frac{d}{dt}\varphi_{Y_2|Y_1}(t|y))^2}{\varphi_{Y_2|Y_1}(t|y)}dtdy\\
&=\int_{\R^2} \frac{(\sqrt{1-\lambda^2}\varphi'_1(u)\varphi_2(v)-\lambda \varphi'_2(v)\varphi_1(u))^2}{\varphi_1(u)\varphi_2(v)}dudv\\
&=(1-\lambda^2)J(X_1)+\lambda^2J(X_2).
\end{aligned}
\end{equation}
\end{IEEEproof}

Now we are ready to prove Lemma \ref{Fisher-information-basic-Hadamard-transform}. Suppose the distributions of $X_1$ and $X_2$ are given by \eqref{equal-general-initial-distribution-mixed-representation} and \eqref{general-initial-distribution-mixed-representation}. On the one hand,
\begin{equation}
\begin{aligned}
\hat{J}(Y_1) =\rho^0J(C^0)&\overset{(a)}{\leq} (1-\rho_1)\rho_2J(\bar{D}_1+\bar{C}_2)+ \rho_1(1-\rho_2)J(\bar{C}_1+\bar{D}_2) + \rho_1\rho_2J(\bar{C}_1+\bar{C}_2)\\
&\overset{(b)}{\leq} 2(1-\rho_1)\rho_2J(C_2) + 2\rho_1(1-\rho_2)J(C_1)+ \rho_1\rho_2(J(C_1)+J(C_2))/2\\
&\leq \frac{5}{2}(\hat{J}(X_1) + \hat{J}(X_2)),
\end{aligned}
\end{equation}
where $(a)$ follows from Proposition \ref{H2}, and $(b)$ holds due to Proposition \ref{H1} and Proposition \ref{H3}. On the other hand,
\begin{equation}
\begin{aligned}
\hat{J}(Y_2|Y_1)& = \E_{y\sim Y_1}[d(Y_2|Y_1=y)J(\langle Y_2|Y_1=y\rangle_c)] \\
&= \rho^0\E_{y\sim C^0}[d(Y_2|Y_1=y)J(\langle Y_2|Y_1=y\rangle_c)] + (1-\rho^0)\E_{y\sim D^0}[d(Y_2|Y_1=y)J(\langle Y_2|Y_1=y\rangle_c)]\\
&\overset{(a)}{=}\rho^0\int_\R \frac{F_3(y)}{F(y)}J(\bar{C}_1-\bar{C}_2|\bar{C}_1+\bar{C}_2=y)\frac{F(y)}{\rho^0}dy \\
&=\rho_1\rho_2 J(\bar{C}_1-\bar{C}_2|\bar{C}_1+\bar{C}_2)\\
&\overset{(b)}{=}\rho_1\rho_2(J(C_1) + J(C_2))/2 \\
&\leq \frac{1}{2}(\hat{J}(X_1)+\hat{J}(X_2)),
\end{aligned}
\end{equation}
where $(a)$ holds because $\langle Y_2|Y_1=y\rangle_c = \langle \bar{C}_1-\bar{C}_2|\bar{C}_1+\bar{C}_2=y\rangle$ if $y\notin \spt(D^0)$ and $d(Y_2|Y_1=y)=0$ when $y\in\spt(D^0)$, and $(b)$ follows from Proposition \ref{H4}.

\section{Proof of Lemma \ref{Fisher-information-EPI}}\label{appendix I}
Our proof is based on the following lemma.
\begin{lemma}\label{I1}
Let $X$ be a continuous random variable with $\E X^2<\infty$ and $J(X)<\infty$, then
\begin{equation}\label{I1-eq}
h(X) \geq  \frac{1}{2}\log(2\pi\e J(X)^{-1}).
\end{equation}
\end{lemma}
\begin{IEEEproof}
\eqref{I1-eq} is the corollary of EPI and de Bruijn's identity. We refer to \cite{Stam1959,CC1984,Courtade2018} for its proof and related contents.
\end{IEEEproof}

Let $(\Gamma,C,D)$ be the mixed representation of $\langle U|V\rangle$. According to Lemma \ref{I1},
\begin{equation}
\begin{aligned}
\hat{h}(U|V) &= \E_{V} [d(U|V=v)h(C|V=v)] \\
&\geq\frac{1}{2}\E_V[d(U|V=v)\log(2\pi\e J(C|V=v)^{-1})]\\
&=-\frac{d(U|V)}{2}\E_V\left[\frac{d(U|V=v)}{d(U|V)}\log((2\pi\e)^{-1} J(C|V=v))\right].
\end{aligned}
\end{equation}
Define the probability measure $\widetilde{\P}$ as
\begin{equation}
\widetilde{\P}(A) = \frac{1}{d(U|V)}\E_{V}[d(U|V=v)\mathbf{1}_{\{v\in A\}}].
\end{equation}
Clearly $\widetilde{\P}\ll P_V$ and the Randon-Nikodym derivative is given by $\frac{d\widetilde{\P}}{dP_V}(v) = \frac{d(U|V=v)}{d(U|V)}$. It follows that
\begin{equation}
\begin{aligned}
\E_V\left[\frac{d(U|V=v)}{d(U|V)}\log((2\pi\e)^{-1} J(C|V=v))\right]&= \E_{\widetilde{\P}}[\log((2\pi\e)^{-1} J(C|V=v))]\\
&\leq\log((2\pi\e)^{-1}\E_{\widetilde{\P}}[J(C|V=v)]),
\end{aligned}
\end{equation}
where the final inequality follows from Jensen's inequality. The proof is completed by
\begin{equation}
\begin{aligned}
\E_{\widetilde{\P}}[J(C|V=v)] =\frac{1}{d(U|V)}\E_V[d(U|V=v)J(C|V=v)] =\frac{\hat{J}(U|V)}{d(U|V)}.
\end{aligned}
\end{equation}

\bibliographystyle{IEEEtran}
\bibliography{reference.bib}

\end{document}